\documentclass[11pt]{preprint}              % for a regular run
\DeclareMathOperator{\dom}{Dom}
\DeclareMathOperator{\intr}{int}
\DeclareMathOperator{\rint}{rel-int}

\DeclareMathOperator{\ran}{Ran}

\DeclareMathOperator{\tangt}{Tan}
\DeclareMathOperator{\ctangt}{CoTan}
\DeclareMathOperator{\diam}{Diam}
\DeclareMathOperator{\mmd}{MMD}

\newcommand{\scparam}{\rho_*}

\newcommand{\p}{\mathbb{P}}
\newcommand{\Q}{\mathbb{Q}}
\newcommand{\one}{\mathbf{1}}
\newcommand{\Rb}{\bar{\R}}

\newcommand{\D}{\mathcal D}
\newcommand{\Yc}{\mathcal Y}
\newcommand{\Wc}{\mathcal W}

\newcommand{\txs}[1]{\hspace{1.5em} \text{#1} \hspace{1.5em}}

\newcommand{\pinner}{L_f}
\newcommand{\pouter}{\Pc_{\mu}\left(f\right)}
\newcommand{\douter}{\D_{\mu}\left(f\right)}

\renewcommand{\[}{\left[}

\usepackage{pgf}

\title{\bfseries Frank-Wolfe Methods in Probability Space}

\author{Carson Kent \\
\textsf{Stanford University} \\
\texttt{crkent@stanford.edu} \\
\and
Jose Blanchet\footnote{J. Blanchet gratefully acknowledges support
   from the Air Force Office of Scientific Research under award number
   FA9550-20-1-0397, as well as NSF grants 1915967, 1820942 and 1838576.} \\
\textsf{Stanford University}\\
\texttt{jose.blanchet@stanford.edu}\\
\and
Peter Glynn \\
\textsf{Stanford University} \\
\texttt{glynn@stanford.edu} \\
}

\date{}

\begin{document}
%%%%%%%%%%%%%%%%
\maketitle
\thispagestyle{empty}

\begin{abstract}
  We introduce a new class of Frank-Wolfe algorithms for minimizing
  differentiable functionals over probability measures. This framework can be
  shown to encompass a diverse range of tasks in areas such as artificial
  intelligence, reinforcement learning, and optimization. Concrete computational
  complexities for these algorithms are established and demonstrate that these
  methods enjoy convergence in regimes that go beyond convexity and require
  minimal regularity of the underlying functional. Novel techniques used to
  obtain these results also lead to the development of new complexity bounds and
  duality theorems for a family of distributionally robust optimization
  problems. The performance of our method is demonstrated on several
  nonparametric estimation problems.
\end{abstract}

\newpage

\section{Introduction.}\label{sec:intro}
Problems in artificial intelligence,
statistics, and optimization often find a common root as an infinite
dimensional optimization problem in the form
\begin{equation}
\inf\left\{J\left(  \mu\right)  :\mu\in\mathcal{P}\left(  \mathbb{R}^{d}\right)
\right\},\label{P}%
\end{equation}
for the space $\mathcal{P}\left( \mathbb{R}^{d}\right)$ of Borel probability
measures over $\R^{d}$. In recent years, quantitative statistical and
algorithmic treatments of these formulations have produced insights into modern
computational methods-- resulting in novel approaches to difficult, open
problems. Recent works in robust optimization \cite{BKorig, Esfahani18, Sinha18,
  Staib19}, probabilistic fairness \cite{Taskesen20, Scetbon20}, reinforcement

learning \cite{MWang20, ZhangGAN20}, and generalized adversarial networks
\cite{Mahdian19, Chu19, Chu20} highlight these gains and are linked by the
following theme: problems in the form of \eqref{P} provide access to rich
infinite dimensional structure that sidesteps brittle artifacts of finite
dimensional formulations. In this paper, we develop a Frank-Wolfe algorithm for
\eqref{P} that operates from this infinite dimensional perspective and provides
concrete convergence and complexity guarantees for a sub-family of \eqref{P}
which are well-behaved with respect to the Wasserstein distance of order 2.

Development of our Frank-Wolfe method is inspired by efforts in distributionally
robust optimization \cite{Esfahani18, BKorig, Gao16, Sinha18} which have
considered variants of \eqref{P} in the form
\begin{equation}
  \label{eq:droprob}
\sup \left\{\int f \, d\mu  : D_c(\mu, \mu_{0}) \leq \delta \right\},
\end{equation}
where $D_c(\mu, \mu_0)$ is the optimal transport cost between $\mu$ and $\mu_0$
(a reference measure) under some cost function $c$. The form of
\eqref{eq:droprob}, itself, immediately suggests the basis of an infinite
dimensional Frank-Wolfe procedure since it provides a ``linear'' objective
subject to a local, ``trust-region'' constraint-- centered at $\mu_0$. More
generally, one can even consider variants of \eqref{P} in the form
\begin{equation}
  \label{eq:psiwithot}
\inf \left\{\int f \, d\mu + \psi \left(D_c(\mu, \mu_0)\right) : \mu\in \Pc(\R^{d}) \right\},
\end{equation}
where $\psi : \bar{\R} \to \bar{\R}$ is a convex penalty function. The benefit
of this formulation is suggested by its finite dimensional analogue
\begin{equation}\label{eq:psifinite}
  \left\{\inf_{y \in \R^{d}} s^Ty + \widetilde{\psi}(y) : y \in \R^d \right\}
\end{equation}
where common instantiations of $\widetilde{\psi}$ (including powers of norms,
Bregman divergences, and indicator functions of convex sets) allow one to
express an array of first-order methods and account for a variety of non-trivial
geometries. By appropriately configuring the cost $c$ and penalty $\psi$ in
\eqref{eq:psiwithot}, similar benefits can be realized in the context of \eqref{P}.

These considerations, motivated by the extent to which \eqref{P} proliferates
data-related fields, give rise to the following investigation for this work.
First, to what extent can a Frank-Wolfe method for \eqref{P} be formulated
within the framework of \eqref{eq:droprob}-- such that quantitative bounds on
complexity and convergence can be obtained. Second, how can problems in the form
of \eqref{eq:droprob} or \eqref{eq:psiwithot} be efficiently solved-- subject to
assumptions that are compatible with an infinite-dimensional,
first-order framework?

%%% Local Variables:
%%% mode: latex
%%% TeX-master: "main"
%%% End:

\subsection{Previous work.}\label{subsec:prev}
The relevance of \eqref{eq:droprob} in distributionally robust optimization (DRO) and
mathematical finance results in notably more literature for the latter of these
issues than for the former. Indeed, \cite{Esfahani18, Kuhn19, Li19, Sinha18,
  Xie21} all highlight computational schemes for solving \eqref{eq:droprob} that
are similar in objective to this work. What makes such efforts notable and
solution of \eqref{eq:droprob} non-trivial is: without particular assumptions,
\eqref{eq:droprob} can disguise an NP-hard problem-- despite being convex in the
usual Banach sense on $\Pc(\R^{d})$. In fact, even in the case where the cost is
the squared Euclidean norm $c(x,y) = \norm{x-y}^2$ (the case of primary concern
for this work), computational trouble can lie dormant-- an artifact of
inherently difficult problems in unconstrained optimization \cite{Cheng02}.
These issues are discussed in further detail in
Section \ref{subsec:iproc}, but this should not be surprising given specters of
computational hardness dating back to early formulations of DRO \cite{Delage10}.

Such computational pitfalls are not realized in practice, however, and two
relevant approaches have emerged for removing these concerns from quantitative
analyses. First, is to consider particular instances of \eqref{eq:droprob} where
the objective and constraints are sufficiently structured to preclude
computational intractability and permit solution via methods adapted to the
provided structure. Early work with this line \cite{Goh10, Delage10,
  Wiesemann14}, has recently been supplemented by approaches \cite{Chen18CC,
  Ghosh18, Fan18, NamDuchi16, Li19, ZhangZhou20, Levy20} which focus directly on
DRO formulations from particular contexts in machine learning and operations
research. Unfortunately, the techniques offered by these efforts require
assumptions which are too restrictive for this work. These assumptions typically
relate to a specific form for the objective function or constraints in
\eqref{eq:droprob} (e.g. linear/convex functions/piecewise-convex objectives or
constraints with support or density requirements, see \cite{Hanasusanto18,
  Zheng20, Esfahani18, Kuhn19, Guan18, Bartl20, Parys} for additional examples).
In this instance, such limitations preclude their applicability since, in
general, a ``gradient object'' for a functional $J$ (see
Section \ref{prop:wassprop}) need not satisfy these
conditions. A second, more relevant, approach to perform quantitative analyses
of DRO problems \eqref{eq:droprob} is to restrict the level of robustness for
which the problem is solved. In the context of \eqref{eq:droprob}, this reduces
to preventing $\delta$ from being too large. Such an approach is substantially
more befitting of our purposes since a Frank-Wolfe procedure need only solve a
sequence of \textit{local} problems.

This technique has been used by works such as \cite{BKorig, Sinha18} and the
approach presented in this work (for establishing computationally tractability
of \eqref{eq:psiwithot}) is similar to ideas appearing in \cite{Sinha18}. In
that work, smoothness of the objective in \eqref{eq:droprob} is used to
(qualitatively) argue that a sufficiently small $\delta$ will regularize the
dual of \eqref{eq:droprob} sufficiently strongly to produce a
computationally-tractable optimization problem. In contrast, however, we provide
quantification of the level of robustness required to achieve such a goal and do
so in the scope of a more general problem class \eqref{eq:psiwithot}.

Formulation of a Frank-Wolfe method for \eqref{P} with quantitative bounds on
complexity and convergence has, to the best of the authors' knowledge, failed to
appear in previous literature. Perhaps the most closely related effort is
\cite{Liu20} where similar, infinite dimensional conditions to those appearing
in this work (Section \ref{subsec:smoothandloja}) are used to study a
particle-based methods for computing Nash equilibria of zero-sum games. It
should be noted that, as a special case, our Frank-Wolfe method can produce a
particle-based optimization procedure and this hints at possible connections
with other particle techniques \cite{LiuSV16, Futami19, Carrillo19,
  Carrillo19PD}. Such connections are beyond the scope of this work, however,
and left for future consideration.

%%% Local Variables:
%%% mode: latex
%%% TeX-master: "main"
%%% End:

\section{Main result.}\label{sec:overview}
This work considers the problem
\begin{equation}\label{eq:genobj}
  \min_{\nu \in \Pc_2(\R^{d})} J(\nu)
\end{equation}
for functionals $J : \Pc_2 (\R^{d}) \to \bar{\R}$ over \eqref{eq:finite2norm}
that possess a ``gradient object'' (Definition \ref{def:wassdiff}) with
respect to $\Wc$-- the Wasserstein distance of order 2
\begin{equation}
  \label{eq:wassdef}
  \Wc^2(\mu, \nu) := \inf_{\pi \in \Pi(\mu, \nu)}\int_{\R^{d}} \norm{x-y}^2 \, d\pi(x,y)
\end{equation}
Our main result (Theorem \ref{thm:outerconvres}) provides a Frank-Wolfe
algorithm for \eqref{eq:genobj} which operates on $\Pc_2(\R^{d})$ and obtains
quantitative iteration and sample complexities. This yields an
intuitive, non-parametric algorithm for \eqref{eq:genobj} with the
guarantee:
\begin{theorem}[Informal; see Theorem \ref{thm:outerconvres}]\label{thm:convinformal}
  For a differential functional $J$ whose ``gradient'' $F_\nu$ provides an
  approximation that is slightly more than first order accurate
\begin{equation}\label{eq:holderinformal}
\min_{\Wc(\mu, \nu)\leq\delta}J\left(  \mu\right)
=\min_{\Wc(\mu, \nu) \leq\delta}\iprod{F_\nu}{\mu - \nu}+O\left(
\delta^{1+\alpha}\right)  , \hspace{ 0.2in } 0 < \alpha \leq 1
\end{equation}
and obeys the domination condition
\begin{equation}\label{eq:plinformal}
  \tau \left( J(\mu) - \inf_{\nu \in \Pc_2(\R^{d})} J(\nu) \right)^\theta \leq \,\,
  \norm{F_\nu}, \hspace{ 0.2in } \tau, \theta \,\in \,\R_*
\end{equation}
there is a Frank-Wolfe procedure which obtains an $\epsilon$-optimal solution of
\eqref{eq:genobj} in $O\left(\epsilon^{1-\alpha^* \theta}\right)$ iterations
where $\alpha ^* = (1+\alpha) / \alpha$ is the dual exponent.
\end{theorem}
When $J$ is convex in a Wasserstein sense (Definition \ref{def:geoconv}) and has
at least one minimizer, \eqref{eq:plinformal} holds with $\theta=1$. Hence, for
smooth $J$ ($\alpha = 1$ in \eqref{eq:holderinformal}), Theorem
\ref{thm:convinformal} recovers a intuitive $O(k^{-1})$ convergence rate
(accelerated rates are difficult in this context due to
the difficulty of averaging in Wasserstein spaces, see Remark
\ref{remrk:accel}). A highlight of Theorem \ref{thm:convinformal} is
that the assumptions needed for quantitative convergence are relatively
weak. Indeed, the condition \eqref{eq:plinformal} (properly known as a
{\L}ojasiewicz inequality; Section \ref{subsec:smoothandloja}) is generally
broader than convexity. The condition \eqref{eq:holderinformal} is less
stringent than smoothness, particularly as utilized in other literature \cite{Chu20,
  Liu20, Arbel19, Chewi20a}.

Supplementary to Theorem \ref{thm:convinformal}, we also construct a scalable
implementation of our Frank-Wolfe method and demonstrate it's performance on
several non-parametric estimation problems (Section
\ref{sec:experiments}). We also detail algorithms with novel
complexity guarantees for \eqref{eq:droprob} and \eqref{eq:psiwithot}
(Theorem \ref{thm:primaldual}) and provide a new strong duality result for
\eqref{eq:psiwithot} (Theorem \ref{thm:sduality}). These results are of
independent interest due to the relevance of \eqref{eq:droprob} and
\eqref{eq:psiwithot} for distributionally robust optimization, mathematical
finance, and stochastic processes \cite{Bartl20, BartlRWalks20, BKorig}. All
technical proofs of these results are given in the appendix.
%%% Local Variables:
%%% mode: latex
%%% TeX-master: "main"
%%% End:

\section{Preliminaries on Wasserstein geometry.}\label{sec:wassgeom}
\subsection{Notation and terminology}\label{sec:notation}
Denote the set of real numbers by $\R$, the set of extended real numbers by
$\bar{\R}$, and their respective subsets of non-negative numbers by $\R_+$ and
$\bar{\R}_+$. For a general function $f$, $\dom(f)$ and $\ran(f)$
denote the domain and range (respectively) while, for a convex function
$t : X \to \bar{\R}$ over some vector space $X$, the notation is overloaded so
that $\dom(t)$ denotes the \textit{effective domain} of $t$. That is,
\begin{equation*}
  \dom(t) = \left\{ x \in X : t(x) < \infty \right\}
\end{equation*}
We further say that the convex function $t$ is \textit{proper} if
$-\infty < t(x)$ for all $x \in X$ and $t(y) < \infty$ for some $y \in X$. For a
concave function $z : X \to \bar{\R}$, these terms are likewise defined by
considering the convex function $-z$. A convex function $t: X \to \Rb$ is
called \textit{closed} if it is lower-semicontinuous with respect to the
topology on $X$. Likewise, a concave function $z: X \to \Rb$ will be called
closed if it is upper-semicontinuous.

Unless otherwise specified, $\norm{\cdot}$ denotes the Euclidean norm on
$\R^{d}$ and a function $\phi : \R^d \to \bar{\R}$ is called \textit{semiconvex} (or
\textit{weakly convex}) if
\begin{equation}\label{eq:semiconv}
  x\, \longrightarrow \,\, \phi(x) + \frac{\lambda}{2} \norm{x-x_0}^2
\end{equation}
is convex for some for some $\lambda \geq 0$. The choice of $x_0$ in
\eqref{eq:semiconv} is largely irrelevant: if \eqref{eq:semiconv} is convex for
one such $x_0$, it is convex for all $x_0 \in \R^{d}$. It is clear that a
semiconvex function possesses a minimal $\lambda \geq 0$ such that
\eqref{eq:semiconv} is a convex function. This value will be denoted by
$\scparam$ and a semiconvex function with such a value will be termed a
$\scparam$-semiconvex function. Clearly, any convex function is $0$-semiconvex.

A continuously differentiable function $\phi : \R^d \to \bar{\R}$ will be called
\textit{$\alpha$-H{\"o}lder smooth} with parameter $T$ if it has H{\"o}lder
continuous gradients with parameter $T$ and exponent $\alpha$. That is:
\begin{equation}\label{eq:smoothness}
  \norm{\nabla \phi (y) - \nabla \phi (x)} \leq T \norm{y-x}^{\alpha}
\end{equation}
When \eqref{eq:smoothness} holds for $\alpha = 1$, $\phi$ will simply be called
$T$-smooth. Further, the notation $\Pc(\R^{d})$ denotes the set of Borel probability measures on
$\R^{d}$ while
\begin{equation}\label{eq:finite2norm}
  \Pc_2(\R^{d}) := \left\{ \mu \in \Pc(\R^{d}) : \int_{\R^{d}} \norm{x}^2 \,
    d\mu(x) < \infty \right\}
\end{equation}
The expression $C_c^\infty(\R^{d})$ denotes the space of all compactly supported,
smooth functions on $\R^{d}$.

\subsection{Functionals on probability measures}
Before providing a rigorous specification of a ``gradient'' with respect to
Wasserstein distance consider the
following possible instances of $J$, for illustrative purposes.
\begin{example}[Divergences]
  A common functional on $\Pc\left(\R^{d}\right)$ is KL-divergence with respect
  to a fixed, reference measure on $\nu$:
  \begin{equation}
    \label{eq:kldiv}
    J(\mu) := D_{KL}(\mu || \nu) = \int_{\R^d} \log\left( \frac{d \mu }{d \nu }
    \right) \, d \mu
  \end{equation}
  More generally, for any convex, lower-semicontinuous function $f : \R_+ \to
  \R$ such that $f(1) = 0$, one can consider a ``$f$-divergence'' of the form
  \begin{equation}
    \label{eq:divergencedef}
    J(\mu) := D_{f}(\mu || \nu) = \int_{\R^d} f \left( \frac{d \mu }{d \nu }
    \right) \, d \nu
  \end{equation}

  Canonical dual formulations show that such functionals
  \eqref{eq:divergencedef} are lower-semicontinuous with respect to the weak
  topology on $\Pc\left(\R^{d}\right)$ \cite{Santambrogio15}. This helps make
  these functionals amenable to our analyses-- since lower-semicontinuity is at
  least necessary for an iterative optimization procedures (such as a
  Frank-Wolfe algorithm) to converge to an optimizer. As the Wasserstein
  topology on $\Pc_2(\R^{d})$ is finer than the weak topology, this means that
  weak lower-semicontinuity is at least sufficient for our purpose.

  In many cases, divergences can also be supplemented with a potential
  $v : \R^{d} \to \R$ and interaction function $w : \R^{d} \times \R^{d} \to \R$
  \begin{equation}
    \label{eq:energyfunc}
    J(\mu) := \int_{\R^{d}} v(x) \, d\mu(x) + \int_{\R^{d}} w(x,y) \,
    d\mu(x)d\mu(y) + D_{f}(\mu || \nu)
  \end{equation}
  to yield ``energy functionals'' on $\Pc\left(\R^{d}\right)$
  \cite{Santambrogio15}.
\end{example}
\begin{example}[Integral Probability Metrics]
  For a set of real valued functions $F$ on $\R^{d}$ one can define the
  discrepancy
  \begin{equation}
    \label{eq:ipm}
    J(\mu) := \text{IPM} \left(\mu,\nu \right) = \sup_{f \in F} \,\, \int_{\R^{d}} f \, d\mu - \int_{\R^{d}} f \, d\nu
  \end{equation}
  for $\mu, \nu \in \Pc\left(\R^{d}\right)$, where $\nu$ is a fixed, reference
  measure. Such discrepancies are termed Integral Probability Metrics
  (IPMs), although they may not strictly satisfy the requirements of a metric--
  say, by failing to distinguish all pairs of measures. Instead, for a pair of
  measures $\mu, \nu \in \Pc\left(\R^{d}\right)$, IPMs can be interpreted as
  measuring the extent to which $\mu$ and $\nu$ differ on functions in $F$-- or,
  rather, measuring the extent to which $\mu$ and $\nu$ can be distinguished by
  $F$.
\end{example}
\begin{example}[Markov Decision Process]
  Consider a set of states $S = \R^{m}$ and a set of actions
  $A = \R^{n}$. At a denumerable set of times $t=1,2,3,\ldots$ an agent
  which occupies state $s_{t-1}$ chooses an action $a_t$ and randomly
  transitions to a new state $s_t$, while receiving a reward $r_t \in \R$. For
  transitions which are Markovian, this process can be described by a set of
  Markov transition kernels $p_t(s_t,r_t | s_{t-1}, a_t)$ which give the
  probability of obtaining state $s_t$ and reward $r_t$ for an agent which was
  most recently in state $s_{t-1}$ and chose action $a_t$.

  The goal of the agent to choose a distribution
  $\mu ^* \in \Pc \left(\R^{m + n}\right)$, termed a \textit{policy,} so as to
  maximize his or her expected reward:
  \begin{equation}
    \label{eq:mdprfunc}
    \mu ^* := \underset{\mu \in \Pc\left(\R^{m +
          n}\right)}{\argmin} \, J(\mu) = \underset{\mu \in \Pc\left(\R^{m +
          n}\right)}{\argmin} \, \Ep_{\mu, p_t |_{t=1}^\infty}\left[\sum_{i=1}^{\infty} r_i\right]
  \end{equation}
  Here, the expectation is taken with the transition kernels $p_t$ and a an
  agent that chooses actions which are distributed according to the conditional
  distribution of $\mu$. Note that, in most works, the \textit{policy} is
  specified in terms of a (potentially infinite) set of conditional
  distributions over actions: $\mu(a | s)$. Hence, the expected reward is,
  instead, a functional over the product space $\otimes_{s \in
    \R^{m}}\mu(a | s)$. However, by choosing an arbitrary distribution
  $\alpha \in \Pc(\R^{m})$ and considering the joint distribution $\mu(s,a) =
  \alpha(s) \mu(a | s)$, this formulation can be seen to be equivalent to
  \eqref{eq:mdprfunc}-- see \cite{Chu19} for further details.
\end{example}
\subsection{Properties of Wasserstein space}
Under Wasserstein distance, $\Pc_2(\R^{d})$ is a Polish space \cite{Villani08}
and, via to it's kinematic characterizations (Proposition \ref{prop:wassprop}),
provides a natural structure for studying stochastic optimization. For a
Frank-Wolfe method to meet a stated goal of minimizing \textit{local, linear
  approximations}, one requires an appropriate definition of a ``gradient.'' This
requires providing rigorous meaning to the expression
\begin{equation}
  \label{eq:formalwassgrad}
    \lim_{\alpha \to 0} \frac{J \left(\mu_\alpha \right) - J(\mu)}{\alpha}
\end{equation}
where $\mu_\alpha$ denotes a (purely formal) perturbation from $\mu$ of
Wasserstein distance $\alpha$. To this end, consider the following properties of
Wasserstein space that are essential for this work-- a basic proof is given in
Appendix \ref{sec:proofwassprop}.
\begin{proposition}[Properties of Wasserstein space]\label{prop:wassprop}
  \
  \begin{itemize}
  \item Under the Wasserstein metric $\Wc$, $\Pc_2(\R^{d})$ is a \textit{geodesic}
    space. That is, for every $\mu, \nu \in \Pc_2(\R^{d})$, there exists a
    constant-speed geodesic curve $\mu_t : [0,1] \to \Pc_2(\R^{d})$ where
    $\mu_0  = \mu$, $\mu_1  = \nu$ and
  \begin{equation}\label{eq:csgeodef}
    \Wc(\mu_t, \mu_s ) = |t - s| \Wc(\mu_0, \mu_1)
  \end{equation}
  Moreover, there is a bijection between constant-speed geodesics and optimal
  transport plans. Every geodesic corresponds to a unique, optimal transport plan
  $\gamma \in \Pi \left(\mu,\nu \right)$
  \begin{equation}\label{eq:omapdef}
    \Wc(\mu, \nu)^2 = \int \norm{x-y}^2 \, d \gamma(x,y)
  \end{equation}
  such that
  \begin{equation}\label{eq:geobiject}
    \mu_t  = \left((1-t) x + t y\right)_{\#}\gamma
  \end{equation}
  Conversely, every optimal transport plan gives rise to a unique geodesic via \eqref{eq:geobiject}.
\item For a constant-speed geodesic $\mu_t : [0,1] \to \Pc_2(\R^{d})$, there
  exists a ($\mu_t$-almost surely) unique Borel vector field
  $v_t  : [0,1] \times \R^{d} \to \R^{d}$ which satisfies
  \begin{equation}\label{eq:minvectorfield}
    \Wc(\mu_0, \mu_1)^2 = \int_{0}^1 \int_{\R^{d}} \norm{v_t(x)}^2 \, d \mu_t(x)
    \, dt = \min_{v_t \in V_{\mu}} \int_{0}^1 \int_{\R^{d}} \norm{v_t(x)}^2 \, d \mu_t(x)
  \end{equation}
  for
  \begin{equation}\label{eq:continuity}
    V_{\mu} := \left\{ v_t  : \frac{d\mu_t}{dt} + \nabla \cdot \left(v_t \mu_t\right) = 0 \right\}
  \end{equation}
  defined as the set of all Borel vector fields which solve the continuity equation for
  $\mu_t$. The continuity equation is understood in duality with
  $C_c^{\infty}(\R^{d})$.
\item For any constant-speed geodesic $\mu_t$, the corresponding optimal
  transport plan $\gamma \in \Pi(\mu_0 , \mu_1 )$ and the corresponding
  vector field $v_t$ (given by \eqref{eq:minvectorfield}) satisfy the relation
  \begin{equation}\label{eq:velexpr}
    v_t((1-t) x + t y) = y-x, \hspace{ 0.2in } \text{$\gamma$-almost surely}
  \end{equation}
  for Lebesgue-almost every $t$.
\item The space $\Pc_2(\R^{d})$ is positively curved under $\Wc$ and
  at each point $\mu \in \Pc_2(\R^{d})$, the tangent space
  \begin{equation}\label{eq:tangdef}
    \tangt(\mu) := \overline{\left\{ \nabla \psi \,\, : \,\, \psi \in C_c^{\infty}(\R^{d}) \right\}}^{L^2(\mu)}
  \end{equation}
  is the closure in $L^2(\mu)$ of the gradients of smooth functions with
  compact support. Via the
  Riesz isomorphism, $\ctangt(\mu) = \tangt(\mu)$ where
  $\ctangt(\mu)$ denotes the cotangent space.
  The tangent and cotangent bundles will be denoted
  $\tangt_{\Pc_2(\R^{d})}$ and $\ctangt_{\Pc_2(\R^{d})}$, respectively.
  \end{itemize}
\end{proposition}

\subsection{Differentiability in Wasserstein space}
Proposition \ref{prop:wassprop} clarifies that $\Pc_2(\R^{d})$ has a
non-Euclidean geometry with respect to $\Wc$. Unfortunately, this complicates
the notion of a ``gradient'' in the sense of \eqref{eq:formalwassgrad}. Since
the tangent space \eqref{eq:tangdef} varies from point to point, the notion of
linear approximation varies from point to point. Hence, one must define
gradients in terms of a selections in the cotangent bundle. Despite these
complications, however, the theory of Proposition \ref{prop:wassprop} now yields
a direct expression of the ``gradients'' that our Frank-Wolfe algorithm will
utilize.
\begin{definition}[Geodesic convexity]\label{def:geoconv}
  A set $S \subseteq \Pc_2(\R^{d})$ is said to be convex or geodescially convex
  if for any $\mu, \nu \in S$ one has $\mu_t \in S$ for any geodesic curve
  $\mu_t$ between $\mu$ and $\nu$. Similarly, a functional
  $J : \Pc_2(\R^{d}) \to \bar{\R}$ is said to be convex or geodescially convex
  if, for any $\mu, \nu \in S$ in a convex set $S$,
  \begin{equation}\label{eq:geoconv}
    J(\mu_t) \leq t J(\nu) + (1-t) J(\mu)
  \end{equation}
  for all geodesics $\mu_t$ between $\mu$ and $\nu$.
\end{definition}
\begin{definition}[Wasserstein differentiability]\label{def:wassdiff}
  Let $S$ be a geodescially convex set. A functional $J$ is
  Wasserstein differentiable on $S$ if there is a
  map $F : \Pc_2(\R^{d}) \to \ctangt_{\Pc_2(\R^{d})}$ such
  that for all $\mu, \nu \in S$ and any constant-speed geodesic
  $\mu_t : [0,1] \to \Pc_2\left(\R^{d}\right)$ between $\mu$ and $\nu$, one has
  \begin{equation}\label{eq:wassderiv}
    \lim_{\alpha \to 0} \frac{J \left(\mu_\alpha \right) - J(\mu)}{\alpha} =
    \int_{\R^{d}  \times \R^{d}} F(\mu;x)^T(y-x)  \, d \gamma(x,y)
  \end{equation}
  where $\gamma$ is the unique optimal transport plan
  \eqref{eq:geobiject} corresponding to $\mu_t$. Note that $F(\mu;x) =
  \left(F(\mu)\right)(x)$ provides a more aesthetic way of
  representing the evaluation at $x \in \R^d$ of the output of $F$ at $\mu$.
  The map $F$ will be called the Wasserstein derivative of $J$.
\end{definition}
\begin{remark}
  The description of differentiability provided by Definition \ref{def:wassdiff}
  falls within the general framework of metric derivatives and Wasserstein
  gradient flows, originally codified in \cite{Ambrosio05}. This framework is
  now a well-established component of the theory of Wasserstein spaces, while
  the relation \eqref{eq:wassderiv}, itself, presents only a narrow structuring
  of ideas from this framework. Definition \ref{def:wassdiff}, however, is often
  how works in statistical and algorithmic fields interact with this broader
  area \cite{Taghvaei19, Chizat18, Liu19, Liu20}. Moreover, this literature
  demonstrates the most motivating feature of \eqref{eq:wassderiv}: a large
  number of functionals of interest for machine learning and statistical
  inference exhibit Wasserstein gradients in the sense of \eqref{eq:wassderiv}.
  The curious reader is referred to \cite{Ambrosio05, Santambrogio15, Carmona18}
  for precise statements of conditions under which \eqref{eq:wassderiv} is
  guaranteed. However, let it suffice to say that $F$ typically arises from the
  Gateaux differential for $J$ \cite{Santambrogio15, Taghvaei19}. Recall, the
  Gateaux differential for a functional $J$ exists when there is an appropriate,
  dual space $D^* \supseteq C_b(\R^{d})$ on a closed subspace
  $D \subseteq \Pc(\R^{d})$ such that
  \begin{equation}
    \label{eq:gateaux}
    \left\langle dJ(\mu), \nu - \mu \right\rangle = \lim_{\alpha \to 0} \frac{J
      \left(\mu + \alpha (\nu - \mu)\right) - J(\mu)}{\alpha}
  \end{equation}
  for some $dJ(\mu) \in D^*$ and all $\mu$ in some set $S$ such that
  $S - S \subseteq D$. In instances where the Gauteaux differential $dJ(\mu)$
  exists, the Wasserstein derivative $F$ will often also exist and be given by
  $\nabla dJ(\mu) \in \tangt(\mu)$. Here, we use the gradient operator formally,
  and omit a rigorous exposition on this operation in the context of $\tangt(\mu)$.
\end{remark}
\begin{remark}
  The notion of geodesic convexity given in Definition \ref{def:geoconv} is
  standard for Wasserstein spaces and dates back to at least \cite{mccann97}. It
  has appeared ubiquitously in subsequent works \cite{Ambrosio05, Craig17}. What
  is surprising, however, is that functionals which are non-convex with respect
  to canonical vector space structure on $\Pc(\R^{d})$ are convex in the
  sense of Definition \ref{def:geoconv}.
\end{remark}

It should be noted that computation of the Wasserstein derivative might be
difficult. Indeed, for a $J$ in a variational form such as \eqref{eq:ipm},
computation of the Wasserstein derivative is equivalent to finding a
\textit{witness} function that achieves the supremum \cite{Santambrogio15}. In
the case of a pathological $F$ (in \eqref{eq:ipm}), such a task might be
intractable. To resolve this issue, this work utilizes the existence of an
oracle for the computation of a Wasserstein gradient. This oracle permits a unified
description of our Frank-Wolfe algorithm and
abstracts away variation in functional-specific computational cost.
\begin{definition}[Wasserstein Derivative Oracle]\label{def:wassdervoracle}
  Let $J : \Pc_2(\R^{d}) \to \R$ be a Wasserstein differentiable functional on a
  set $S$ with Wasserstein derivative $F: \Pc_2(\R^{d}) \to \tangt_{\Pc_2(\R^{d})}$.
  A $L$-smooth Wasserstein derivative oracle over $S$ is an oracle
  which, given sample access to a distribution
  $\mu \in S$ and an error parameter $\epsilon$, returns an $L$-smooth function
  $\widehat{\phi}_\mu \in C^1(\R^{d})$ satisfying
  \begin{equation}\label{eq:gradoraclerror}
    \norm{\nabla \widehat{\phi}_\mu - F(\mu)}_{L^2(\mu)} \leq \epsilon
  \end{equation}
\end{definition}
\begin{remark}
  The qualification that the Wasserstein derivative oracle return an $L$-smooth
  function is necessary to exclude the, aforementioned, possibility of a
  pathological Wasserstein derivative-- which would be intractable for use in a
  computational procedure. In some ways, this is representative of the fact that
  the cotangent space $\ctangt(\mu)$ at a point is too large; the $L^2(\mu)$
  closure of gradients of smooth, compactly supported functions still contains
  vector fields that are stubbornly complex. Such a condition is common in other
  variational methods \cite{Arbel19, Chu20, ZhangGAN20, Cohen20} and is
  relatively superficial-- when coupled with the degree of approximation
  afforded by $\epsilon$. Indeed, via smoothing techniques \cite{Sanjabi18,
    Carrillo19, Liu19}, functionals can often be assumed to have Wasserstein
  derivatives which are $C^1(\R^{d})$ or are well-approximable by $C^1(\R^{d})$
  functions.
\end{remark}
\subsection{Smoothness and {\L}ojasiewicz inequalities}\label{subsec:smoothandloja}
In finite dimensions, iterative, gradient-based methods typically require the
specification of two conditions in order to achieve convergence.
\begin{itemize}
\item The accuracy of local, linear approximations that are provided by
  the gradient.
\item The extent to which local descent makes global progress on the objective.
\end{itemize}
Here, we state these conditions in the context of functionals over Wasserstein
space.
\begin{definition}[$\alpha$-Holder smoothness]\label{def:holdersmooth}
  Let $S$ be a geodesically convex set and let
  $J : \Pc_2\left(\R^{d}\right) \to \R$ be a functional which is continuously
  Wasserstein differentiable on the set $S$. $J$ is said to be locally
  $\alpha$-Holder smooth on $S$ with parameters $T$ and $\Delta $ if for all
  $\mu \in S$ and all $\nu \in S$ such that $W(\mu, \nu) \leq \Delta $, there
  exists an optimal transport plan $\gamma \in \Pc_2(\R^{d} \times \R^{d})$ such
  that
    \begin{equation}\label{eq:holdersm}
      J(\nu) \leq J(\mu) + \int_{\R^{d} \times \R^{d}} F(\mu; x)^T(y-x) \, d \gamma(x,y)
      + \frac{T}{1+ \nu} \Wc^{1+ \alpha} \left(\nu, \mu \right)
    \end{equation}
\end{definition}
\begin{definition}[{\L}ojasiewicz inequality]\label{def:lojaineq}
  A Wasserstein differentiable functional $J$ on a set
  $S \subseteq \Pc_2(\R^{d})$ is said to satisfy a \textit{{\L}ojasiewicz
    inequality} with parameter $\tau$ and exponent $\theta$ if for all
  $\mu \in S$ and $J_* := \inf_{\mu \in S}J(\mu)$
  \begin{equation}\label{eq:lojaineq}
    \tau \left(J(\mu) - J_*\right)^\theta \leq \norm{F(\mu)}_{L^2(\mu)}
  \end{equation}
  where $F$ is the Wasserstein derivative \eqref{eq:wassderiv} of $J$.
\end{definition}
\begin{remark}
  More restrictive versions of both \eqref{eq:holdersm} and \eqref{eq:lojaineq}
  commonly appear in previous literature \cite{Arbel19, PLIneqFromGeodesic19, Liu20,
    Chu20, Chewi20a}. In most cases, the $\alpha$-H{\"o}lder smoothness
  condition \eqref{eq:holdersm} is stated for $\alpha = 1$ and required to hold
  globally ($\Delta = \infty$). This smoothness criterion is considerably weaker
  since it requires that the Wasserstein gradient only provide a local
  approximation that is slightly more than first-order accurate. Further,
  statement of the {\L}ojasiewicz inequality \eqref{eq:lojaineq} is broader than
  canonical treatments due to the presence of the auxiliary power
  $\theta$. Most often, the specific instances of either $\theta=1/2$ or
  $\theta=1$ are considered, since they are implied by various forms
  \cite{Ambrosio05} of geodesic convexity \eqref{eq:geoconv}-- for instance, see
  Lemma \ref{lem:linearlb}.
\end{remark}

\section{The Frank-Wolfe algorithm.}\label{sec:fwalgo}
Algorithm \ref{alg:frankwolfe} provides our Frank-Wolfe procedure along with
its associated convergence guarantees and sample complexities (Theorem \ref{thm:outerconvres}).
\begin{algorithm}
  \caption{Frank Wolfe for \eqref{eq:genobj}}
  \label{alg:frankwolfe}
  \begin{algorithmic}
    \Input{Wasserstein derivative oracle $\Theta$, initial distribution $\mu_0$,
      smoothness parameter $\alpha$,
      gradient error $\hat{\epsilon}$, estimation error $\bar{\epsilon}$,
      iterate error $\widetilde{\epsilon}$, stopping threshold $r$, step sizes
      $\left(\beta_1,\beta_2,\beta_3\right)$, number of iterations $k$}
    \For{$1 \leq i \leq k$}
    \State Let
    $\widehat{\phi}_{\mu_{i-1}} \leftarrow \Theta(\mu_{i-1}, \hat{\epsilon})$
    $\Comment{\norm{\nabla\hat{\phi }_{\mu_{i-1}} -
        F(\mu_{i-1})}_{L^2(\mu_{i-1})} \leq \hat{\epsilon}}$ \vspace{0.05in}
    \State Compute
    $\norm{\nabla\hat{\phi }_{\mu_{i-1}}}_{L^2(\mu_{i-1})} - \bar{\epsilon} \leq s \leq
    \norm{\nabla\hat{\phi }_{\mu_{i-1}}}_{L^2(\mu_{i-1})}$ \vspace{0.05in}
    \If{$s \leq r$,} \textbf{break} \Else \hspace{ 0.05in }
    $\delta \leftarrow \min\left(\beta_1, \beta_2 s, \beta_3
      s^{1/\alpha}\right)$, $\,\zeta \leftarrow \delta\widetilde{\epsilon}$
    \EndIf
    \State Compute $\mu_{i}$ satisfying
    $W(\mu_{i}, \mu_{i-1}) \leq \delta$ and
    $\Comment{\text{using Algorithm \ref{alg:primaldual}}}$
  \begin{equation}\label{eq:algotrustregion}
    \int \phi_{\mu_{i-1}} \, d \mu_{i} - \inf_{W(\nu, \mu_{i-1}) \leq \delta} \int
    \phi_{\mu_{i-1}} \, d \nu \leq \zeta
  \end{equation} \vspace{-0.2in}
  \EndFor
  \Return $\mu_i$
  \end{algorithmic}
  \end{algorithm}
  To obtain these guarantees, we require the following assumptions on the
  objective $J$--phrased in the language of the previous theory.
\begin{assumption}[Smoothness assumption]\label{asm:fwsmooth}
  The functional $J:\Pc_2(\R^d) \to \bar{\R}$ is Wasserstein differentiable
  (Definition \ref{def:wassdiff}) and locally \textit{$\alpha$-Holder smooth}
  \eqref{def:holdersmooth} on a set $S\subseteq \Pc_2(\R^d)$ with parameters $T$
  and $\Delta_1 > 0$ (Definition \ref{def:holdersmooth}). Further, a $L$-smooth
  Wasserstein derivative oracle (Definition \ref{def:wassdervoracle}) for $J$
  exists.
\end{assumption}
\begin{assumption}[Local richness]\label{asm:rich}
  The set $S$ is rich enough to contain the solution to
  \eqref{eq:trustregionprob} for $\mu \in S$, $L$-smooth $f$, and
  $\delta \leq \Delta_2$.
\end{assumption}
\begin{assumption}[{\L}ojasiewicz assumption]\label{asm:fwloja}
  The functional $J$ satisfies a \textit{{\L}ojasiewicz inequality} \eqref{eq:lojaineq}
  on $S \subseteq \Pc_2(\R^d)$ with parameters $\tau > 0$ and $\theta$.
\end{assumption}
  \begin{theorem} \label{thm:outerconvres}
    Under Assumptions \ref{asm:fwsmooth}, \ref{asm:rich}, \ref{asm:fwloja},
    and an appropriate choice of input parameters, Algorithm \ref{alg:frankwolfe}
    computes a distribution $\mu^*$ satisfying
  \begin{equation}\label{eq:fwdesiredresid}
    r(\mu ^*) := J(\mu ^*) - \inf_{\mu \in S} J(\mu) \leq \epsilon
  \end{equation}
  in at most
  \begin{equation}\label{eq:fwcomplexity}
    k = \widetilde{O}\left(r(\mu_0)^{p_+}\epsilon^{-p_-}\right)
  \end{equation}
  iterations, where $\mu_0$ is the initial iterate and $p_+, p_-$ denote the
  positive and negative parts of $p= 1-\alpha ^*\theta$ for the dual exponent
  $\alpha^* = (1+\alpha)/\alpha$. Further, each iteration of Algorithm
  \ref{alg:frankwolfe} can be performed using at most
  $\widetilde{O}\left(\epsilon^{-2\alpha^*\theta}\right)$ independent samples from the
  initial distribution $\mu_0$. Note that the notation $\widetilde{O}(\cdot)$
  obscures logarithmic factors in it's arguments.
\end{theorem}
\begin{remark}
  Since the computation of the Frank Wolfe step \eqref{eq:algotrustregion} is
  performed using Algorithm \ref{alg:primaldual}, the result of Algorithm
  \ref{alg:frankwolfe} is a bi-level procedure with inner and outer iteration
  loops. Further, since Algorithm \ref{alg:primaldual} requires only sample
  access to it's inputs and can return an oracle providing sample access to it's
  output, all operations in Algorithm \ref{alg:frankwolfe} can be implemented
  with only sample access to the underlying distributions $\mu_i$. Via a simple
  induction argument, it also follows that all operations in Algorithm
  \ref{alg:frankwolfe} can be implemented using only sample access to the
  initial distribution $\mu_0$; this analysis provides the stated sample
  complexity of Theorem \ref{thm:outerconvres}. Further, since the chief
  consumer of these samples (Algorithm \ref{alg:primaldual}), uses them to
  compute $O(\log \epsilon^{-1})$ sample averages, it is clear that nearly all
  of the $\widetilde{O}\left(\epsilon^{-2\alpha^*\theta}\right)$ samples in
  Theorem \ref{thm:outerconvres} can be drawn in parallel. That is,
  \eqref{eq:algotrustregion} can be computed with \textit{low parallel depth}.

  Practically, it is often more efficient to directly maintain approximations to
  the $\mu_i$ via a non-parametric estimator-- as opposed to a exact sampling
  oracle. When this is done, it results in an additional, additive error in the
  residual \eqref{eq:fwdesiredresid} at each step of Algorithm
  \ref{alg:frankwolfe}. However, so long as this error is on the order of the
  additive error produced by the Wasserstein derivative oracle $\Theta$, the
  iteration complexity \eqref{eq:fwcomplexity} remains unaffected. Moreover
  since analysis of the error induced by a particular approximation of the
  $\mu_i$ is highly problem dependent, we do not consider it in the context of
  these results.
\end{remark}
\begin{remark}\label{remrk:accel}
  The dependence on the dual exponent $\alpha ^*$ in \eqref{eq:fwcomplexity} can
  be rather punishing for small $\alpha$. It is natural to ask if this exponent
  could be improved within the scope of Assumptions \ref{asm:fwsmooth},
  \ref{asm:rich}, and \ref{asm:fwloja}-- perhaps under the auspice of the class
  of first order methods presented in Section \ref{sec:dualandcomp}. Moreover,
  in finite dimensions, it is well known that first-order methods for convex and
  $\alpha$-H{\"o}lder smooth functions (also known as weakly smooth functions)
  can obtain $\epsilon$-optimal solutions in $O(\epsilon^{-2/(1+3\alpha)})$
  iterations \cite{Nesterov85}. Hence, it could even be considered whether,
  given geodesic-convexity assumptions on $J$, a better iteration
  complexity for Algorithm \ref{alg:frankwolfe} would be obtainable.

  We conjecture that such improvements are unlikely, however. Particularly those
  that would draw on analogy from finite dimensional techniques; the motivation
  for this is as follows. A common approach to establishing improved iteration
  complexities for convex, $\alpha$-H{\"o}lder smooth functions in finite
  dimensions is to consider their gradient oracles as inexact oracles for
  convex, $1$-H{\"o}lder smooth functions \cite{Devolder14}. Using either
  averaging arguments or accelerated methods, more rapid progress on an
  underlying objective can then be made with these inexact oracles. Our
  Frank-Wolfe method already utilizes an inexact step
  \eqref{eq:algotrustregion}, thus it is conceivable that such an approach could
  be applied to Algorithm \ref{alg:frankwolfe}.

  Unfortunately, this finite dimensional analogy fails due to the fact that
  averaging is difficult is Wasserstein space. Indeed to prevent error
  accumulation from outpacing objective progress, averaging iterates is
  crucial-- either directly or in the form of an accelerated method. Since
  Wasserstein space is positively curved (Proposition \ref{prop:wassprop})
  computing analogous convex combinations of the $\mu_i$ in Algorithm
  \ref{alg:frankwolfe} is itself a variational problem and could be as
  expensive to compute.

  % used to either convex fall flat for
  % investigation of these questions is left for future work, we believe that such
  % improvements are unlikely. While encouraging, these observations Note that,
  % simple improvements in the iteration complexity for Algorithm
  % \ref{alg:frankwolfe} would naturally result in improvements in the sample
  % complexity since the sample complexity is primarily the result of error
  % accumulation.

  % While we leave investigation of these questions for future work, we
  % conjecture the following. Further improvements to the iteration complexity
  % \eqref{eq:fwcomplexity} that draw inspiration from finite
  % dimensional techniques such as \cite{Devolder14}, likely for the following reasons. First

\end{remark}

%%% Local Variables:
%%% mode: latex
%%% TeX-master: "main"
%%% End:

\section{Computational experiments.}\label{sec:experiments}
In this section, we demonstrate the application our Frank-Wolfe algorithm to
several non-parametric estimation problems in statistics and machine learning.
\subsection{Gaussian deconvolution}
A classical task in nonparametric statistics \cite{Carroll88, Caillerie11} is to
infer a latent, data-generating distribution $\nu \in \Pc_2 \left(\R^d\right)$
from a set of observations that are corrupted by independent, additive Gaussian
noise. For observations $Y_1, \ldots, Y_n$ such that
\begin{equation}\label{eq:deconvobs}
  Y_i = X_i + Z_i \hspace{ 0.2in } \text{where} \hspace{ 0.2in } X_i \sim \nu
  ,\,\,\,\, Z_i \sim N(0, \sigma^2)
\end{equation}
one seeks to compute a non-parametric estimate of $\nu$-- the variance of the
noise $\sigma^2$ is considered known. Since $Z_i$ is independent of
$X_i$, this task amounts to ``deconvolving'' $\nu$ from the distribution of
$Z_i$. A natural candidate for $\nu$ is the maximum-likelihood estimator (MLE)
\begin{equation}
  \label{eq:deconvmle}
  \widehat{\nu} := \argmax_{\mu \in \Pc_2 \left(\R^d\right)} \,\,\, \sum_{i=1}^{n}\log \left(\phi_{\sigma}\,
    \ast\, d\mu(Y_i)\right) \hspace{ 0.2in } \text{where} \hspace{ 0.2in }
  \phi \,\ast \,d\mu(Y_i) = \int_{\R^d} \phi_{\sigma }\left(Y_i - x\right) \, d\mu(x)
\end{equation}
where $\phi_{\sigma}$ is the density of $Z_i$. In \cite{Rigollet18}, it was
shown that $\widehat{\nu}$ has an equivalent characterization as
\begin{equation}
  \label{eq:otdeconv}
  \hat{\nu} = \argmin_{\mu \in \Pc_2 \left(\R^d\right)} \,\,\, \Wc_\sigma^2(\mu, \hat{P}_{Y})
\end{equation}
where
\begin{equation}
  \label{eq:semiot}
  \Wc_\sigma^2(\mu_1, \mu_2) := \inf_{\pi \in \Pi(\mu_1, \mu_2)} \frac{1}{2}\int
  \norm{x-y}^2 \, d\pi(x,y) + \sigma^2 D(\pi \,|| \, \mu_1 \otimes \mu_2)
\end{equation}
is the entropic optimal transportation distance \cite{Cuturi13} and $\hat{P}_Y$
is the empirical distribution of the $Y_i$. The problem \eqref{eq:otdeconv}
readily lies within the framework of \eqref{eq:genobj} for
$J(\mu) := \Wc_\sigma^2(\mu, \hat{P}_{Y})$. Moreover, it is known \cite{Luise19}
that the Wasserstein derivative \eqref{eq:wassderiv} of
$\Wc_\sigma^2(\mu, \hat{P}_{Y})$ with respect $\mu$ is given by
\begin{equation}
  \label{eq:wassder}
  \phi_{\mu}(x) = \sigma^2 \log \left(\frac{1}{n}\sum_{i=1}^{n} \exp\left(
      \left( v^*_i - \norm{x-y_i}^2/2\right)/\sigma^2\right)\right)
\end{equation}
where $v ^* \in \R^{d}$ is dual variable (corresponding to
$\hat{P}_{Y}$) which is optimal for $\Wc_\sigma^2(\mu, \hat{P}_{Y})$. This
provides a Wasserstein derivative oracle for
\eqref{eq:otdeconv} as the vector $v ^*$
can be readily approximated using stochastic gradient methods \cite{Bach16}.

\begin{figure}
  \centering
  \resizebox{0.7\columnwidth}{!}{%
        \input{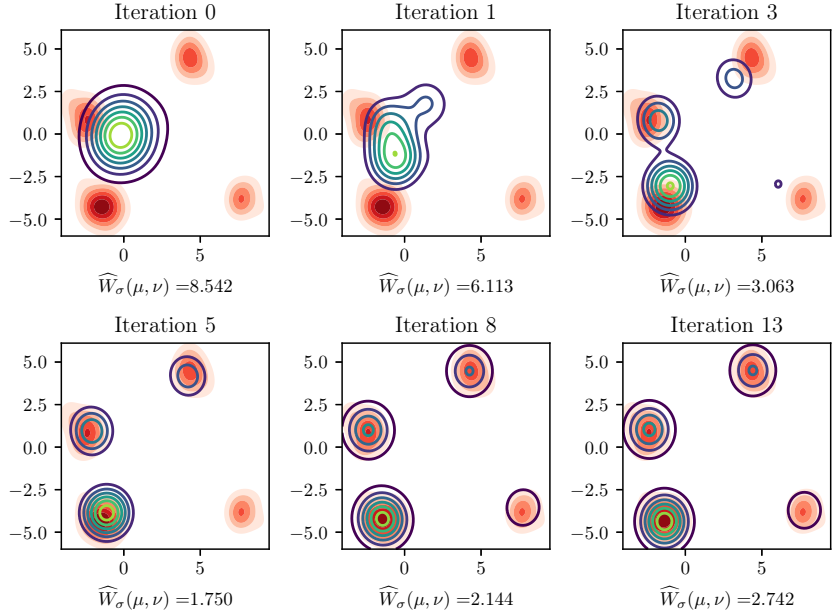}
  }
    \caption{Deconvolution of a multi-modal dataset via the Frank-Wolfe
      algorithm. Background contours (in red) provide an illustration of the
      underlying data distribution, while the foreground contours provide the
    density of the iterate maintained by Algorithm \ref{alg:frankwolfe}.}
  \label{fig:deconv_dens}
\end{figure}

\begin{figure}
  \centering
  \resizebox{0.6\columnwidth}{!}{%
        \input{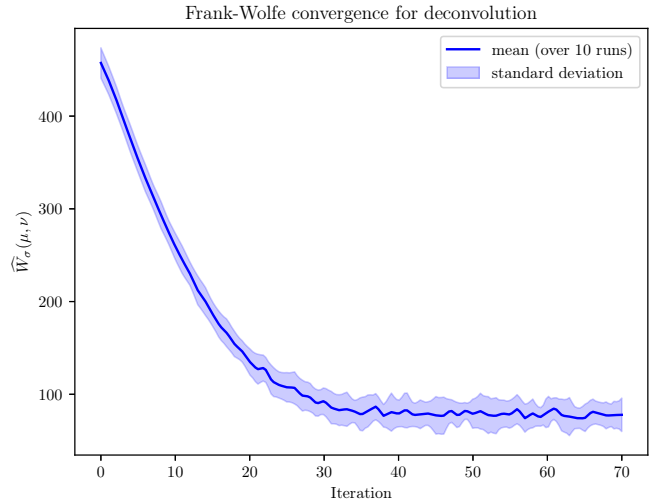}
  }
  \caption{Estimated entropic Wasserstein distance computed at each iteration of
    the Frank-Wolfe algorithm (\ref{alg:frankwolfe}) for a 64-dimensional
    deconvolution problem with multi-modal data. Displayed is the average
    distance over 10 independent runs with random initializations.}
  \label{fig:deconv_error}
\end{figure}

A simple, two dimensional instance of this problem is shown in Figure \ref{fig:deconv_dens} on
a dataset $Y_i$ of 50 samples with four distinct modes-- illustrated by the
kernel density estimator of the $Y_i$, shown in red. The behavior of Algorithm
\ref{alg:frankwolfe} is depicted over the course of several iterations, where
the foreground contours provide the density of the iterate, $\mu_i$, that is
maintained by the algorithm. In this setting, $\mu_i$ is approximated as a
mixture of $N$-gaussians of fixed bandwidth (for $N=200$); as opposed to
maintaining a full sampling oracle for each $\mu_i$. This approximation induces
an additional, additive error in the residual of each iterate. So long as this
error is of the same order as the error in the Wasserstein gradient, however,
the analysis of Theorem \ref{thm:outerconvres} is unaffected. Moreover, empirically, this is
consistent with the performance of the Frank-Wolfe algorithm. Instead,
performance appears to be dominated by the accuracy of the Wasserstein
derivative computation; which consumes the majority of the computational time
for this problem. Figure \ref{fig:deconv_error} provides a quantitative demonstration of the
convergence of Algorithm \ref{alg:frankwolfe} for a similar,
multi-modal data in 64 dimensions.


\subsection{Maximum mean discrepancy}
For a reproducing kernel Hilbert space (RKHS) $H$ on a space on a space $X$, the
maximum mean discrepancy (MMD) \cite{Gretton12} is the integral probability
metric (IPM) between distributions $\mu, \nu \in \Pc(X)$ generated by
the unit ball of $H$. That is,
\begin{equation}
  \label{eq:mmddef}
  \mmd(\mu, \nu) := \sup_{\norm{f}_H \leq 1} \,\, \int_X f(x) \, d\mu - \int_X f(x) \, d\nu
\end{equation}
where $\mmd(\mu, \nu)$ quantifies the degree to which $\mu$ and $\nu$ can be
distinguished by functions in $H$. Indeed, for an $H$ which is
\textit{universal} and an $X$ which is compact, MMD provides a metric on
$\Pc(X)$ \cite{Gretton12}. The rise of generalized adversarial networks (GANs)
\cite{Goodfellow14} and efforts connecting neural networks and kernel regression
\cite{Chizat19}, have generated interest in MMD, particularly with respect to
it's role in constructing high-dimensional, distributional embeddings
\cite{Cohen20,Mroueh19}. This development is predicated on the observation that
any neural network $(x, \theta) \to \psi(x,\theta)$, which produces an output
$\psi(x,\theta) \in \R^d$ from input data $x \in X \subseteq \R^{d}$ and parameters
$\theta \in \Theta \subseteq \R^{m}$, yields a kernel on the parameter
set $\Theta$:
\begin{equation}
  \label{eq:neuralkern}
  k(\theta_1, \theta_2) := \Ep_x\left[\psi(x,\theta_1)^T\psi(x,\theta_2)\right]
\end{equation}
where the expectation over $x$ is taken with respect to a data generating
distribution. Via MMD, $k$ induces a natural discrepancy measure between
distributions over network parameters $\theta$ and, therefore, learning of
a generative image model can be expressed as minimizing \eqref{eq:mmddef} with
respect to latent, generative distribution for $\nu$. We refer to
\cite{Mroueh19, Arbel19} for further descriptions of these applications.

With respect to the variational framework of this paper \eqref{eq:genobj}
minimization of \eqref{eq:mmddef} against a latent, target distribution $\nu$
provides a natural fit for \eqref{eq:genobj}. Indeed, for
\begin{equation}
  \label{eq:mmdJ}
  J(\mu) := MMD^2(\mu, \nu)
\end{equation}
the Wasserstein derivative (Definition \ref{def:wassdiff}) of $J$ is
the unique \textit{witness} function $f_\mu^*$ achieving \eqref{eq:mmddef}
\cite{Arbel19}. Moreover, $f^*$ has a natural expression as the
difference between the mean embeddings of $\mu$ and $\nu$
\begin{equation}
  \label{eq:mmdderiv}
  f_{\mu}^*(x) = \Ep_{z \sim \mu}\left[k(z,x)\right]  - \Ep_{z \sim \nu}\left[k(z,x)\right]
\end{equation}
and can be computed via sampling methods, even when $\mu$ or $\nu$ are
continuous or are large, discrete distributions \cite{Gretton12}. Perhaps the most advantageous
consequence of \eqref{eq:mmdderiv}, however, is that the Wasserstein gradient
directly inherits regularity present in $k$. Indeed, should $\nabla_xk(x,y)$ be
$L$-Lipschitz in $x$ (uniformly for all $y$), $J$ \eqref{eq:mmdJ} is naturally
$L$-smooth \cite{Arbel19}. This has led to the
development of several variational or particle-based methods for minimizing
\eqref{eq:mmdJ} \cite{Arbel19,Mroueh19,Cohen20}.

\begin{figure}
% \hspace{-1.5in}
  \centering
  \resizebox{\columnwidth}{!}{%
        \input{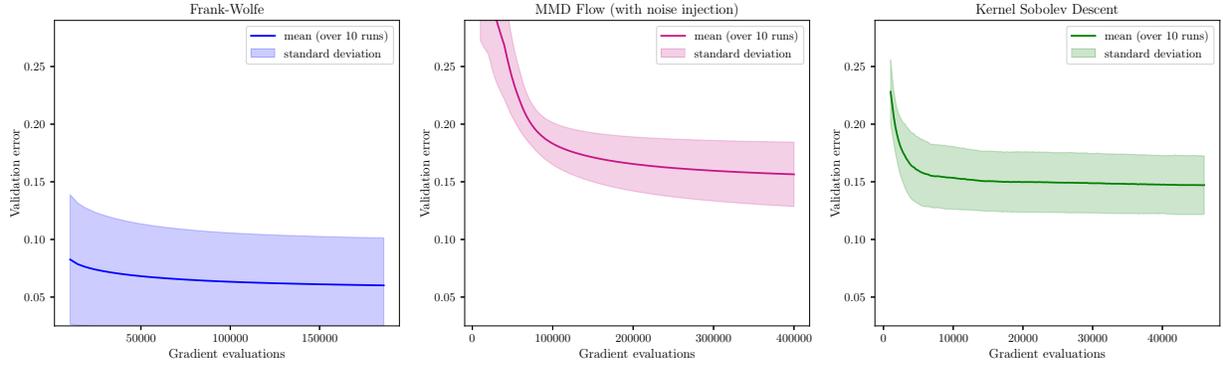}
  }
  \caption{Maximum mean discrepancy (with respect to a validation dataset)
    between a latent, ``teacher'' neural network and a distribution over
    ``student'' networks as computed by each algorithm. The discrepancy, also
    referred to as validation error, is shown as a function of the total number
    of gradient evaluations performed by each algorithm.}
  \label{fig:mmd}
\end{figure}

Figure \ref{fig:mmd} contrasts the performance of our Frank-Wolfe algorithm with two of
these methods on the student-teacher network problem showcased in
\cite{Arbel19}. Our method is shown on the left, the center plot shows the ``MMD
gradient flow'' algorithm from \cite{Arbel19}, and the right plot provides the
``Sobolev Descent'' algorithm of \cite{Mroueh19}. Performance is evaluated in
terms of MMD error on a validation dataset and is shown as a function of the total
gradient evaluations performed by each method. This provides a better proxy for
relative performance and convergence since an iteration of Algorithm
\ref{alg:frankwolfe} performs multiple solves that are, each, similar in
terms of gradient complexity to a single iteration of MMD gradient flow or Sobolev
descent. Further, the total number of gradient evaluations should not be viewed
as a proxy for wall-time as, for each gradient evaluation, the number of
operations performed by each method can vary widely. Indeed, for each gradient
evaluation in Sobolev descent an entire linear system solve is performed.
Also, note that, as both MMD gradient flow and Sobolev descent are
particle-based, Algorithm \ref{alg:frankwolfe} was, for the purposes of
comparison, instantiated with a particle distribution of equal size.

% The illustration of Figure \ref{fig:mmd} is that our Frank-Wolfe method is at least
% competitive with these other, infinite-dimensional, decent-based algorithms in
% terms of both absolute performance and convergence rate. Indeed, over ten
% replications with randomly chosen initializations, our Frank-Wolfe method
% consistently achieved lower overall-validation error and lower validation error
% for a given number of gradient evaluations-- relative to either MMD gradient
% flow or Sobolev descent, which were run using reference
% implementations\footnote{\url{https://github.com/MichaelArbel/MMD-gradient-flow}}.
% This is partly explained by the finding that our Frank-Wolfe method located an
% extremely good solution after relatively few iterations. Note that each method
% utilizes gradient evaluations in a fundamentally different computational
% procedure and all methods were run until convergence was observed-- leading to
% the wide range of gradient evaluations between the plots in Figure \ref{fig:mmd}.
% Additionally, since validation error was only measured once during each
% Frank-Wolfe iteration, rather than during each sub-iteration, the
% per-gradient-evaluation performance of the Frank-Wolfe method is rather more
% granular than might be illustrated by Figure \ref{fig:mmd} and does not account for the first 10000 evaluations.

%%% Local Variables:
%%% mode: latex
%%% TeX-master: "main"
%%% End:

\section{Duality and computational procedures.}\label{sec:dualandcomp}
The focus of this section is to provide a complete analysis of the Frank Wolfe
method in Section \ref{sec:fwalgo} by furnishing a concrete, computational
procedure (and complexity guarantee) for the subroutine in Algorithm \ref{alg:frankwolfe}: compute a
$\nu^* \in \Pc_2\left(\R^{d}\right)$ such that $\Wc(\nu^*, \mu) \leq \delta$ and
\begin{equation}\label{eq:trustregionprob}
\int f \, d \nu ^* - \inf_{\Wc \left(\nu, \mu\right) \leq \delta} \,\, \int f \, d
\nu \leq \epsilon
\end{equation}

This problem and it's computational solution are,
themselves, of independent interest since they frequently arise in
distributionally robust optimization (DRO) \cite{BKorig, Guan18, Esfahani18, Sinha18}--
typically, phrased as a maximization problem. One can take an even broader view,
however, that \eqref{eq:trustregionprob} is a particular instance
of
\begin{equation}\label{eq:primallite}
  \inf_{\pi \in \Pi(\mu)}\,\, \int f \, d\pi + \psi \left(\int c \, d\pi \right)
\end{equation}
where $\Pi(\mu)$ is the set of couplings whose first marginal is given by $\mu$,
$c: \R^{d} \times \R^{d} \to \bar{\R}_+$ is now an arbitrary, non-negative,
Borel-measurable cost function (having replaced the Wasserstein cost
$\norm{\cdot}^2$), and $\psi: \bar{\R}_+ \to \bar{\R}_+$ is now a proper, closed,
and convex function (having replaced the trust-region constraint
$W(\mu, \nu) \leq \delta$).

Beyond the expanded relevance that \eqref{eq:primallite} has for stochastic
processes and gradient flows \cite{Bartl20, Ambrosio05, BartlRWalks20}, the
purpose of this consideration is two-fold. First, \eqref{eq:primallite} provides
a template for a wide class of infinite-dimensional, first-order optimization
methods. In finite dimensional optimization, first-order procedures are often
expressed as solving a sequence of problems in the form
\begin{equation}
  \label{eq:fwfinite}
  \inf_{y \in \R^{d}} s^Ty + \psi(y)
\end{equation}
for a convex, lower-semicontinuous, function $\psi: \R \to \bar{\R}$. Under this
same token, usage of \eqref{eq:trustregionprob} in Algorithm
\ref{alg:frankwolfe} could be replaced with another instance of
\eqref{eq:primallite} for, say, a problem-specific cost function $c$. This would
yield an alternate variational procedure that could be better suited for a
particular problem at hand. Second, the
tool enabling a computational procedure for \eqref{eq:trustregionprob},
\textit{duality}, exists with the same level of utility for \eqref{eq:primallite}
and yields the same structures that facilitate computation: supergradients.

In the hope that these considerations elucidate how further variational
procedures could be derived from our techniques, we resolve a computational
procedure for \eqref{eq:trustregionprob} in the following manner. In Section
\ref{subsec:dual}, we show that \eqref{eq:primallite} exhibits a dual formulation
that makes it approachable for computation. We do this under more general
assumptions than are available in previous works \cite{BKorig,Gao16,Bartl20} to
highlight the breadth of possible extensions to our Frank Wolfe procedure. In
Section \ref{subsec:iproc}, we then specialize our techniques to
\eqref{eq:trustregionprob} and provide a sampling-based algorithm for
\eqref{eq:trustregionprob} with complexity bounds.

\subsection{Duality}\label{subsec:dual}
The full generalization of \eqref{eq:trustregionprob} to be considered is
\begin{equation}\label{eq:primal}
  \pouter := \inf_{\pi \in \Pi(\mu)} \pinner(\pi) = \inf_{\pi \in \Pi(\mu)}\,\,
  \int_{S_1} f \, d\pi + \psi \left(\int_{S_0 \times S_1} c \, d\pi \right)
\end{equation}
where $S_0, S_1$ are Polish spaces, $\Pi(\mu)\in \Pc(S_0 \times S_1)$ is the set
of joint couplings with first marginal given by $\mu$, $f:S_1 \to \bar{\R}$ and
$c: S_0 \times S_1 \to \bar{\R}_+$ are Borel-measurable, and
$\psi: \bar{\R}_+ \to \bar{\R}_+$ is proper, closed, and convex. The objective
\eqref{eq:primal} follows the convention that $\infty - \infty = \infty$ and has
a value of $\infty$ if $\int f \, d \pi$ is not defined. The dual of \eqref{eq:primal} is
\begin{equation}\label{eq:dual}
  \douter := \sup_{\lambda \in \R}\,\, \int_{S_0} f^{\lambda c} \,\, d\mu - \psi
  ^* (\lambda), \hspace{ 0.3in }
   f^{\lambda c}(x) := \inf_{y \in S_1} f(y) + \lambda c(x,y)
\end{equation}
where $f^{\lambda c} : S_0 \to \bar{\R}$ is canonically called the
``c-transform'' of $f$ \cite{Villani08}. Note, \eqref{eq:primal} induces the
convention $f(y) + \lambda c(x,y) = \infty$ if $c(x,y) = \infty$.
\begin{remark}
Rigorously, the dual of \eqref{eq:primal} is better defined as
\begin{equation}\label{eq:dualrig}
  \douter := \sup_{\lambda \in \R} \,\, \left(\,\,\sup_{\phi \in
      \Lambda_{\mu} \,\, \left(f + \lambda c\right)}\int_{S_1}\phi \,
    d\mu - \psi ^* (\lambda)\right)
\end{equation}
where $\psi ^*$ is the convex conjugate of $\psi$ and, for any
 $g: S_0 \times S_1 \to \bar{\R}$,
\begin{equation*}
  \Lambda_{\mu}(g) := \left\{\phi \in L^1(\mu) : \phi \left(x\right) \leq
    g(x,y) \,\,\,\, \forall y\in S_1 \right\}
\end{equation*}
This definition side steps the technicality that $f^{\lambda c}$ is \textit{not
  necessarily} Borel-measurable and keeps the dual variables within the space of
 integrable functions. However, under conditions for
strong duality (Theorem \ref{thm:sduality}), the formulations \eqref{eq:dual} and \eqref{eq:dualrig}
are equivalent and the lack of Borel-measurability in $f^{\lambda c}$ is a
formality since $f^{\lambda c}$ is universally measurable-- therefore it is
measurable with respect to the completion of $\mu$. These details are
discussed in greater length in Appendix \ref{ap:reglr}.
\end{remark}

Define the functionals $\tau_c : \Pi\left(\mu\right) \to \bar{\R}_+$ and
$\tau_f: \Pi\left(\mu\right) \to \bar{\R}$
  \begin{equation}\label{eq:taudef}
    \tau_c \left(\pi \right) := \int c \, d \pi \txs{and} \tau_f \left(\pi \right) := \int f \, d \pi
  \end{equation}
  where $\tau_f(\pi)$ is set to be $\infty$ if the integral is undefined;
  $\tau_c$ is always well-defined by the non-negativity of $c$. Since both
  functionals are linear on $\Pi\left(\mu\right)$ there is flexibility in
  defining their effective domains (Section \ref{sec:notation}). For
  the sake of Theorem \ref{thm:sduality},
  the effective domains of $\tau_c$ and $\tau_f$ are defined by regarding them to be
  convex.
\begin{theorem}[Strong Duality] \label{thm:sduality}
  Let
  \begin{equation}\label{eq:edom}
    D : = \dom \left(\tau_f\right) \cap \dom \left(\tau_c\right) \hspace{ 0.2in
    }  \text{and} \hspace{ 0.2in } t_c(D) := \left\{ t_c(\pi)  : \pi \in D \right\}
  \end{equation}
  If
  \begin{equation}
    \label{eq:domconds}
    \hspace{-0.4in}\dom\left( \psi \right) \,\, \cap \,\,
    \rint\left(\tau_c \left(D\right) \right) \neq \emptyset\hspace{ 0.2in
    }  \text{and} \hspace{ 0.2in }0 \in \dom\left(\psi\right)
  \end{equation}
  where $\rint(\cdot)$ denotes the relative interior of a set, then
  \begin{equation}\label{eq:strDuality}
     \inf_{\pi \in \Pi(\mu)}\,\, \int f \, d\pi + \psi \left(\int c \, d\pi \right) = \sup_{\lambda \in \R} \,\, \int_{S_0} f^{\lambda c} \,\, d\mu - \psi
  ^* (\lambda)
  \end{equation}
\end{theorem}
\begin{remark}
  The key consequence of Theorem \ref{thm:sduality} that facilitates the
  development of computational methods for \eqref{eq:trustregionprob} is: the
  primary decision variable of an equivalent dual problem \eqref{eq:dual} is a
  single, scalar number. Granted, \eqref{eq:dual} also depends on the
  c-transform $f^{\lambda c}$. However, $f^{\lambda c}$ is given by an
  optimization problem (on the ambient spaces $S_0 \times S_1$) which is
  regularized by $\lambda$ and $c$. This is a setting which is now significantly
  more amenable to computation using iterative procedures.
\end{remark}
\begin{remark}
  Strong duality of the form \eqref{eq:strDuality} has been previously noted in
  \cite{Bartl20}, under more stringent conditions and assumptions. Most notably,
  \cite[Section 2]{Bartl20} requires the cost function to be
  lower-semicontinuous, satisfy growth conditions, and approach certain values
  on subsets of $S_0 \times S_1$. Additional restrictions are also placed on
  $\psi$. Related work \cite[Theorem 1]{BKorig}(a special case of Theorem
  \ref{thm:sduality} in this work) makes similar assumptions: the cost
  function must attain a specific value on a subset of $S_0 \times S_1$, and
  $f$ and $c$ must be upper and lower-semicontinuous, respectively. Theorem
  \ref{thm:sduality} eliminates all of these assumptions and replaces
  them with a natural, Fenchel-type condition \eqref{eq:domconds}. Stated
  simply, \eqref{eq:domconds} requires the objective is finite on a set with
  suitable ``interior.'' This is essentially what one would anticipate
  from analogs in finite dimensional optimization. Moreover, Fenchel-type
  are often more precise because the primarily tend to fail when the primal
  is already infinite/infeasible or when it is a pathological limit of
  infinite/infeasible problems.
\end{remark}
\begin{example}
  \ \\
  \begin{itemize}
  \item When $\psi(x) = \infty \one_{(\delta, \infty]}(x)$ for $\delta > 0$
    ($\psi$ is zero on $[0,\delta]$ and $\infty$ outside), then
    $\psi ^*(\lambda) = \left(\delta \lambda\right)_+$ and Theorem
    \ref{thm:sduality} gives
    \begin{equation}\label{eq:drodual}
       \inf \left\{ \int f \, d \pi : \pi \in \Pi(\mu), \int c \, d\pi \leq \delta \right\} =
      \sup_{\lambda \in \R } \,\, \int f^{\lambda c } \, d\mu(x) - \left(\delta \lambda \right)_+
    \end{equation}
    provided that there exists a $\pi \in \Pi(\mu)$ such that
    $\int c \, d \pi < \delta$ and $\int f \, d \pi < \infty$. Note, when $c$ is
    lower-semicontinuous, the infimum in \eqref{eq:drodual} can be taken over
    optimal couplings between $\mu$ and any Borel measure $\nu \in \Pc(S_1)$--
    resulting in the optimal-transport-based, robust optimization problem
    \eqref{eq:droprob}.
   \item When $\psi(x) = x^{1 + \alpha} / (1+\alpha)$ for $\alpha \geq 0$,
    \begin{equation}\label{eq:holddual}
      \inf_{\pi \in \Pi(\mu)} \int f \, d \pi + \frac{1}{1+ \alpha}\left(\int c \, d\pi\right)^{1+\alpha} =
      \sup_{\lambda \in \R } \,\, \int f^{\lambda c } \, d\mu(x) - \frac{\alpha }{1+\alpha }\left(\lambda \right)^{(1+\alpha)/\alpha}_+
    \end{equation}
    provided there exists a $\pi \in \Pi(\mu)$ such that
    $\int c \, d \pi < \infty$ and $\int f \, d \pi < \infty$. Duality holds for other, commonly-used,
    smooth penalties (such as $x \mapsto e^x$) under the same condition.
  \end{itemize}
\end{example}

\subsection{Computational procedures.}\label{subsec:iproc}
The dual \eqref{eq:dual} can be re-expressed as
\begin{equation}
  \label{eq:gfunc}
  \D_{\mu}(f) = \sup_{\lambda \in \R} \,\, g(\lambda) - \psi ^* (\lambda),
  \hspace{ 0.2in }
  g(\lambda) := \Ep_{x \sim \mu}\left[f^{\lambda c}(x)\right]
\end{equation}
where, henceforward, sufficient conditions for strong duality
\eqref{eq:domconds} are assumed. The function $g$ is concave, non-decreasing,
and upper-semicontinuous (Lemma \ref{lem:subgrad}). Therefore, \eqref{eq:gfunc}
makes sense as a one-dimensional, stochastic, convex optimization problem. A
standard approach to solve \eqref{eq:gfunc} is to notice that
supergradients/subgradients of $g$ and $\psi ^*$ exist at every point in
$\rint \left(\dom(g)\right)$ and $\rint\left(\dom(\psi ^*)\right)$
\cite{Rockafellar70}. If one can compute estimates of these supergradients, then
a supergradient-ascent procedure in $\lambda$ will provide a suitable
algorithm for computing \eqref{eq:gfunc}. See Appendix \ref{sec:subgradproof} for
a more detailed description of the supergradients of $g$.
\subsubsection{Computing supergradients}
Computation of supergradients for $g$ is where meaningful computational
difficulty arises. This difficulty is the result of
the inner minimization problem defining the c-transform $f^{\lambda c}$.
Estimating $f^{\lambda c}$, at even a single point $x \in S_0$, suggests the
need to solve
\begin{equation}
  \label{eq:inner-opt-prob}
  \inf_{y \in S_1} f(y) + \lambda c(x,y)
\end{equation}
which, without additional regularity in $f$ and $c$, could be NP-hard-- even for
relatively simple $f$ and $c$. Indeed, consider the case $S_0=S_1=\R^d$,
$c(x,y) = \one_{\Delta_d}(y)$ is the indicator function of the simplex
$\Delta_d$, and $f(y) = y^T\left(I + A\right)y$ for the adjacency matrix $A$ of
any graph $G$. Then, for any $x \in \R^d$ and $\lambda > 0$,
\eqref{eq:inner-opt-prob} is the maximum independent set problem for the graph $G$
\cite{Motzkin65}.

In the interest of developing a computational procedure for
\eqref{eq:trustregionprob}, we consider computation of supergradients for
\eqref{eq:gfunc} when $S_0,S_1 \subseteq \R^d$ and $c(x,y) = \norm{x-y}^2/2$. In
this case, the dual \eqref{eq:gfunc} becomes
\begin{equation}\label{eq:2normdual}
  \sup_{\lambda \in \R } \,\,\int \inf_{y \in \R^{d}} f(y) + \frac{\lambda
  }{2}\norm{y-x}^2 \, d \mu - \psi ^*(\lambda)
\end{equation}
and \eqref{eq:inner-opt-prob} provides the Moreau-Yosida envelope for the
function $f$ \cite{Yoshida80}. If $f$ is semiconvex \eqref{eq:semiconv}, then
\eqref{eq:inner-opt-prob} is computationally tractable for large enough
$\lambda$.
\begin{definition}[Supergradient oracle with high probability]\label{def:strorprob}
  A function $\theta_g : \R \to \R$ is called a
  ($\epsilon, \delta$)-supergradient oracle \textit{with high probability} for
  $g$ (on the interval $[l,u]$) if, when queried with a $\lambda \in [l,u]$, it
  returns an independent random sample $\theta_g(\lambda)$ satisfying
  \begin{equation}
    \label{eq:proboracle}
    \p\left(\left[ \min_{z \in \partial g (\lambda)}\left|\theta_g(\lambda) - z
        \right| \geq \frac{\epsilon}{\max \left(\lambda - l, 1\right)} \right]\right) \leq \delta
  \end{equation}
\end{definition}
\begin{algorithm}[H]
  \caption{Supergradient oracle \eqref{def:strorexp}}
  \label{alg:sorc}
  \begin{algorithmic}
    \Input{Distribution $\mu$, point $\lambda$, semi-convexity parameter
      $\scparam$, smoothness parameter $L$, error tolerance $\epsilon$}
    \State Sample $x \sim \mu$
    \State $y_0 \leftarrow x$, $\kappa \leftarrow \sqrt{(\lambda + L)/( \lambda - \scparam )}$
    \State $k \leftarrow \max \left( \left\lceil 4\kappa \log(12\kappa\norm{\nabla
          f(x)}/\epsilon) \right \rceil, 0\right)$
  \For{$1 \leq i \leq k$}
  \State $z_i = y_{i-1} - \frac{1}{\kappa} \left(\nabla f(y_{i-1}) + \lambda (y_{i-1} - x) \right)$
  \State $y_i = z_i + \frac{\kappa - 1}{\kappa+ 1} \left(z_i - z_{i-1}\right)$
  \EndFor
  \Return $\theta = \frac{1}{2}\norm{y_k - x}^2$
\end{algorithmic}
\end{algorithm}
\begin{proposition}\label{prop:sorcalgp}
  For a $\scparam$-semiconvex function $f : \R^{d}\to \R$, which is also $L \geq
  \scparam$ smooth \eqref{eq:smoothness}, the mean of
  \begin{equation}
    \label{eq:kconst}
    K \geq \frac{64\Ep_{\mu}\left[\norm{\nabla f(x)}^4\right]}{(\lambda-\scparam)^2\min \left((\lambda - \scparam)^2,1\right)\delta\tilde{\epsilon}^2}
  \end{equation}
  independent calls to Algorithm \ref{alg:sorc} with inputs $\lambda > \scparam$ and
  $\epsilon = \tilde{\epsilon}/(2\max(\lambda - \scparam, 1))$,
  provides a $\left(\tilde{\epsilon}, \delta\right)$-supergradient
  oracle with high probability \eqref{def:strorprob} for $g$ in \eqref{eq:gfunc}
  on the interval $(\scparam, \infty)$.
\end{proposition}
\subsubsection{A primal-dual algorithm}\label{subsec:pdalgo}
The supergradient oracle of Proposition \ref{prop:sorcalgp} provides a mechanism
to perform ascent steps in $\lambda$ to solve \eqref{eq:gfunc} (for
$c=\norm{x-y}^2/2$). Previous work \cite{NamDuchi16,Ghosh18}, regarding related,
distributionally robust optimization problems, has focused on mirror ascent and
bisection search to perform these ascent steps. For completeness, these
algorithms (along with their complexities) are provided in the context of
\eqref{eq:gfunc} in Appendices \ref{ap:mirror} and \ref{sec:proof-sbisec}.

The caveat to these procedures is that they only provide well-founded ascent
methods for \eqref{eq:gfunc} when $\lambda$ is sufficiently large. Previous
works \cite{NamDuchi16, Blanchet18Iter} have noted this in the context of
\eqref{eq:droprob}; that it results in \eqref{eq:droprob} only being computable
for \textit{small to moderate} values of $\delta$. The following
condition will be used to actually quantify these values.
\begin{definition}
  For a proper, closed, and convex function $\psi: \bar{\R}_+ \to \bar{\R}$, let
  $\partial_- \psi$ denote it's left derivative. The function $\psi$ is said to
  provide $R$-regularization at $x \in \R$ if
  \begin{equation} \label{eq:dualgradcond2}
    \partial_- \psi ^*(x) \leq R^{-1}
  \end{equation}
\end{definition}
\begin{remark}
  In the context of problem \eqref{eq:drodual} with
  $\psi = \infty \one_{(\delta, \infty]}$, one has
  $\psi ^*(x) = \left(\delta x\right)_+$ and therefore
  $\partial_- \psi ^*(x) = \delta \one_{(0, \infty)}(x)$. Hence,
  $R$-regularization at $x > 0$ induces the requirement $\delta \leq R^{-1}$ and
  asserts that the level of robustness in \eqref{eq:droprob}, $\delta$, is
  moderate. A broader understanding of \eqref{eq:dualgradcond2}, results from
  considering: since $x \in \partial \psi(y)$ for $y = \partial_- \psi ^*(x)$
  (where $\partial \psi(y)$ denotes the subgradient set at $y$), enforcement of
  \eqref{eq:dualgradcond2} for large values of $x$ and $R$ constrains $\psi$ to
  attain large subgradients on small neighborhoods of 0. Thus,
  \eqref{eq:dualgradcond2} quantifies the degree of regularization provided by
  $\psi$ in \eqref{eq:2normdual} and ensures that the level of regularization meets a
  given threshold.
\end{remark}

With this mechanism, an algorithm with concrete computational
guarantees for solving \eqref{eq:2normdual} can be furnished. This algorithm
performs bisection ascent, using the supergradient oracle provided by Algorithm
\ref{alg:sorc}. For the sake of our Frank-Wolfe procedure, it is of importance
that the algorithm implicitly maintains a primal-feasible iterate for
\begin{equation}\label{eq:2normprimal}
  \inf_{\pi \in \Pi(\mu)} \int f \, d \pi + \psi \left(\int \frac{1}{2}
    \norm{y-x}^2 \, d \pi\right)
\end{equation}
and that the algorithm makes progress on the primal-dual gap between
\eqref{eq:2normprimal} and \eqref{eq:2normdual}. For this reason, we title the
algorithm a ``primal-dual'' algorithm.
\begin{algorithm}[H]
  \caption{Primal-dual algorithm}
  \label{alg:primaldual}
  \begin{algorithmic}
    \Input{Supergradient oracle
      $\theta_{g}$, error tolerance $\epsilon$, termination width $B$}
    \State $\eta \leftarrow \infty$, $b \leftarrow l$
    \While{$u-l>\epsilon/B$}
    \State $\lambda \leftarrow \left(l + u\right)/2$
    \State $\eta \leftarrow \theta_g(\lambda)$, $\eta \leftarrow \left(\eta -
      \left(\psi ^*\right)'(\lambda)\right)$
    \If{$\eta < - \epsilon/\max \left(\lambda - b, 1\right)$} $u \leftarrow \lambda$
    \Else{ $l \leftarrow \lambda$}
    \EndIf
    \EndWhile \\
  \Return $u$
  \end{algorithmic}
  \end{algorithm}
\begin{remark}
  The primal iterate that this algorithm maintains can be clarified by remarking
  that Assumption \ref{asm:regularization} enforces \eqref{eq:dualgradcond2}
  with appropriate constants to guarantee that $\lambda ^* > \scparam$ for a
  $\scparam$-semiconvex $f$ in \eqref{eq:2normprimal} and optimal $\lambda ^*$
  \eqref{eq:2normdual}. Since the function
  $y \mapsto f(y) + \lambda/2 \norm{x-y}^2$ is strictly convex for
  $\lambda > \scparam$, the distribution $\pi_{\lambda, \mu} \in \Pi(\mu)$
  given by
\begin{equation}\label{eq:pilamdef}
  (X,m(X)) \sim \pi_{\lambda, \mu}, \hspace{ 0.4in }  X \sim \mu,\hspace{ 0.1in }   m_\lambda(x) = \argmin_{y \in \R^{d}} f(y) + \frac{\lambda}{2}\norm{y-x}^2
\end{equation}
is the unique distribution in $\Pi(\mu)$ such that
\begin{equation}\label{eq:primalval3}
  \int f(y) + \frac{\lambda ^*}{2} \norm{y-x}^2 \, d\pi_{\lambda, \mu} =
  \int \min_{y \in \R^{d}} \left(f(y) + \frac{\lambda^*}{2}\norm{y-x}^2\right) \, d\mu(x)
\end{equation}
Hence, $\pi_{\lambda, \mu}$ is the implicit distribution that is maintained by
Algorithm \ref{alg:primaldual}. The criterion that is used for bisection of an
interval in Algorithm \ref{alg:primaldual} is designed to make progress on the
primal-dual gap between the current dual iterate $\lambda_i$ and
$\pi_{\lambda_i, \mu}$:
\begin{equation}\label{eq:basicpdgap}
  G(\lambda_i) := \int f \, d \pi_{\lambda_i, \mu} + \psi \left(\int \norm{y-x}^2 \, d
    \pi_{\lambda_i, \mu}\right) - \left( g(\lambda_i) - \psi^*(\lambda_i) \right)
\end{equation}
This stands contrary to the sequence of iterates that are maintained by, say,
Algorithm \ref{alg:bsearch} where, $\pi_{\lambda_i, \mu}$ need not
even be primal feasible for a dual feasible $\lambda_i$.
\end{remark}
\begin{assumption}\label{asm:regularization}
  The function $f : \R^{d} \to \R$ is $L$-smooth \eqref{eq:smoothness},
  $\scparam$-semiconvex, and $\psi : \bar{\R} \to \bar{\R}$ provides
  $C \bigr / \left(\Ep_{\mu}\left[\norm{\nabla
        f(x)}^2\right]\right)$-regularization \eqref{eq:dualgradcond2} at
  $\scparam + 1$, for some $C \geq 8L^2$ Further, $\psi$ is minimized at 0 and
  $\psi ^*$ is $M$-smooth ($\left(\psi^*\right)'$ exists and is $M$-Lipschitz)
  on the interval $[l,u]$ where
  \begin{equation}\label{eq:intvbnds2}
    l := \scparam + 1 \hspace{ 0.2in } \text{and} \hspace{ 0.2in } u := \scparam + 1
    + \sqrt{2C}
  \end{equation}
\end{assumption}
\begin{theorem}\label{thm:primaldual}
  Under Assumption \ref{asm:regularization} and a correct configuration of it's
  inputs, Algorithm \ref{alg:primaldual} returns a $\lambda ^*$ such that the
  primal-dual gap \eqref{eq:basicpdgap} satisfies $G(\lambda ^*) \leq \epsilon$
  with probability $1 - \delta$. Moreover, the algorithm draws at most
  \begin{equation}\label{eq:pdsamplebound}
  \widetilde{O}\left(\frac{\scparam^2\Ep_{\mu}\left[\norm{\nabla f(x)}^4\right]}{\delta\epsilon^2}\right)
  \end{equation}
  independent samples from $\mu$ and performs
  $\widetilde{O}\left(\scparam^2L^{1/2}\Ep_{\mu}\left[\norm{\nabla
        f(x)}^4\right]/\left(\delta\epsilon^2\right)\right)$
  expected gradient evaluations of $f$-- where $\widetilde{O}$ suppresses logarithmic
  factors in $\scparam, L,C, M, \Ep_{\mu}\left[\norm{\nabla f(x)}^2\right]$ and
  $\epsilon$.
\end{theorem}
\begin{corollary}\label{cor:fwcompv2}
  If $\mu \in \Pc_2\left(\R^{d}\right)$, $f$ is $L$-smooth
  \eqref{eq:smoothness}, and
  $\delta \leq \norm{\nabla f}_{L^2(\mu)}/\left(2L\right)$,
  there exists a stochastic algorithm which (for any probability $\gamma < 1$)
  computes a $\lambda^*$ such that $\Wc \left(\nu_{\lambda ^*}, \mu\right)
  \leq \delta$ and
  \begin{equation}\label{eq:trerror}
  \int f \, d \nu_{\lambda ^*} - \inf_{\Wc \left(\nu, \mu\right) \leq
    \delta} \,\, \int f \, d \nu \leq \epsilon
  \end{equation}
  where $\nu_{\lambda ^*}$ is second marginal of $\pi_{\lambda ^*, \mu}$ in
  \eqref{eq:pilamdef}. This algorithm requires at most
  $\widetilde{O}(L^2\norm{\nabla f}^4_{L^4(\mu)}/((1-\gamma)\epsilon^2))$ independent
  samples from $\mu$ and executes $
  \widetilde{O}(L^{5/2}\norm{\nabla f}^4_{L^4(\mu)}/((1 - \gamma )\epsilon^2))$
  gradient evaluations of $f$ in expectation.
\end{corollary}
\begin{remark}
  The conclusion of Theorem \ref{thm:primaldual} is: regularization from
  $\psi$ enables the computational solution of \eqref{eq:2normprimal} when it
  occurs at the level specified by Assumption \ref{asm:regularization}. When
  this conclusion is specialized to the instance \eqref{eq:trustregionprob}, it
  results in Corollary \ref{cor:fwcompv2} and a bound on the magnitude of
  $\delta$. Such a result is quite befitting of our purposes, however, since the
  Frank-Wolfe procedure (Algorithm \ref{alg:frankwolfe}) need only solve
  \textit{local} problems, not \textit{global} ones. Further, since the
  instruction of these results is that an appropriate $\delta$ should
  necessarily depend on $\mu$ and $f$ \eqref{eq:trustregionprob}, Algorithm
  \ref{alg:frankwolfe} adapts it's choice of $\delta$, per iteration.

  It should also be noted that restriction of $\delta$ to provide computational
  tractability for \eqref{eq:droprob} has been used both qualitatively
  \cite{BKorig} and quantitatively \cite{Sinha18} in previous works. Indeed, the
  techniques presented in this work most closely resemble ideas from
  \cite{Sinha18}, where smoothness \eqref{eq:smoothness} was used similarly. In
  contrast, however, Assumption \ref{asm:regularization} and Theorem
  \ref{thm:primaldual} provide actual quantification of the level of robustness
  required to achieve tractability (through \eqref{eq:dualgradcond2}) and they do so
  for a more general set of problems \eqref{eq:psiwithot}. Moreover, Theorem
  \ref{thm:primaldual} provides guarantees with respect to the primal-dual gap
  of these problems-- a more elusive criterion than considered in previous work.
\end{remark}

%%% Local Variables:
%%% mode: latex
%%% TeX-master: "main"
%%% TeX-engine: xetex
%%% End:

%%% Local Variables:
%%% mode: latex
%%% TeX-master: "main"
%%% End:

%%% Local Variables:
%%% mode: latex
%%% TeX-master: "main"
%%% End:

\appendix

\section{Proof of weak duality}\label{ap:wduality}
\begin{proposition}[Weak Duality] \label{prop:wduality}
  Weak duality always holds for the pair \eqref{eq:primal} and \eqref{eq:dualrig}. That is,
  \begin{equation}\label{eq:wduality}
     \douter \leq \pouter
  \end{equation}
\end{proposition}
\proof
  It is sufficient to show that, for any primal variable $\pi \in
  \Pi(\mu)$ and any dual variables $\lambda \in \R$ and
  $\phi \in \Lambda_{\mu}(f+\lambda c)$
  \begin{align*}
    \int f \, d\pi + \psi \left(\int c \, d \pi \right) \geq \int \phi \, d\mu
    - \psi^*(\lambda)
  \end{align*}
  This nearly follows by definition:
  \begin{align}
    \int f \, d\pi + \psi\left(\int c \, d\pi\right) &= \int f \, d\pi +
                                                       \sup_{\eta \in \R} \,\,
    \eta\int c \, d\pi - \psi^*(\eta) \notag\\
    &= \sup_{\eta\in \R} \int \left(f +
      \eta c\right) \, d\pi - \psi^*(\eta) \notag\\
    &\geq \int \phi \, d\mu - \psi^*(\lambda) \label{eq:wkineq}
  \end{align}
  where the first line is justified by the fact that $\psi$ is convex and closed with
  $\dom(\psi) \subseteq \R$. Therefore, $\psi(x) = \psi^{**}(x)$ for all $x \in
  \R_+ \cup \left\{ \infty \right\}$.

\endproof

\section{Properties of the dual \eqref{eq:dualrig}}\label{ap:reglr}
This section establishes properties of the dual problem \eqref{eq:dualrig}
that are necessary to prove Theorem \ref{thm:sduality}. Define
\begin{equation}\label{eq:dualinner}
  \Kc(g, \mu) := \sup_{\phi \in
    \Lambda_{\mu}\left(g\right)} \,\, \int_{S_1}\phi \, d\mu \txs{where} \douter = \sup_{\lambda \in \R} \,\, \Kc\left(f + \lambda c, \mu\right) - \psi ^* (\lambda)
\end{equation}
where we begin with the ansatz
\begin{equation}\label{eq:hdef}
  \Kc(g, \mu) = \int h_g \, d \mu, \hspace{ 0.3in } h_g \left(x\right) := \inf_{y \in S_1} g \left(x, y\right)
\end{equation}
A small technicality that occurs when writing the relation \eqref{eq:hdef}:
the function $h_g$ need not be Borel measurable even when $g$ is Borel
measurable. This arises from the fact that the sets
\begin{equation*}%\label{eq:hlvlset}
  h_g^{-1} \left((-\infty,a)\right) = \left\{ x : g(x,y) < a \right\}
\end{equation*}
are projections of Borel sets and therefore not necessarily
Borel. The
sets $h_g^{-1} \left((-\infty,a)\right)$ are analytic, however, which
makes them universally measurable and therefore
measurable with respect to the completion of
$\mu$ or any other Borel measure \cite{Srivastava08}. For our purposes,
this means that the lack of Borel measurability is superfluous. One can always
define the right-hand side of \eqref{eq:hdef} to be the integral of $h_g$
under the completion of $\mu$-- assuming the integral is well-defined.

The following lemmas establish \eqref{eq:hdef} and the conditions under which
it's right-hand side is well defined.
\begin{lemma} \label{lem:unimeasdecom}
  For any universally measurable set $U$ and Borel measure $\mu$, there exist Borel sets $B,N$ and a set
  $T$ such that
  \begin{equation*}
  U = B \cup T, \,\,\, \, T \subseteq N \txs{and} \mu(N) = 0
  \end{equation*}
\end{lemma}
\proof
  Since $U$ is contained in the completion of the Borel
  $\sigma$- algebra under $\mu$, we have
  \begin{equation}\label{eq:outermeas}
    \mu(U) = \inf_{\substack{S_n \text{ Borel } \\ U \subseteq \,\, \bigcup_{n \in \N } S_n}} \sum_{n \in
      \N } \mu(S_n)
  \end{equation}
  This implies that there exists a Borel set $S$ such that $U \subseteq S$ and
  $\mu(U) = \mu(S)$. Defining the universally measurable set $D := S
  \setminus U$ and noticing that $\mu(D) = 0$, one can again apply
  \eqref{eq:outermeas} to obtain a Borel measurable $N$ such that $D \subseteq
  N$ and $\mu(N) = 0$. Setting $B = S \setminus N$ it is easy to
  that $B \subseteq U$ and that this set is Borel. Moreover, for $T = U \setminus
  B$ we have $T \subseteq N$.
\endproof
\begin{lemma} \label{lem:unimeas}
  Let $g:S_0 \times S_1 \to \bar{\R}$ be any Borel measurable function and let
  $g_+$ be it's non-negative part. If there exists a $\pi \in
  \Pi\left(\mu\right)$ such that $\int g_+ \, d \pi <
  \infty$, the integral $\int h_g \, d \mu$ is well defined and
  \begin{equation}\label{eq:pmax}
    \Kc(g,\mu) = \sup_{\phi \in \Lambda_{\mu}(g)} \int \phi  \, d \mu
    = \int h_g \, d \mu
  \end{equation}
\end{lemma}
\hfill
\proof
  Note the following trivial inequality
  \begin{equation}\label{eq:ordering}
    g(x,y) \geq h_g(x) \geq \phi(x) \hspace{ 0.2in }  (x,y) \in S_0 \times S_1,
    \,\, \phi \in \Lambda_{\mu}(g)
  \end{equation}
  and consider the functions
  \begin{equation*}
    p_k := \max \left(g, -k\right) \hspace{ 0.2in } z_k(x) := \Ep_{\pi}\left[p_k \,\, | \,\, x\right]
  \end{equation*}
  Clearly, $z_k$ exists $\mu$ almost everywhere and is integrable for all $k
  \in \N$ since $\int g_+ \, d \pi < \infty$ for some $\pi \in \Pi(\mu)$.
  Notice that \eqref{eq:ordering} implies
  \begin{equation*}
   z_k(x) \geq h_g(x) \;\;\;\; \mu \text{
      a.s }, \,\,\, \forall k \in \N
  \end{equation*}
  Thus, $\int h_g \, d \mu$ is well defined and, by Fatou's lemma
  \begin{equation}\label{eq:intlwrbd}
    \int g \, d \pi \geq \limsup_{k \to \infty} \int z_k \, d \mu \geq \int
    h_g \, d \mu
  \end{equation}
  Now, observe that $\Lambda_{\mu}(g) = \emptyset$ implies that
  $\int g \, d \pi = -\infty$. Hence, without loss of generality,
  we can assume that $\Lambda_{\mu}(g) \neq \emptyset$ and consider a sequence $\phi_n \in \Lambda_{\mu}(g)$ such that
  \begin{equation*}
    \int \phi_n  \, d \mu \geq \sup_{\phi \in \Lambda_{\mu}(g)} \int \phi  \, d
    \mu - \frac{1}{n}
  \end{equation*}
  From \eqref{eq:ordering} and the fact that $\int g_+ \, d \pi <
  \infty$, it follows that for $\phi^* = \sup_{n \in
    \N } \phi_n $ one has $\phi^* \in \Lambda_{\mu}(g)$. Thus, the
  supremum in \eqref{eq:dualinner} is
  achieved for some $\phi ^* \in \Lambda_{\mu}(g)$. Since, \eqref{eq:ordering} implies
  \begin{equation*}
    \int h \, d\mu \geq \sup_{\phi \in \Lambda_{\mu}(g)} \int \phi \, d \mu
  \end{equation*}
  if it can shown that $\int h_g \, d
  \mu = \int \phi ^*  \, d \mu$, the desired conclusion \eqref{eq:pmax}
  will hold.

  Let $\epsilon > 0$ and consider the universally measurable set
  \begin{equation*}
    A_\epsilon := \left\{ x \in S_0: \phi^*(x) < h_g(x) - \epsilon \right\}
  \end{equation*}
  By Lemma \ref{lem:unimeasdecom}, there exist Borel measurable $B_\epsilon$ and $N_\epsilon$
  such that $B_\epsilon \subseteq A_\epsilon$, $A_\epsilon \subseteq B_\epsilon
  \bigcup N_\epsilon$ and $\mu(N_\epsilon) = 0$. Additionally, observe that
  $\mu\left(B_\epsilon\right) = 0$ by the optimality of $\phi ^* $. Thus,
  \begin{equation*}
    \mu(Z_0) = 0 \txs{where} Z_0 := \bigcup_{k \in \N } \left(B_{1/k} \cup N_{1/k}\right)
  \end{equation*}
  and $\phi ^* = h_g$ on the complement of the Borel measurable set $Z_0$. This gives
  \begin{equation*}
    \int \phi ^*  \, d \mu = \int h_g \, d \mu
  \end{equation*}
\endproof

One can now derive an approximation property
for $\Kc(g,\mu)$ using Lemma \ref{lem:unimeas}.
\begin{lemma}\label{lem:piapprox}
  Let $g:S_0 \times S_1 \to \bar{\R}$ be Borel measurable such that
  there exists a $\pi \in \Pi\left(\mu\right)$ for which
  $\int g_+ \, d \pi < \infty$.
  Then, there exists a sequence of $\pi_n \in \Pi(\mu)$ such that
  \begin{equation}
    \label{eq:dapprox}
    \Kc(g,\mu) = \int h_g \, d \mu = \lim_{n \to \infty} \int g \, d \pi_n
  \end{equation}
\end{lemma}
\proof We give a proof following the design of Lemma 8 in
\cite{BKorig}. First, observe that, since $g$ dominates $h_g$, it is sufficient
to show that there exists a sequence of distributions $\pi_n \in \Pi(\mu)$ such
that
  \begin{equation*}
     \limsup_{n \to \infty} \int g \, d \pi_n \leq \int h_g \, d \mu
  \end{equation*}
  Again, one can consider the notation overloaded so that $\int g
  \, d \pi_n$ denotes both the integral of $g$ with respect to $\pi_n$ and the
  integral with respect to the completion of $\pi_n$.

  Let $n \in \N$ and for any $i \leq
  2n^2$, define the sets
  \begin{equation*}
    G^{(n)}_k := \left\{ (x,y) \in S_0 \times S_1 : \frac{k-1}{n} - n \leq g(x,y) \leq \frac{k}{n} - n \right\}\\
  \end{equation*}
  Also, define
  \begin{equation*}
    G^{(n)}_0 := \left\{ (x,y) \in S_0 \times S_1 : g(x,y) \leq -n \right\}
    \txs{and}  G^{(n)}_{2n^2+1}:= \left\{ (x,y) \in S_0 \times S_1 : g(x,y) \geq n \right\}
    % H_\infty := \left\{ x \in S_1 : h_g(x) = \infty \right\}
    % \txs{and}  G^{(n)}_n:= \left\{ (y,x) \in S_0 \times S_1 : g(y,x) > n \right\}
    % L_n := \left\{ x \in S_1 : n \leq g(y,x) \,\,\, y \in S_0\right\}
    % \txs{and} L_{\infty} := \bigcap_{n \in \N } L_n
  \end{equation*}
  Denoting the projection operation onto $S_0$ by $\text{Proj}_{S_0}(\cdot)$, set
  \begin{equation*}
    Z^{(n)}_i := G^{(n)}_i \setminus \left(\bigcup_{j < i} G^{(n)}_j\right) \txs{and} A^{(n)}_i := \text{
      Proj}_{S_0} \left(Z^{(n)}_i\right)
  \end{equation*}
  and notice that the $Z^{(n)}_i$ are Borel and, therefore, the $A^{(n)}_i$ are universally
  measurable. Also, notice that the $Z_i$ form a partition of $S_0 \times S_1$ and
  the $A_i$ form a partition of $S_0$.

  From the von-Neumann selection theorem
  \cite{Srivastava08}, it follows that, for each $i \leq 2n^2+2$, there exists a universally
  measurable selection $\xi_i : A^{(n)}_i \to S_1$ such that $(\xi_i(x), x) \in Z^{(n)}_i$
  for all $x \in A^{(n)}_i$.
  Since the $A^{(n)}_i$ form a partition of $S_0$, define $\gamma_n: S_0 \to S_1$ to
  be the unique, universally measurable extension of the $2n^2+2$ selections
  $\xi_i$ to all of $S_0$.

  Now, notice that for
  \begin{equation*}
    D_n := \bigcup_{i=0}^{2n^2} A^{(n)}_i
  \end{equation*}
  we have $D_j \subseteq D_k$ for $j \leq k$ and
  \begin{equation}\label{eq:uniapprox}
    h_g(x) \leq g(x,\gamma_{n}(x)) \leq \max \left(h_g(x), -n\right) + \frac{1}{n}
    \hspace{ 0.2in } \forall x \in D_n
  \end{equation}
  Moreover, let $(X,Y) \sim \pi$ (where, without loss of generality, we assume
  that $\pi$ is complete) and consider the law $\pi_n$ of the random variable given by
  \begin{equation*}
    (X_n,Y_n) =
    \begin{cases}
     (X,\gamma_n(X))  & \text{if } X \in D_n \\
      (X,Y) & \text{otherwise} \\
    \end{cases}
  \end{equation*}
  Observe that $\pi_n$ induces a unique Borel measure in $\Pi(\mu)$, which
  we also denote by $\pi_n$.

  By \eqref{eq:uniapprox} and the construction of $(Y_n,X_n)$, we have,
  for all $n \in \N$,
  \begin{equation} \label{eq:ftouineq}
    g(X_n(\omega),Y_n(\omega)) \leq \max \left(g(X(\omega),Y(\omega)), -n\right) + \frac{1}{n}
  \end{equation}
  Moreover, taking \eqref{eq:uniapprox} in the limit as $n \to \infty$, we get
  \begin{equation} \label{eq:ftliminf}
    \limsup_{n \to \infty} \,\, g(X_n(\omega),Y_n(\omega)) \leq h_g(X(\omega))
  \end{equation}
  Technically, it should be noted that taking \eqref{eq:uniapprox} in the limit as $n \to \infty$ does
  not cover $X(\omega) \in
  D_{\infty}$, where $D_{\infty} := \left(\bigcup_{n \in \N} D_n\right)^C$.
  However, this is a trivial technicality since
  \begin{equation*}
    x \in D_{\infty} \txs{$\Rightarrow$} g(x,y) = \infty \,\,\,\,\, \forall y \in S_1
  \end{equation*}
  and therefore \eqref{eq:ftliminf} still holds for $X(\omega) \in D_{\infty}$.
  Since \eqref{eq:ftouineq} implies that the $g(X_n,Y_n)$ have a common,
  integrable upper bound, Fatou's Lemma applies and one obtains
  \begin{equation*}
    \limsup_{n \to \infty} \int g \, d \pi_n \leq \int h_g \, d \mu
  \end{equation*}
\endproof

As a trivial consequence of Lemma \ref{lem:piapprox}, one has the aesthetic result:
\begin{corollary}\label{cor:probsup}
  If $g:S_0 \times S_1 \to \bar{\R}$ is Borel measurable such that
  there exists a $\pi \in \Pi \left(\mu\right)$ where $\int g \, d \pi <
  \infty$, then
  \begin{equation*}
    \sup_{\phi \in \Lambda_{\mu}(g)} \int \phi \, d \mu = \inf_{\pi \in \Pi(\mu)} \int g \, d \pi
  \end{equation*}
\end{corollary}
\proof
  The equality follows from Lemma \ref{lem:piapprox} and the fact that
  \begin{equation*}
    \int h_g \, d \mu \leq \int g \, d \pi
  \end{equation*}
  for all $\pi \in \Pi(\mu)$ such that $\int g \, d \pi < \infty$.
\endproof
This culminates in the chief regularity result needed to establish strong duality:
\begin{proposition}\label{prop:exchange}
  If there exists a $\pi \in \Pi(\mu)$ such that $\int f \, d \pi < \infty$
  and $\int c \, d \pi < \infty$ then
  \begin{equation}\label{eq:dualinfsup}
   \douter = \sup_{\lambda \in \R} \inf_{\pi \in \Pi(\mu)} \underset{S_0
     \times S_1}{\int} \left(f(y) + \lambda c(x,y)\right) \, d \pi(x,y) - \psi ^* (\lambda)
  \end{equation}
\end{proposition}
\proof
  Since, for each $\lambda \in \R$,
  \begin{equation*}
    \int f(y) + \lambda c(x,y)  \, d \pi(x,y) < \infty
  \end{equation*}
  applying Corollary \ref{cor:probsup} to the function $g(x,y) = f(y) + \lambda c(x,y)$ gives the result.
\endproof
\section{Proof of Theorem \ref{thm:sduality}}\label{ap:strduality}
\proof
  Define the function $\widetilde{f}: \R \to \bar{\R}$
  \begin{equation*}
    \widetilde{f}(x) = \inf_{\pi \in M(x)} \tau_f(\pi)
    \txs{where} M(x) := \left\{ \pi \in \Pi(\mu) : \tau_c(\pi) = x \right\}
  \end{equation*}
  and notice that $\widetilde{f}$ is convex. Indeed, for any $\pi_x, \pi_y \in
  \Pi(\mu)$ such that $\tau_c (\pi_x) = x$ and $\tau_c(\pi_y) = y$, one can
  construct
  \begin{equation*}
    \pi_{\alpha x + (1-\alpha) y} := \alpha \pi_x + (1- \alpha) \pi_y, \,\,\,\, \alpha \in [0,1]
  \end{equation*}
  such that $\pi_{\alpha x + (1-\alpha) y}\in \Pi(\mu)$ and
  $\tau_c(\pi_{\alpha x + (1-\alpha) y}) = \alpha x + (1-\alpha
  ) y$. Hence,
  \begin{align}\label{eq:fconc}
  \widetilde{f}\left(\alpha x + (1- \alpha) y\right) \leq \alpha \int f \, d\pi_x + (1- \alpha) \int f \, d\pi_y = \int f \, d\pi_{\alpha \pi_x + (1-\alpha) \pi_y}
  \end{align}
  and convexity follows by taking an infinimum of the right-hand side of
  \eqref{eq:fconc}. Note that the effective domain of
  $\widetilde{f}$ is $\dom(\widetilde{f}) = \tau_c \left(D\right)$ where $D$ is
  as defined in \eqref{eq:edom}.

  As a first step, we will show that if $f(y) = -\infty$ for some $y \in \R_+$
  then \eqref{eq:strDuality} holds. Let $\pi_n \in \Pi(\mu)$ be a sequence
  such that $\tau_c(\pi_n) = y$ for all $n$ and $\lim_{n \to \infty}
  \tau_f(\pi_n) = -\infty$. By the hypothesis of Theorem \ref{thm:sduality}, there also exists a $\pi ^*  \in \Pi(\mu)$ such that
  \begin{equation}\label{eq:pistr}
    x = \tau_c\left(\pi ^* \right) \in S \txs{and} \tau_f\left(\pi ^* \right) < \infty
  \end{equation}
  Moreover, if $y \neq x$, then $\rint \left(\tau_c(D)\right)$ is an open interval; and combined with the fact that $0 \in \dom
  \left(\psi\right)$ and $S \neq \emptyset$, $S$ must contain an open interval. Hence, no matter if $y = x$ or $y
  \neq x$, there exists an $\alpha ^* \neq 0$ such that
  \begin{equation*}
    \tau_c \left(\pi_{\alpha ^* \pi ^* + (1- \alpha ^* ) \pi_n}\right) = \alpha
    ^* x + (1- \alpha ^* )y \in S \txs{and} \lim_{n \to \infty} \tau_f
    \left(\pi_{\alpha ^* \pi ^* + (1- \alpha ^* ) \pi_n}\right) = -\infty
  \end{equation*}
  This gives $\widetilde{f}(\alpha ^* x + (1- \alpha ^* )y) = -\infty$ and
  $\psi(\alpha ^* x + (1- \alpha ^* )y) < \infty$; or, in other words,
  $\Pc(f,\mu,c) = -\infty$. In this case, weak duality
  \eqref{eq:wduality} implies strong duality. Thus, without loss
  of generality, one can assume that $-\infty < f(y)$ for all $y \in
  \R_+$. Additionally, since $\tau_f \left(\pi ^* \right) < \infty$,
  $\widetilde{f}$ is finite at the point $x$. Therefore, $\widetilde{f}$ is a
  proper, convex function; this also implies that the function $-\widetilde{f}$
  is a proper, concave function.

  Now, to finish the proof, observe that, since $\pi ^* $ satisfies the conditions of
  Proposition \ref{prop:exchange}, the dual problem \eqref{eq:dualinfsup} can be rewritten as
  \begin{align*}
   \douter &= \sup_{\lambda \in \R} \inf_{\pi \in \Pi(\mu)} \underset{S_0 \times S_1}{\int} \left(f(y) + \lambda c(x,y)\right) \, d
    \pi(x,y) - \psi ^* (\lambda)\\
    &= \sup_{\lambda \in \R} \,\, \inf_{x \in \ran \left(\tau_c\right)} \,\,
      \widetilde{f}(x) + \lambda x - \psi ^* (\lambda)\\
    &= \sup_{\lambda \in \R} \widetilde{f}_{-}^*(\lambda) - \psi ^* (\lambda)
  \end{align*}
  where $\widetilde{f}_{-}^*: \R \to \bar{\R}$ denotes the concave conjugate of
  the concave function $-\widetilde{f}$. Since the primal problem trivially has the expression
  $\pouter = \inf_{x \in \R} \psi(x) - \left(-\widetilde{f}(x)\right)$, showing \eqref{eq:strDuality} is
  equivalent to
  \begin{align}\label{eq:rckdual}
   \inf_{x \in \R} \psi(x) - \left(-\widetilde{f}(x)\right) = \sup_{\lambda \in
    \R} \widetilde{f}_-^*(\lambda) - \psi ^* (\lambda)
  \end{align}
  By Fenchel-Rockafeller duality \cite[Theorem 31.1]{Rockafellar70}, \eqref{eq:rckdual} holds
  if $\psi$ is proper convex, $-\widetilde{f}$ is proper concave, and
  \begin{equation}\label{eq:frint}
    \rint \left(\dom \left(\psi\right)\right) \cap \rint \left(\dom
    \left(-\widetilde{f}\right)\right)\neq \emptyset
  \end{equation}
  However, since $0 \in \dom \left(\psi \right)$ but $0 \not \in
  \rint (\tau_c(D)) \subseteq \R_+$, one has
  \begin{equation*}
    \dom \left(\psi \right) \cap \rint \left(\tau_c(D)\right) \neq \emptyset
    \txs{$\Rightarrow$} \rint \left(\dom\left(\psi \right)\right) \cap \rint
    \left(\tau_c(D)\right) \neq \emptyset
  \end{equation*}
  As $\tau_c(D) = \dom \left(-\widetilde{f}\right)$, it follows that \eqref{eq:frint}
  holds, giving strong duality via \eqref{eq:rckdual}.
\endproof

%%% Local Variables:
%%% mode: latex
%%% TeX-master: "main"
%%% End:

\section{Charaterization of gradients for $g$ \eqref{eq:gfunc} and primal-dual
  gap bounds}\label{sec:subgradproof}
This section establishes key properties of \eqref{eq:gfunc} that are used
to provide the results of sub-section \ref{subsec:pdalgo}. These
results are of independent interest, however, since (through the similar
arguments) they could be used in establishing analogous computational bounds for
other cost functions in \eqref{eq:gfunc}.

Recalling \eqref{eq:dual}, define the set of $\delta$-optimizers of the
$c$-transform as
  \begin{equation*}
    \Yc_\delta(\lambda,x) := \left\{ y \in S_1 : f(y) + \lambda c(x,y) \leq
      f^{\lambda c}(x) + \delta \right\}
  \end{equation*}
and let $z_x(\lambda) := f^{\lambda c}(x)$ denote the $c$-transform of $f$ as a
function of $\lambda$. Clearly, this function is concave, upper-semicontinuous,
and non-decreasing.
\begin{lemma}\label{lem:gradint}
  If $z_x$ is proper, then the right derivative of $z_x$ satisfies
  \begin{equation}\label{eq:zsupgrad}
    \partial_+z_x(\lambda) = \lim_{ \delta \to 0 }\,\, \inf_{y \in \Yc_\delta (\lambda, x) } c(x,y)
  \end{equation}
\end{lemma}
\proof
Let
\begin{equation*}
  T := \left\{ (x,y) \in S_0 \times S_1 : f(y) \text{ and } c(x,y) \text{ are finite}\right\}
\end{equation*}
Since $z_x$ is proper, one can write
\begin{equation*}
  z_x(\lambda) = \inf_{(x,y) \in T} w_{x,y}(\lambda) \txs{where}
  w_{x,y}(\lambda) := f(y) + \lambda c(x,y)
\end{equation*}
Moreover, the $w_{x,y}$ are closed/upper-semicontinuous so
one has $z_x = z_x ^{**} = \inf_{(x,y) \in T} w_{x,y}^{**}$, where
$\left(\cdot\right)^{**}$ denotes the biconjugate in the concave sense.

From this, Theorem 4 in \cite{Lopez08} permits the characterization:
\begin{equation}\label{eq:spgradchar}
  \partial z_x(\lambda) = \bigcap_{\delta > 0} \text{cl conv}
  \left(\bigcup_{(x,y) \in T_\delta(\lambda)} \partial_\delta w_{x,y}(\lambda)
    - N_{\dom (z_x)}(\lambda)\right)
\end{equation}
where $\text{cl conv} \left(\cdot\right)$ denotes the convex closure, $N_{\dom
  (z_x)}$ is the normal cone of $\dom (z_x)$, $\partial_\delta w_{x,y}(\lambda)$
denotes the $\delta$-superdifferential of $w_{x,y}(\lambda)$, and
\begin{equation*}
  T_\delta(\lambda) = \left\{ (x,y) \in T : w_{x,y}(\lambda) \leq z_x(\lambda) +
  \delta \right\}
\end{equation*}
is the set of $(x,y) \in T$ which are $\delta$ optimal. Recall that
$\partial_\delta w_{x,y}(\lambda)$ is the set of $t \in \R$ such that
\begin{equation*}
  w_{x,y}(\xi) \leq w_{x,y}(\lambda) + t(\xi - \lambda) + \delta, \hspace{ 0.2in
  } \forall \,\,\xi \in \R
\end{equation*}
Since $w_{x,y}$ is affine for all $(x,y) \in T$, this set is identical to
the usual superdifferential of $w_{x,y}$ and one has
\begin{equation}\label{eq:wspgrad}
  \partial_\delta w_{x,y}(\lambda) = c(x,y), \hspace{ 0.15in } \forall \, \delta \geq 0
\end{equation}

Additionally, observe that either $\dom (z_x)
= \R$ or $\dom (z_x) = [a, \infty)$ for some $a \in \R$,
since $z_x$ is proper, closed, concave, and non-decreasing. Hence,
\begin{equation}\label{eq:nspgrad}
  N_{\dom \left(z_x\right)}(\lambda) = \begin{cases} \R_- & \text{if } \lambda
    = a \\ 0 & \text{otherwise}\end{cases}
\end{equation}
Using \eqref{eq:wspgrad} and \eqref{eq:nspgrad} to simplify
\eqref{eq:spgradchar}, we obtain
\begin{equation}\label{eq:spgradchar2}
  \partial z_x(\lambda) = \bigcap_{\delta > 0} \text{cl conv} \left(A_\lambda\right)
\end{equation}
where
\begin{equation}
  \label{eq:spgradchar3}
  A(a) := \bigcup_{y \in \Yc_\delta (a, x)} \bigr[ c(x,y), \infty \bigr)
  \hspace{ 0.16in } \text{and otherwise} \hspace{ 0.16in }A(\lambda) := \bigcup_{y \in \Yc_\delta (\lambda, x)} \left\{ c(x,y)\right\}
\end{equation}
Since $\partial_+ z_x(\lambda) = \min \partial z_x(\lambda)$, \eqref{eq:zsupgrad} directly follows from
\eqref{eq:spgradchar2} and \eqref{eq:spgradchar3}.
\endproof
\begin{lemma}\label{lem:subgrad}
  If sufficient conditions for strong duality \eqref{eq:domconds} hold,
  $g(\lambda)$ \eqref{eq:gfunc} is upper-semicontinuous. Further, for any
  $\lambda \in \R$ at which the right derivative of $g$ exists (denoted
  $\partial_+ g(\lambda)$):
  \begin{equation}\label{eq:sgradlb}
    \partial_+ g(\lambda) = \Ep_{x \sim \mu}\left[\lim_{\delta \to 0} \, \inf_{y \in
        \Yc_\delta(\lambda,x)} c(x,y)\right]
  \end{equation}
\end{lemma}
\proof
Since $g$ is non-decreasing, it is sufficient to show that $g$ is continuous
from the right in order to prove that it is upper-semicontinuous. To this end,
observe that duality conditions \eqref{eq:domconds} imply
\begin{equation}\label{eq:gprop}
  g(\lambda) = \Ep_{\mu}\left[z_x(\lambda)\right] < \infty
\end{equation}
for all $\lambda \in \R$. Hence, for any $\lambda_n \downarrow a$, the monotone
convergence theorem applies to the sequence $z_x(\lambda_n) \downarrow z_x(a)$.
This gives $\lim_{\lambda \to a+}g(\lambda) = g(a)$.

To show \eqref{eq:sgradlb}, define
\begin{equation*}
  D_{-\infty} := \left\{ x : \forall \lambda \in \R, \,\, z_x(\lambda) = -\infty
  \right\} \txs{and} D_\infty := \left\{ x : \exists \lambda \in \R, \,\, z_x(\lambda) = \infty \right\}
\end{equation*}
The first claim is that $D_{-\infty}$ and $D_\infty$ are universally measurable.
Indeed, since $z_x(\lambda)$ is point-wise non-decreasing in $\lambda$:
\begin{equation*}
  D_{-\infty} = \bigcap_{q \in \Q}\left\{ x : z_x(q) = -\infty
  \right\} \txs{and} D_\infty = \bigcup_{q \in \Q}\left\{ x : z_x(q) = \infty \right\}
\end{equation*}
Thus, the universal measurability of $z_x(\lambda)$ (see Appendix
\ref{ap:reglr}) gives measurability of $D_{-\infty}$ and $D_\infty$.

Now, without loss of generality, assume $\dom \left(g\right) \neq \emptyset$.
Since $g(\lambda) > -\infty$ for some $\lambda \in \R$,
$\mu \left(D_{-\infty}\right) = 0$. Further, by using \eqref{eq:gprop}, one has
$\mu\left(\left\{ x : z_x(q) = \infty \right\}\right) = 0$ for all $q \in
\Q$, implying $\mu\left(D_\infty\right) = 0$. Thus,
\begin{equation*}
  \mu \left(D_{-\infty} \cup D_{\infty}\right) = 0
\end{equation*}
Since $z_x(\cdot)$ is a proper, concave function for
$x \in D_{-\infty}^C \cap D_{\infty}^C$, this implies that $z_x$ is a proper,
concave function in $\lambda$ for $\mu$-almost every $x$.

To complete the proof, notice that Lemma \ref{lem:gradint} can now be used to
conclude that $\partial_+ z_x(\lambda)$ is given
by \eqref{eq:zsupgrad} for $\mu$-almost every $x$. Moreover, from
\cite{Strassen65}, one has
\begin{equation}\label{eq:strassres}
  \hspace{0.1in}\partial_+ g(\lambda) = \Ep_{\mu}\left[\partial_+ z_x(\lambda)\right],
  \hspace{ 0.1in }\lambda \in \intr \left(\dom (g)\right)
\end{equation}
since $g$ is finite on $\intr \left(\dom (g)\right)$. This gives the conclusion
\begin{equation*}
  \partial_+ g(\lambda) = \Ep_{\mu}\left[\lim_{\delta \to 0} \inf_{y \in
      \Yc_\delta (\lambda, x)} c(x,y)\right],
  \hspace{ 0.1in }\lambda \in \intr \left(\dom (g)\right)
\end{equation*}

Finally, to show the desired conclusion \eqref{eq:sgradlb} on the boundary
$\partial \left(\dom (g)\right)$, let $a \in \partial \left(\dom (g)\right)$ and
assume $\partial_+ g(a)$ exists. Since $g$ is upper-semicontinuous and concave,
\begin{equation}\label{eq:strassres2}
  \partial_+ g(a) = \lim_{\lambda \downarrow a} \partial_+ g(\lambda) = \lim_{\lambda
    \downarrow a} \Ep_{\mu}\left[\partial_+ z_x(\lambda)\right]
\end{equation}
where the second equality follows from \eqref{eq:strassres} and
$\intr \left(\dom (g)\right) = (a, \infty)$.
Noticing that $z_x(\lambda)$ is concave
and non-decreasing in $\lambda$, it follows that $\partial_+ z_x(\lambda)$
is non-negative and non-increasing in $\lambda$. Hence, \eqref{eq:strassres2}
and monotone convergence now give the desired result \eqref{eq:sgradlb}.
\endproof
  \begin{lemma}\label{lem:pdgap}
    Let $\lambda \in \dom(g)$ and assume that there exists a $\pi \in
    \Pi(\mu)$ which satisfies
    \begin{equation*}
      \int_{S_0 \times S_1} f(y) + \lambda c(x,y) \, d\pi = \int_{S_0} h_{f + \lambda c}(x) \, d\mu(x)
    \end{equation*}
    Then, for any $\lambda ^* \in \partial \psi \left(\int c \, d \pi \right)$
    and any $t \in \partial \psi ^*(\lambda)$, one has
    \begin{equation}\label{eq:pdgap1}
      \int f \, d\pi + \psi\left( \int c \, d\pi \right) -
      (\lambda - \lambda ^*) \left(t - \int c \, d \pi\right)
      \leq g(\lambda) - \psi ^*(\lambda)
    \end{equation}
    Additionally, if $\psi$ is $M$-Holder smooth with exponent $\nu$ then
    \begin{equation}\label{eq:pdgap2}
      \int f \, d\pi + \psi\left( \int c \, d\pi \right) - M \left|t - \int c \,
        d\pi \right|^{1 + \nu}
      \leq g(\lambda) - \psi ^*(\lambda)
    \end{equation}
  \end{lemma}
  \proof Define $z := \int c \, d \pi$ and let
  $\lambda \in \dom(g)$ and $\lambda ^* \in \partial \psi(z)$. Since
  $\lambda ^*$ is a subgradient at $z$, one has the identity
\begin{equation*}
  \int \lambda ^* c(x,y) \, d\pi = \psi ^*(\lambda^*) + \psi(z)
\end{equation*}
From this, it follows that
\begin{align*}
  g(\lambda) - \psi ^*(\lambda) &= \int f(y) + \lambda c(x,y) \, d\pi - \psi
                                  ^*(\lambda) \\
                                &= \int f \, d\pi + \psi(z) + \psi ^*(\lambda^*)
                                  + (\lambda - \lambda^*)z  - \psi ^*(\lambda)\\
                                &= \int f \, d\pi + \psi(z) -
                                  \left(\psi ^*(\lambda) - \psi ^*(\lambda
                                  ^*)-z (\lambda - \lambda ^*)\right)\\
                                &\geq \int f \, d\pi + \psi(z) -
                                  (\lambda - \lambda^*)
                                  \left(t-z\right)
\end{align*}
which gives \eqref{eq:pdgap1}. If $\psi$
is also $M$-Holder smooth with exponent $\nu$, then $\lambda = \psi '(t)$ and
$\lambda ^* = \psi'(z)$. Hence, one obtains \eqref{eq:pdgap2} from
\eqref{eq:pdgap1} and the inequality
\begin{equation*}
  |\psi '(t) - \psi '(z)| \leq M |t - z|^\nu \\[-1em]
\end{equation*}  \endproof

\section{Optimality conditions for \eqref{eq:2normdual}}
This section provides bounds \eqref{eq:intvbnds} on the magnitude of a
near-optimal decision variable for \eqref{eq:2normdual}. These bounds are used
to establish the computational complexities of Theorem \ref{thm:primaldual}.
  \begin{lemma} \label{lem:minlb2}
    If $f$ is $L$-smooth \eqref{eq:smoothness} and $L < \lambda$ then,
    for any $\epsilon >0$ there exists $\delta > 0$, such that all
    $\delta$-optimizers
  \begin{equation*}
    f(y_\delta) + \frac{\lambda}{2} \norm{y_\delta-x}^2 \leq \inf_{y \in \R^{d}}  f(y) + \frac{\lambda}{2} \norm{y-x}^2 + \delta
  \end{equation*}
  satisfy $\norm{y_\delta -x} \geq \frac{\norm{\nabla f(x)}}{2 \lambda} - \epsilon$.
\end{lemma}
\proof
    From $L$-smoothness and the fact $\lambda > L$, the function
    \begin{equation*}
      v(y) := f(y) + \frac{\lambda}{2}\norm{y-x}^2
    \end{equation*}
    is ($\lambda - L$)-strongly convex. Thus, it has a unique minimizer $y ^*$ and
    for any $\epsilon$ there exists a $\delta >0$ such that
    \begin{equation*}
      \norm{y ^* - y_\delta} \leq \epsilon
    \end{equation*}
    for any $\delta$-optimizer $y_\delta$. Hence, it is sufficient to show that
    \begin{equation*}
      \frac{\norm{\nabla f(x)}}{2 \lambda} \leq \norm{y ^* - x}
    \end{equation*}
    to prove the desired result. To do this, notice that
    \begin{equation}\label{eq:yfocond}
      \nabla f( y ^*) + \lambda (y ^* - x) = 0
    \end{equation}
    by first-order optimality conditions for $y ^*$. Combining
    \eqref{eq:yfocond} with the $L$-smoothness of $f$, one obtains
    \begin{align}
      \norm{\nabla f(x) - \nabla f(y ^*)}^2 &\leq L^2 \norm{x - y ^*}^2 \notag\\
      \Rightarrow \norm{\nabla f(x)}^2 + \left(\lambda^2 - L^2\right)\norm{x-y ^*}^2 &\leq 2
      \lambda \nabla f(x)^T(x - y ^*) \leq 2 \lambda \norm{\nabla f(x)}\norm{x-y ^*}\label{eq:distlb}
    \end{align}
    Using the fact that $\lambda > L$, the desired result then follows directly
    from \eqref{eq:distlb}.
    \endproof
\begin{proposition}\label{prop:tractcond}
  Let $S_0,S_1=\R^{d}$ and $c(x,y) = \norm{x-y}^2/2$. If $f$ is differentiable and
  $\scparam$-semiconvex then,
  for any $\epsilon > 0$, there exists a
  $\lambda_\epsilon \leq \scparam + \Ep_{\mu}\left[\norm{\nabla f(x)}^2\right]/(2
  \epsilon)$ such that
  \begin{equation}\label{eq:lamub}
    \left(\sup_{\lambda \in \R}\, g(\lambda) - \psi ^* (\lambda)\right) -
    \left(g(\lambda_\epsilon) - \psi ^* (\lambda_\epsilon)\right) \leq \epsilon
  \end{equation}
  Further, if $f$ is $L$-smooth and
  \begin{equation} \label{eq:dualgradcond}
    \partial_- \psi ^*(\scparam) = \frac{\Ep_{\mu}\left[\norm{\nabla f(x)}^2\right]}{C}
  \end{equation}
  for $C \geq 8L^2$, then $\lambda_\epsilon$ can be chosen in the interval
  $[l, u] \subseteq \R$ for
  \begin{equation}\label{eq:intvbnds}
    l = \scparam \hspace{ 0.2in } \text{and} \hspace{
      0.2in } u = \min \left(\beta, \scparam + \sqrt{2C}\right)
  \end{equation}
  where $\beta = \scparam + \Ep_{\mu}\left[\norm{\nabla f(x)}^2\right]/(2 \epsilon)$
\end{proposition}
    \proof
    For any $\hat{\lambda} \geq \scparam$, $\scparam$-semiconvexity of $f$ provides the lower bound
    \begin{equation*}
    g(\hat{\lambda}) = \Ep_{\mu}\left[\inf_{y \in \R^{d}} f(y) +
    \frac{\hat{\lambda}}{2} \norm{y-x}^2\right] \geq \Ep_{\mu}\left[f(x)\right]
    - \frac{1}{2(\hat{\lambda} - \scparam)}\Ep_{\mu}\left[\norm{\nabla
        f(x)}^2\right]
  \end{equation*}
  Since
  \begin{equation*}
    g(\lambda) = \Ep_{\mu}\left[\inf_{y \in \R^{d}} f(y) + \frac{\lambda}{2}
    \norm{y-x}^2\right] \leq \Ep_{\mu}\left[f(x)\right] \,\,\, \forall
  \lambda \in \R
  \end{equation*}
  one obtains the identity
  \begin{equation}\label{eq:objsmoothlb}
    g(\hat{\lambda}) - \psi ^*(\hat{\lambda}) \geq \left(g(\lambda) - \psi
      ^*(\lambda)\right) - \frac{1}{2(\hat{\lambda} - \scparam)}\Ep_{\mu}\left[\norm{\nabla f(x)}^2\right] + \left(\psi
      ^*(\lambda) - \psi ^*(\hat{\lambda})\right)
  \end{equation}
  for any $\hat{\lambda} \geq \scparam$ and $\lambda \in \dom(\psi ^*)$. Via
  \eqref{eq:objsmoothlb}, Proposition \ref{prop:tractcond} can be easily established;
  indeed let us first show \eqref{eq:lamub}.

  Define $\lambda_n \in \R$ be an optimizing sequence for \eqref{eq:gfunc}
\begin{equation*}
  \lim_{n \to  \infty}g(\lambda_n) - \psi ^* (\lambda_n) = \sup_{\lambda \in \R} g(\lambda) - \psi ^* (\lambda)
\end{equation*}
and set
$\beta := \scparam + \Ep_{\mu}\left[\norm{\nabla f(x)}^2\right]/(2\epsilon)$. Since
$\psi ^* $ is lower-semicontinuous and $g$ is upper-semicontinuous ($L$-smoothness
of $f$ \eqref{eq:smoothness} guarantees that Lemma \ref{lem:subgrad} applies),
it is sufficient to show that there exists a $\lambda_\epsilon \leq \beta$
satisfying \eqref{eq:lamub} if
$\beta < \underset{n \to \infty}{\liminf} \,\, \lambda_n$.

Since $\beta < \liminf_{n \to \infty} \lambda_n$, one can assume without loss of
generality that $\beta < \lambda_n$ for all $n \in \N$. As $\psi ^*$ is non-decreasing
(the domain of $\psi$ is $\R_+$), this gives
\begin{equation}\label{eq:psibd}
  \psi ^* (\beta) \leq \psi ^* (\lambda_n) \hspace{
    0.2in } \forall \, n \in \N
\end{equation}
Substituting $\hat{\lambda} = \beta$ and $\lambda = \lambda_n$ in
\eqref{eq:objsmoothlb}, \eqref{eq:psibd} and algebraic simplification provide
  \begin{equation}
    g(\beta) - \psi ^* (\beta) \geq g(\lambda_n) - \psi ^* (\lambda_n) - \epsilon \label{eq:limlb}
  \end{equation}
  Taking the limit in \eqref{eq:limlb} and setting $\lambda_\epsilon = \beta$
  gives the desired result \eqref{eq:lamub}.

  To show the second half of Proposition \ref{prop:tractcond}, observe that the
  previous result implies one can assume
  $\liminf_{n \to \infty} \lambda_n \leq \beta$ for an optimizing sequence
  $\lambda_n$. Otherwise,
  $\beta$ is
  $\epsilon$-optimal and the second half of Proposition \ref{prop:tractcond} is
  trivially true. The immediate consequence of this assumption is that an
  optimizer $\lambda ^* $ of \eqref{eq:gfunc} exists. Indeed,
  $L$-smoothness of $f$ provides $g(\lambda) = -\infty$ for any $\lambda < -L$
  and, combined with $\liminf_{n \to \infty} \lambda_n \leq \beta$, the
  optimizing sequence $\lambda_n$ can be assumed to be bounded. Via
  Bolzano-Weierstrauss, the sequence is therefore convergent to some
  $\lambda ^* \leq \beta$ and upper-semicontinuity of $g$ along with
  lower-semicontinuity of $\psi ^* $ then imply that $\lambda ^* $ is an
  optimizer of \eqref{eq:gfunc}.

  The main consequence of the existence of $\lambda ^*$ is that,
  in combination with \eqref{eq:objsmoothlb}, one has the upper bound
  \begin{align*}
    \psi^*(\lambda ^*) -
    \psi^*(\lambda) - \frac{1}{2(\lambda - \scparam)_+}
    \Ep_{\mu}\left[\norm{\nabla f(x)}^2\right] + g(\lambda ^*) -
    \psi^*(\lambda ^*) &\leq g(\lambda) - \psi^*(\lambda)\\
    \Rightarrow \hspace{ 0.1in }  \psi^*(\lambda ^*) - \psi^*(\lambda) &\leq \frac{1}{2(\lambda - \scparam)_+}
    \Ep_{\mu}\left[\norm{\nabla f(x)}^2\right]
  \end{align*}
  for any $\lambda \in \dom(\psi^*)$, where $\left(\cdot\right)_+$ denotes the
  non-negative part. If $\lambda \leq \lambda ^*$, then the convexity of
  $\psi ^*$ gives
  \begin{align}
    (\lambda - \scparam)_+\psi^*(\lambda ^*) - \psi^*(\lambda) &\leq \frac{1}{2}
    \Ep_{\mu}\left[\norm{\nabla f(x)}^2\right] \notag\\[0.4em]
    \Rightarrow \hspace{ 0.1in }  (\lambda - \scparam)_+(\lambda ^* - \lambda) \partial_+ \psi ^*(\lambda) &\leq \frac{1}{2}
    \Ep_{\mu}\left[\norm{\nabla f(x)}^2\right] \label{eq:dualupperbd}
  \end{align}
  Taking
  $\lambda = (\lambda ^* + \scparam)/2$ in \eqref{eq:dualupperbd} will lead to
  the desired conclusion of Proposition \ref{prop:tractcond}-- so long as
  $\scparam \leq \lambda ^*$. To show that \eqref{eq:dualgradcond} implies
  $\scparam \leq \lambda ^*$, observe that, in the notation of Lemma
  \ref{lem:subgrad},
    \begin{equation}\label{eq:costoptlb}
      \frac{\norm{\nabla f(x)}^2}{2(2 \lambda)^2} \leq \lim_{\delta \to 0}
      \inf_{y \in \Yc_\delta(\lambda,x)} c(x,y), \,\,\,\,\, \, \lambda > L
    \end{equation}
    by Lemma \ref{lem:minlb2}. In combination with the result of Lemma
    \ref{lem:subgrad}, this yields
    \begin{equation}\label{eq:direcderlb}
      \frac{1}{8\lambda^2}\Ep_{\mu}\left[\norm{\nabla f(x)}^2\right] \leq
      \partial_+g(\lambda), \,\,\,\,\, \lambda \geq L
    \end{equation}
    Indeed, since $g$ is upper-semicontinuous by Lemma
    \ref{lem:subgrad}, $\lim_{\lambda \downarrow L}\partial_+g(\lambda) =
    \partial_+g(L)$ and it is sufficient that \eqref{eq:costoptlb} hold for
    $\lambda > L$ to obtain \eqref{eq:direcderlb} for $\lambda \geq L$.
    Under \eqref{eq:dualgradcond}, \eqref{eq:direcderlb} produces the relation
    \begin{equation}\label{eq:optatL}
      \partial_-\psi ^*(\scparam) \leq \frac{1}{8L^2}\Ep_{\mu}\left[\norm{\nabla
          f(x)}^2\right] \leq \partial_+g(L) \leq \partial_+g(\scparam)
    \end{equation}
    since $\scparam \leq L$. As $g$ is concave and $\psi$ is convex,
    \eqref{eq:optatL} immediately gives
    $g(\scparam) - \psi ^*(\scparam) \geq g(\lambda) - \psi ^*(\lambda)$
    for all $\lambda < \scparam$. Hence, $\lambda ^*$ can be chosen so that
    $\scparam \leq \lambda ^*$.

    Finally, using the fact that $\scparam \leq \lambda ^*$ and substituting
    $\lambda = (\lambda ^* + \scparam)/2$ into \eqref{eq:dualupperbd}, one obtains
  \begin{equation}\label{eq:lmfinalbd}
    \lambda ^* \leq \scparam + \left(\frac{2\Ep_{\mu}\left[\norm{\nabla
            f(x)}^2\right]}{\partial_+\psi ^*((\lambda ^* + \scparam)/2)}\right)^{1/2}
                 \leq \scparam + \sqrt{2C}
  \end{equation}
  where the last inequality is a result of the fact that $\psi ^*$ is convex.
  After combining \eqref{eq:lmfinalbd} with the bounds
  $\scparam \leq \lambda ^*$ and $\lambda ^* \leq \beta$, the final
  conclusion of Proposition \ref{prop:tractcond} follows.
  \endproof

\section{Proof of Proposition \ref{prop:sorcalgp}}
This section establishes the guarantees of Algorithm \ref{alg:sorc} and the
desired result of Proposition \ref{prop:sorcalgp}. Since Algorithm \ref{alg:sorc}
provides a more general oracle than described in Definition
\ref{def:strorprob}, we first give a definition of this oracle. Showing that
Algorithm \ref{alg:sorc} fulfills this broader definition is necessary to
analyze the mirror ascent procedure of Appendix \ref{ap:mirror}.
\begin{definition}[Supergradient oracle in expectation]\label{def:strorexp}
  A function $\theta_g : \R \to \R$ is called a ($\epsilon, V$)-supergradient
  oracle \textit{in expectation} for $g$ (on the interval $[l,u]$) if, when
  queried with a $\lambda \in [l,u]$, it returns an independent random sample
  $\theta_g(\lambda)$ satisfying
  \begin{equation}
    \label{eq:expectoracle}
    \min_{z \in \partial g (\lambda)}\left| \Ep\left[\theta_g(\lambda)\right] -
      z \right| \leq
    \epsilon\hspace{0.2in }\text{and}\hspace{0.2in }\Ep\left[\theta_g(\lambda)^2 \right] \leq V(\lambda)
  \end{equation}
  for $\epsilon \geq 0$ and some function $V : \R \to \R_+$.
\end{definition}
\begin{proposition}\label{prop:sorcalge}
  If $f: \R^{d} \to \R$ is $L$-smooth and $\scparam$-semiconvex
  \eqref{eq:semiconv}, then Algorithm \ref{alg:sorc} implements a
  $\left(\epsilon, V\right)$-supergradient oracle in expectation (Definition
  \ref{def:strorexp}) for $g$ in \eqref{eq:gfunc} on the interval $(\scparam, \infty)$
  where
  \begin{equation}\label{eq:sorceconst}
    V(\lambda) := \frac{256}{(\lambda - \scparam)^4} \Ep_{\mu}\left[\norm{\nabla f(x)}^4\right]
  \end{equation}
\end{proposition}
To prove Proposition \ref{prop:sorcalge}, the following lemma is required.
  \begin{lemma}\label{lem:gdiff}
    If $S_0, S_1 = \R^d$, $c(x,y) = \norm{x-y}^2/2$ and $f$ is differentiable
    and $\scparam$-semiconvex \eqref{eq:semiconv}, then the function $g$
    \eqref{eq:gfunc} is differentiable on $(\scparam, \infty)$ and
    \begin{equation}\label{eq:gradub}
      g'(\lambda) =
      \Ep_{\mu}\left[\frac{1}{2}\norm{y_{\lambda,x}^*-x}^2\right], \hspace{ 0.15in }
      y_{\lambda, x}^* := \, \argmin_{y \in \R^{d}} f(y) + \frac{\lambda}{2}\norm{y-x}^2
    \end{equation}
    where the unique minimizer $y_{\lambda, x}^*$ satisfies
    \begin{equation}\label{eq:umin}
      \frac{1}{2}\norm{y_{\lambda,x}^*-x}^2 \leq
      \frac{2}{\left(\lambda - \scparam\right)^2}\norm{\nabla f(x)}^2
    \end{equation}
    Additionally, for any $\scparam < \lambda_1 \leq \lambda_2$ one has
\begin{equation}\label{eq:gradholder}
  \left(1 - 2\sqrt{\frac{\lambda_2 - \lambda_1}{\lambda_2 - \scparam}}\right)g'(\lambda_1) \leq g'(\lambda_2)
\end{equation}
    This implies that for any $t ^* > \scparam$, $g'$ is $1/2$-Holder continuous
    on $[t ^*, \infty)$ with a constant depending only on $t ^*$ and $\scparam$.
\end{lemma}
\proof
Define the functions
\begin{equation*}
  a_\lambda(y;x) := f(y) + \frac{\lambda}{2}\norm{y-x}^2 \txs{and} z_x(\lambda) := \inf_{y \in \R} a_\lambda(y;x)
\end{equation*}
Since $f$ is $\scparam$-semiconvex \eqref{eq:semiconv}, $a_\lambda(y;x)$ is
$\lambda - \scparam$ strongly convex in $y$ for $\lambda > \scparam$. Therefore,
the minimizer $y_{\lambda,x}$ is unique. Further, semiconvexity and
differentiability of $f$ provide the lower bound
\begin{equation*}
  a_\lambda(y;x) \geq f(x) + l_\lambda(y;x) \txs{where}
  l_\lambda(y;x) := \nabla f(x)^T(y-x) + \frac{\lambda - \scparam}{2}\norm{y-x}^2
\end{equation*}
Noticing $l_\lambda(y;x) > 0$ for any $y \in \R^{d}$ such that $\norm{y-x} > \left(2
  \norm{\nabla f(x)}\right)/\left(\lambda - \scparam\right)$, one obtains \eqref{eq:umin}.

For open subsets $O \subset (\scparam,\infty)$ whose closure does not
contain $\scparam$, \eqref{eq:umin} implies that
the radius of the ball containing $y_{\lambda,x}^*$ is
uniformly bounded for all $\lambda \in O$. Danskin's theorem \cite{Guler10} can,
therefore, be applied to the function $z_x(\lambda) := f^{\lambda c}(x)$ \eqref{eq:dual} to conclude that
$z_x(\lambda)$ is differentiable on $(\scparam, \infty)$ with derivative
\begin{equation*}
  z_x'(\lambda) = \frac{1}{2}\norm{y_{\lambda,x}^*-x}^2
\end{equation*}
Observing that $g(\lambda) = \Ep_{x \sim \mu}\left[z_x(\lambda)\right]$,
the conclusion \eqref{eq:gradub} then follows from
\eqref{eq:umin} and dominated convergence.

Finally, let $\scparam < \lambda_1 \leq  \lambda_2$. Since $z_x(\lambda)$ is
concave in $\lambda$
\begin{equation*}
  \left| z_x'(\lambda_1) - z_x'(\lambda_2)\right| = z_x'(\lambda_1) - z_x'(\lambda_2)
\end{equation*}
and it is enough to show a one-sided bound on the quantity $z_x'(\lambda_1) -
z_x'(\lambda_2)$. To this end, observe
\begin{equation}\label{eq:tribound}
  z_x'(\lambda_1) - z_x'(\lambda_2) \leq \norm{y_{\lambda_1,x}^* - x}\norm{y_{\lambda_2,x}^*-y^*_{\lambda_1,x}}
\end{equation}
Hence, \eqref{eq:gradholder} can be provided by producing a bound on
$\norm{y_{\lambda_2,x}^*-y^*_{\lambda_1,x}}$. Strong convexity of
$a_\lambda(y;x)$ in $y$ yields the identity
\begin{equation*}
  a_{\lambda_2}(y_{\lambda_2,x}^*;x) + \frac{\lambda_2 -
    \scparam}{2}\norm{y_{\lambda_2,x}^*-y ^*_{\lambda_1,x}}^2 \leq
  a_{\lambda_2}(y_{\lambda_1,x}^*;x) = a_{\lambda_1}(y_{\lambda_1,x}^*;x) +
  \frac{\lambda_2 - \lambda_1}{2}\norm{y_{\lambda_1,x}^* - x}^2
\end{equation*}
which, when combined with the fact that
$a_{\lambda_1}(y_{\lambda_1,x}^*;x) \leq a_{\lambda_2}(y_{\lambda_1,x}^*;x)$
($z_x(\lambda)$ is non-decreasing in $\lambda$),
gives
\begin{equation}\label{eq:holdinc}
  \norm{y_{\lambda_2,x}^*-y^*_{\lambda_1,x}} \leq
  \sqrt{\frac{\lambda_2 - \lambda_1}{\lambda_2 - \scparam}}\norm{y_{\lambda_1,x}^* - x}
\end{equation}
Applying \eqref{eq:holdinc} to \eqref{eq:tribound} and rearranging produces
\begin{equation}\label{eq:zsqrtbd}
  \left(1 - 2\sqrt{\frac{\lambda_2 - \lambda_1}{\lambda_2 - \scparam}}\right)z_x'(\lambda_1) \leq z_x'(\lambda_2)
\end{equation}
Taking the expectation with respect to $x$ on both sides of \eqref{eq:zsqrtbd} yields \eqref{eq:gradholder}.

\endproof

\proof[Proof of Proposition \ref{prop:sorcalge}.] Consider the sample $x$ which is
computed by Algorithm \ref{alg:sorc}. In light of Lemma \ref{lem:gdiff}, it is
clear that
\begin{equation*}
  \theta ^* := \frac{1}{2}\norm{y_{\lambda,x}^* - x}^2
\end{equation*}
is an unbiased estimate of $g'(\lambda)$. Hence, to prove the conclusion of
Proposition \ref{prop:sorcalge}, it need only
be shown that the output of Algorithm \ref{alg:sorc}, $\theta$, satisfies
\begin{equation}\label{eq:smdobj}
  |\theta - \theta ^*| \leq \epsilon \txs{and} \theta \leq \left(\frac{4
      \norm{\nabla f(x)}}{\lambda - \scparam}\right)^2
\end{equation}
when $\lambda \in (\scparam, \infty)$.

To this end, notice that Algorithm \ref{alg:sorc} performs Nesterov's
accelerated gradient descent \cite{NesterovConvexBook} on the
$\lambda - \scparam$-strongly convex and $\lambda + L$-smooth function
$a_{\lambda}(y;x)$. Strong convexity yields the identity
\begin{equation}\label{eq:astrconv}
  \frac{\lambda - \scparam}{2}\norm{y ^*_{\lambda, x} - y}^2 \leq a_{\lambda}(y;x) -
  a_{\lambda}(y ^*_{\lambda,x};x)
\end{equation}
while the convergence guarantees of accelerated gradient descent \cite[Theorem
2.2.3]{NesterovConvexBook} give
\begin{equation}\label{eq:agdconv}
  a_{\lambda}(y_k;x) - a_{\lambda}(y^*_{\lambda, x};x) \leq  \left(1 - \kappa\right)^k \left(\lambda + L\right) \norm{y_{\lambda,x}^* - x}^2
\end{equation}
for $\kappa = \sqrt{\left(\lambda + L\right)/ (\lambda - \scparam)}$. Combining these relations
and setting $C = 2\norm{\nabla f(x)}/(\lambda - \scparam)$
\begin{equation}\label{eq:linconv}
  \norm{y ^*_{\lambda, x} - y_k}^2 \leq \frac{2 \left(a_{\lambda}(y_k;x) - a_{\lambda}(y^*_{\lambda, x};x)\right)}{\lambda - \scparam} \leq  2\left(1 - \kappa\right)^k\kappa^2 \norm{y_{\lambda,x}^* - x}^2 \leq 2\left(\frac{\epsilon}{6C}\right)^2
\end{equation}
since
$k \geq 4\kappa\log\left(6\kappa C/\epsilon\right)$ and $\norm{y_{\lambda,x} ^* - x} \leq
C$ via \eqref{eq:umin}. Completing the analysis,
\begin{align}
  |\theta - \theta ^*| &= \frac{1}{2}
                         \left|\norm{y_k - x}^2 - \norm{y_{\lambda,x}^* -
                         x}^2\right| \leq \frac{1}{2}\norm{y_k - y_{\lambda,x}^*}\left(
                          \norm{y_k - x} + \norm{y_{\lambda, x}^* - x}\right) \label{eq:tineq}\\[0.5em]
                       & \leq \frac{3}{2}\norm{y_k -
                         y_{\lambda,x}^*}\norm{y_{\lambda, x}^* - x} \leq \epsilon \label{eq:uniub}
\end{align}
where triangle inequality provides both \eqref{eq:tineq} and
\begin{equation}\label{eq:outub}
  \norm{y_k - x} \leq 2 \norm{y_{\lambda,x} ^* - x} \leq 2 C
\end{equation}
Moreover, \eqref{eq:uniub} is
the desired left-hand inequality of \eqref{eq:smdobj} while \eqref{eq:outub}
contains the desired right-hand inequality-- this completes the proof.

\endproof
With the guarantee on Algorithm \ref{alg:sorc} established by Proposition
\ref{prop:sorcalge}, Proposition \ref{prop:sorcalgp} becomes an immediate corollary.
\proof[Proof of Proposition \ref{prop:sorcalgp}.]
This is a straightforward consequence of Chebyshev's inequality. Indeed,
the proof of Proposition \ref{prop:sorcalge} shows that the output $\theta$ of Algorithm
\ref{alg:sorc} satisfies
\begin{equation}
  \label{eq:errnbnd}
  \left|\Ep\left[\theta\right] - g'(\lambda) \right| \leq
  \frac{\tilde{\epsilon}}{2\max \left(\lambda - \scparam, 1\right)} \txs{and}
  \theta \leq \frac{16}{(\lambda - \scparam)^2}\norm{\nabla f(x)}^2
\end{equation}
when $\epsilon = \tilde{\epsilon}/\left(2\max\left(\lambda - \scparam, 1\right)\right)$.
Letting $\bar{\theta}$ be the average of $K$ independent calls to Algorithm
\ref{alg:sorc}, Chebyshev's inequality gives
\begin{equation}
  \label{eq:cheb}
  \p \left(\left| \bar{\theta} - \Ep\left[\theta\right]  \right| \geq
    \frac{\tilde{\epsilon}}{2\max(\lambda - \scparam, 1)}
  \right) \leq \frac{64\,\,\Ep_{\mu}\left[\norm{\nabla f(x)}^4\right] }{(\lambda
    - \scparam)^2 \min\left(\left(\lambda - \scparam\right)^2, 1\right)\tilde{\epsilon}^2K}\leq \delta
\end{equation} \endproof

\section{Stochastic mirror ascent for \eqref{eq:gfunc}.}\label{ap:mirror}
For completeness with respect to previous approaches \cite{NamDuchi16,Ghosh18},
this section provides an analysis of mirror ascent in the context of
\eqref{eq:gfunc} and \eqref{eq:2normdual}. The main result of this analysis is:
usage of stochastic mirror ascent, under slightly weaker assumptions than those
used to obtain Theorem \ref{thm:primaldual}, provides an output whose expected
objective value (over the randomness of the algorithm) is nearly optimal.
Further, the computational complexity of this procedure has better dependence on
the smoothness of the objective function-- compare \eqref{eq:pdsamplebound} to \eqref{eq:compcplexepsamp}. The
sacrifice is that only an estimate of the optimal value of the
dual \eqref{eq:gfunc} is produced. The output of the algorithm does not provide
a primal-feasible distribution with guarantees on the primal-dual gap
\eqref{eq:basicpdgap}.
\begin{algorithm}[H]
  \caption{Stochastic Mirror Ascent for \eqref{eq:gfunc}}
  \label{alg:innerprob}
  \begin{algorithmic}
    \Input{Supergradient oracle
      $\theta_{g}$, initial iterate $\lambda_1$, step-size $\alpha$,
      number of iterations $k$}
  \For{$1 \leq i \leq k$}
  \State Sample $\eta_i \leftarrow \theta_g(\lambda_i)$
  \State $\xi_i \leftarrow \text{Proj}_{\partial \psi ^*(\lambda_i)} (\eta_i)$ \label{alg:projline}
  \State For $\lambda_{i+1} \leftarrow \text{Proj}_{[l,u]}
  \left(\lambda_i + \frac{\alpha(u-l)}{\sqrt{2k}}\left(\eta_i - \xi_i\right)\right)$
  \EndFor
  \Return $\lambda ^* = \frac{1}{k} \sum_{i=1}^{k}\lambda_i$
  \end{algorithmic}
  \end{algorithm}
  \begin{proposition}[Convergence of Algorithm \ref{alg:innerprob}]\label{prop:smirror}
    For the problem \eqref{eq:gfunc}, let $\theta_{g}$ be a
    ($\epsilon, V$)-supergradient oracle in expectation (Definition
    \ref{def:strorexp}) for $g$ on $[l,u]$. If
    $\sup_{\lambda \in [l,u]} \left|\partial_+ \psi^*(\lambda)\right| \leq D$
    (where $\partial_+$ denotes the right-derivative),
    $\sup_{\lambda \in [l,u]} V(\lambda) \leq C^2$, and
    $\alpha = 1/\sqrt{\left(C^2+D^2\right)}$ then
    \begin{equation}\label{eq:outputg}
      \sup_{\lambda \in [l,u]} g(\lambda) - \psi ^*(\lambda) -
      \Ep\left[g(\lambda ^*) - \psi ^*(\lambda ^*)\right]  \leq \left(u-l\right) \left(\sqrt{\frac{2\left(C^2 +
              D^2\right)}{k}} + \epsilon\right)
    \end{equation}
    where $\lambda ^*$ is the output of Algorithm \ref{alg:innerprob} and the
    expectation is taken with respect to the randomness of the oracle
    $\theta_{g}$.
  \end{proposition}
\proof
    Let $\lambda_i$ be the $i$th iterate computed by Algorithm
    \ref{alg:innerprob} and let $\eta_i$ and $\xi_i$ be the
    corresponding, computed supergradient and subgradients for $g$ and $\psi
    ^*$. By construction, $\lambda_{i+1}$ solves
    \begin{equation*}
      \lambda_{i+1} = \argmax_{\lambda \in [l,u]}\,\, \alpha_k \gamma_i\left(\lambda -
        \lambda_i\right) - \frac{1}{2}\left(\lambda - \lambda_i\right)^2
    \end{equation*}
    where $\alpha_k = \alpha(u-l)/\sqrt{2k}$ and $\gamma_i = \eta_i - \xi_i$.
    From first-order optimality condition
    \begin{equation*}
      \left(\alpha_k \gamma_i - \lambda_{i+1} + \lambda_i \right) \left(\lambda -
         \lambda_{i+1}\right) \leq 0  \hspace{ 0.2in } \forall \lambda \in
      [l,u]
    \end{equation*}
    one obtains
    \begin{equation}\label{eq:3ptineq}
       \alpha_k \gamma_i(\lambda - \lambda_{i+1}) \leq \frac{1}{2}\left(\lambda - \lambda_{i}\right)^2 -
      \frac{1}{2}\left(\lambda - \lambda_{i+1}\right)^2 -
      \frac{1}{2}\left(\lambda_{i+1} - \lambda_i\right)^2
    \end{equation}
    for any fixed $\lambda \in [l,u]$.
    Adding $\alpha_k \gamma_i(\lambda_{i+1} - \lambda_i)$ to both sides of
    \eqref{eq:3ptineq}
    and applying Young's inequality on the right provides the relation
    \begin{equation}\label{eq:yng-ieq}
      \alpha_k \gamma_i(\lambda - \lambda_{i})
      \leq \,\frac{\left(\alpha_k \gamma_i\right)^2}{2} +
      \frac{1}{2}\left(\lambda - \lambda_{i}\right)^2 -
      \frac{1}{2}\left(\lambda - \lambda_{i+1}\right)^2
    \end{equation}
    The equation \eqref{eq:yng-ieq} can then be summed over $i \leq k$ to give
    \begin{equation}\label{eq:sumbd}
      \sum_{i = 1}^{k}\gamma_i(\lambda - \lambda_{i})
      \leq \,\alpha_k\sum_{i = 1}^{k}\frac{\gamma_i^2}{2} +
      \frac{\left(\lambda - \lambda_{1}\right)^2}{2 \alpha_k}
    \end{equation}
    Essentially, what has been obtained is an upper bound on the quantities
    $\gamma_i(\lambda - \lambda_i)$. These quantities, themselves, roughly upper
    bound the difference between the objective value \eqref{eq:gfunc} at
    $\lambda$ and the value at $\lambda_i$. Taking expectations (with respect to
    the randomness of the oracle $\theta_g$) on both sides of \eqref{eq:sumbd},
    \begin{align}
      \Ep\left[\sum_{i = 1}^{k}(\eta_i -\xi_i)(\lambda - \lambda_{i})\right]
      &\leq \,\alpha_k\sum_{i = 1}^{k}\frac{\Ep\left[\gamma_i^2\right] }{2} +
      \frac{\left(\lambda - \lambda_{1}\right)^2}{2 \alpha_k} \notag\\
      &\leq \,k\alpha_k \left(C^2 + D^2\right) +
      \frac{\left(\lambda - \lambda_{1}\right)^2}{2 \alpha_k}\label{eq:optgap}
    \end{align}
    where \eqref{eq:expectoracle},
    $\sup_{\lambda \in [l,u]} V(\lambda) \leq C^2$ and
    $\sup_{\lambda \in [l,u]} \partial_+ \psi^*(\lambda) \leq D$ were used.

    Notice that \eqref{eq:expectoracle} implies there exists
    a $z ^* \in \partial g(\lambda_i)$ such that $\left| \Ep\left[\eta_i \,
        \bigr | \, \eta_j, \, j < i\right] - z ^* \right| \leq \epsilon$.
    Since
    $\xi_i = \text{Proj}_{\partial \psi ^*(\lambda_i)} (\eta_i)$, one has
    $\Ep\left[\xi_i \, \bigr | \, \eta_j, \, j < i\right] \in \partial
    \psi ^*(\lambda_i)$ and this gives
    \begin{align}
      \Ep\left[ (\eta_i -\xi_i)(\lambda - \lambda_{i}) \, \bigr | \, \eta_j,
      \, j < i\right]
      &= \left(\Ep\left[\eta_i\, \bigr | \, \eta_j,
                        \, j < i\right] - \Ep\left[\xi_i\, \bigr | \, \eta_j,
      \, j < i\right]\right)(\lambda - \lambda_{i})
      \label{eq:conde1} \\[0.5em]
                       &\geq  \left(z ^* - \Ep\left[\xi_i\, \bigr | \, \eta_j,
                        \, j < i\right] \right)(\lambda - \lambda_i) - \epsilon
                        (u-l)\notag \\[0.5em]
                       &\geq g(\lambda) - \psi ^*(\lambda) - \left(g(\lambda_i)
                        - \psi ^*(\lambda_i)\right) - \epsilon (u-l) \label{eq:conde2}
    \end{align}
    where \eqref{eq:conde1} is a result of the fact that $\lambda_i$ depends
    only on $\eta_j$ for $j < i$ and \eqref{eq:conde2} follows from the
    concavity of the objective $\lambda \, \mapsto \,g(\lambda) - \psi ^*(\lambda)$.

    Applying the relation \eqref{eq:conde2} to \eqref{eq:optgap},
    \begin{equation*}
      \Ep\left[\sum_{i = 1}^{k}g(\lambda) - \psi ^*(\lambda) - \left(g(\lambda_i)
                        - \psi ^*(\lambda_i)\right)\right]
      \leq \,k\alpha_k \left(C^2 + D^2\right) +
      \frac{\left(\lambda - \lambda_{1}\right)^2}{2 \alpha_k} + \epsilon(u-l)
    \end{equation*}
    Dividing both sides by $1/k$ and substituting $\alpha_k =
    \left(u-l\right)/\sqrt{2k\left(C^2+D^2\right)}$, one obtains
    \begin{equation}\label{eq:fluppbd}
      g(\lambda) - \psi ^*(\lambda) -
      \Ep\left[\frac{1}{k}\sum_{i = 1}^{k}\left(g(\lambda_i) - \psi ^*(\lambda_i)\right)\right]
      \leq \, \left(u-l\right) \left(\sqrt{\frac{2\left(C^2 +
              D^2\right)}{k}} + \epsilon\right)
    \end{equation}
    Since $\lambda \in [l,u]$ is arbitrary and the output of Algorithm
    \ref{alg:innerprob} is given by
    $\lambda ^* = \left(\sum_{i=1}^{k}\lambda_i\right)/k$, the desired result
    \eqref{eq:outputg} follows from \eqref{eq:fluppbd} and concavity of
    $\lambda \mapsto g(\lambda) - \psi ^*(\lambda)$.  \endproof

\begin{proposition}\label{prop:compcplexep}
  Let $f$ be $L$-smooth \eqref{eq:smoothness}, $\scparam$-semiconvex, and assume
  that $\psi$ provides
  $C/\Ep_{\mu}\left[\norm{\nabla f(x)}^2\right]$-regularization
  \eqref{eq:dualgradcond2} at $\scparam + 1$ for $C \geq 8L^2$. If
  $\sup_{\lambda \in [l,u]} \left| \partial_+ \psi^*(\lambda) \right| \leq D$
  for $l$ and $u$ given by \eqref{eq:intvbnds2}, then there
  exists an stochastic algorithm which returns a $\lambda ^*$ such that, recall \eqref{eq:gfunc},
  \begin{equation*}
      \sup_{\lambda \in \R} g(\lambda) - \psi ^*(\lambda) -
      \Ep\left[g(\lambda ^*) - \psi ^*(\lambda ^*)\right]  \leq \epsilon
  \end{equation*}
  where the expectation is taken with respect to the randomness of the
  algorithm. This algorithm draws at most
  \begin{equation}\label{eq:compcplexepsamp}
    O \left(\frac{C\max \left(\Ep_{\mu}\left[\norm{\nabla f(x)}^4\right], D^2\right)}{\epsilon^2} \right)
  \end{equation}
  independent samples from $\mu$ and performs $
    \widetilde{O} \left(L^{1/2}C\max \left(\Ep_{\mu}\left[\norm{\nabla f(x)}^4\right], D^2\right)/\epsilon^2
      \right)$ expected gradient evaluations of $f$--
  where $\widetilde{O}$ suppresses logarithmic factors in
  $L,C, \Ep_{\mu}\left[\norm{\nabla f(x)}\right]$ and $\epsilon$
\end{proposition}
\proof By Proposition
\ref{prop:tractcond}, it is enough for an algorithm to return a value $\lambda^*$
which satisfies
\begin{equation}\label{eq:epbound}
  \sup_{\lambda \in [l,u]} g(\lambda) - \psi ^*(\lambda) - \Ep\left[g(\lambda
    ^*) - \psi ^*(\lambda ^*)\right] \leq \frac{\epsilon}{2}
\end{equation}
for $l$ and $u$ given by \eqref{eq:intvbnds2}. Without loss of generality,
it will be assumed that $u-l > 0$.

Apply Algorithm
\ref{alg:innerprob} to the interval $[l,u]$ with the supergradient oracle given
by Algorithm \ref{alg:sorc}; where the error tolerance in Algorithm \ref{alg:sorc}
is set to $\epsilon/(4(u - l))$. For all $\lambda \in [l,u]$, the
variance bound \eqref{eq:sorceconst} gives
\begin{equation*}
  V(\lambda) \leq 256\, \,\Ep_{\mu}\left[\norm{\nabla f(x)}^4\right]
\end{equation*}
since $\min_{\lambda \in [l,u]} (\lambda - \scparam) \geq 1$. Thus, by Proposition
\ref{prop:smirror}, running Algorithm \ref{alg:innerprob} for
\begin{equation*}
  k = \left \lceil \frac{32\left(u - l\right)^2}{\epsilon^2}
  \left(256\,\Ep_{\mu}\left[\norm{\nabla f(x)}^4\right] +
    D^2\right) \right \rceil
\end{equation*}
iterations will produce a $\lambda^*$ satisfying  \eqref{eq:epbound}. Further, as each
iteration of Algorithm \ref{alg:innerprob} executes a single call to the
supergradient oracle provided by Algorithm \ref{alg:sorc}, it is clear that
\begin{equation*}
    O \left(\frac{C}{\epsilon^2}
      \max \left(\Ep_{\mu}\left[\norm{\nabla f(x)}^4\right], D^2\right)\right)
\end{equation*}
samples are drawn from $\mu$.

To compute a bound on the expected number of gradient evaluations of $f$ that
are performed, observe that each of the $k$ calls to Algorithm
\ref{alg:sorc} (with error tolerance $\epsilon/(4(u - l))$) executes at most
\begin{equation}\label{eq:}
  t = \max \left( \left \lceil 4 \kappa \log \left(\frac{48\kappa\norm{\nabla
          f(x)}(u-l)}{\epsilon}\right) \right \rceil, 0 \right)
\end{equation}
gradient evaluations of $f$; $x$ and $\kappa$ are the random sample and
condition number, respectively, which are used in
Algorithm \ref{alg:sorc}. Both $x$ and $\kappa$
are random variables, but $\kappa = \left((\lambda + L)/(\lambda - \scparam)\right)^{1/2} \leq
(1 + 2L)^{1/2}$ and (due to Jensen)
\begin{equation*}
  \Ep\left[\max \left(\log \left(z\right), 0\right)\right] \leq \log
  \Ep\left[\max \left(z, 1\right)\right]
\end{equation*}
for any non-negative random variable $z$. Hence, the expected number of gradient
evaluations performed by Algorithm \ref{alg:sorc} obeys the bound
\begin{equation}\label{eq:gradevalbnd}
  \Ep_{\mu}\left[ t \right] \leq 4
\left(1 + 2L\right)^{1/2}\log \Ep_{\mu}\left[\left(\max \left(\frac{48 (1 +
      2L)^{1/2}\norm{\nabla
          f(x)}(u-l)}{\epsilon}, 1\right)\right]\right)
\end{equation}
Summing over the $k$ calls to Algorithm \ref{alg:sorc} and using
the identity $u \leq l + \sqrt{2C} \leq \scparam + 1 + (2C)^{1/2}$, one
obtains
\begin{equation*}
  \Ep_{\mu}\left[ t \right] \leq \widetilde{O} \left(\frac{(1 + 2L)^{1/2}C}{\epsilon^2}
    \max \left(\Ep_{\mu}\left[\norm{\nabla f(x)}^4\right], D^2\right)\right)
\end{equation*}
where $\widetilde{O}$ suppresses logarithmic factors in
$L,C,\Ep_{\mu}\left[\norm{\nabla f(x)}\right]$ and $\epsilon$.\endproof

\section{Stochastic bisection for only dual \eqref{eq:gfunc} value
  estimation}\label{sec:proof-sbisec}
This section provides an analysis of a stochastic bisection procedure for
\eqref{eq:gfunc} under slightly weaker assumptions than those used to obtain
Theorem \ref{thm:primaldual}. This demonstrates that, if only estimation of
the value of \eqref{eq:gfunc} is required, then slightly modified version of Algorithm
\ref{alg:primaldual} provides a computational complexities with better
dependence on the smoothness of the objective-- compare \eqref{eq:pdsamplebound}
with \eqref{eq:compcplexpbsamp}.
\begin{algorithm}[H]
  \caption{Stochastic bisection}
  \label{alg:bsearch}
  \begin{algorithmic}
    \Input{Supergradient oracle
      $\theta_{g}$, error tolerance $\epsilon$, termination width $B$}
    \State $\eta \leftarrow \infty$, $b \leftarrow l$
    \While{$|\eta| > \epsilon/\max(\lambda - b,1)$ and $u-l>\epsilon/B$}
    \State $\lambda \leftarrow \left(l + u\right)/2$
    \State $\eta \leftarrow \theta_g(\lambda)$, $\eta \leftarrow \left(\eta - \text{Proj}_{\partial \psi ^* (\lambda)}\left(\eta\right)\right)$
    \If{$\eta > 0$} $l \leftarrow \lambda$
    \Else{ $u \leftarrow \lambda$}
    \EndIf
    \EndWhile \\
  \Return $\lambda$
  \end{algorithmic}
  \end{algorithm}
  \begin{proposition}[Convergence of Algorithm \ref{alg:bsearch}]\label{prop:sbisec}
    For the problem \eqref{eq:gfunc}, let $\theta_{g}$ be a
    ($\epsilon, \delta$)-supergradient oracle with high probability for $g$ on
    $[l,u]$. If
    $\sup_{\lambda \in [l,u]} \left|\partial_- g(\lambda)\right| + \left|\partial_+
      \psi^*(\lambda)\right|\leq B$ (where $\partial_-$ and $\partial_+$ denote the
    left derivative and right derivatives respectively) then the output
    $\lambda ^*$ of Algorithm \ref{alg:bsearch} satisfies
    \begin{equation}\label{eq:outputbsearch}
      \sup_{\lambda \in [l,u]} g(\lambda) - \psi ^*(\lambda) -
      g(\lambda ^*) - \psi ^*(\lambda ^*)  \leq 2\epsilon
    \end{equation}
    with probability at least
    $1-\delta \left(\log_2 \left(B(u-l)/\epsilon\right) + 1\right)$.
  \end{proposition}
\proof Let $\lambda_i,u_i,l_i$ and $\eta_i$ denote the $i$th values of
$\lambda, u, l$ and $\eta$ which are computed by Algorithm \ref{alg:bsearch}--
the indexes $l_0,u_0$ denote the initial values of these variables. Let $k$
denote the total number of iterations performed by the loop of Algorithm
\ref{alg:bsearch}. Since $u_i - l_i = \left(u_{i-1} - l_{i-1}\right)/2$, it is
clear that $k \leq \log_2 \left(B(u_0 - l_0)/\epsilon\right) + 1$. Thus, using
\eqref{eq:proboracle} and the fact that $\lambda_i$ depends only on
$\theta_g (\lambda_j)$ for $j < i$, one obtains the union bound
\begin{equation} \label{eq:unionbd}
  \p\left(\bigcup_{i \leq k}\left[ \min_{z \in \partial g (\lambda_i)}\left|\theta_g(\lambda_i) - z
      \right| \geq \frac{\epsilon}{\max(\lambda_i - l_0, 1)} \right]\right) \leq \delta \left(\log_2
    \left(B(u_0-l_0)/\epsilon\right) + 1\right)
\end{equation}
Hence, it need only be shown that \eqref{eq:outputbsearch} holds when
\begin{equation}\label{eq:errorbd}
  \min_{z \in \partial g (\lambda_i)}\left|\theta_g(\lambda_i) - z
  \right| \leq \frac{\epsilon}{\max \left(\lambda_i - l_0, 1\right)} \hspace{ 0.2in } \forall \, i \leq k
\end{equation}

For brevity, set $\epsilon_{\lambda_i} = \epsilon/ \max(\lambda_i - l_0, 1)$ and
let $z_i ^* = \text{Proj}_{\partial g(\lambda_i)} \theta_g(\lambda_i)$.
Recall
$\eta_i = \theta_g(\lambda_i) - \text{Proj}_{\partial \psi
  ^*(\lambda_i)}\theta_g(\lambda_i)$ and define
$\eta_i ^* := z_i ^* - \text{Proj}_{\partial \psi ^*(\lambda_i)}z_i ^*$ to be
the true supergradient of \eqref{eq:gfunc} which $\eta_i$ approximates.
From \eqref{eq:errorbd}
\begin{equation}\label{eq:biseckeyprop}
   \eta_i \eta_i ^*\leq 0  \hspace{
    0.2in } \Rightarrow \hspace{ 0.2in } \max \left(\left| \eta_i \right|,\left|
      \eta_i ^* \right|\right) \leq \epsilon_{\lambda_i}
\end{equation}
Hence, at all iterations prior to the last iteration (iteration $k$) of the loop
in Algorithm \ref{alg:bsearch}, $\eta_i$ and $\eta_i ^*$ have the same sign.
Since $\lambda \mapsto \left(g(\lambda) - \psi ^*(\lambda)\right)$ is concave,
this gives
\begin{equation}\label{eq:loopinvar}
   \sup_{\lambda \in [l_i,u_i]} g(\lambda) - \psi ^*(\lambda) = \sup_{\lambda \in [l_{i-1},u_{i-1}]} g(\lambda) - \psi ^*(\lambda)
\end{equation}
for all $1< i < k$. Additionally, if $\eta_k \eta_k ^* > 0$ then \eqref{eq:loopinvar}
also holds for $i = k$.

Now, at iteration $k$, either $|\eta_k ^*| \leq 2 \epsilon_{\lambda_k}$ or
$|\eta_k ^* | > 2 \epsilon_{\lambda_k}$.  If $|\eta_k ^*| \leq
2\epsilon_{\lambda_k}$, then
\begin{equation}\label{eq:bisecerror}
  \sup_{\lambda \in [l_0,u_0]} g(\lambda) - \psi ^*(\lambda) - \left(g(\lambda_k) - \psi
  ^*(\lambda_k)\right) \leq \sup_{\lambda \in [l_{k-1},u_{k-1}]}
                         \eta_k ^*(\lambda - \lambda_k)
                       \leq \epsilon_{\lambda_k}( u_{k-1} - l_{k-1} ) \leq 2\epsilon
\end{equation}
where the first inequality of \eqref{eq:bisecerror} is a result of
\eqref{eq:loopinvar} and concavity and the second inequality follows from the
observation $u_{k-1} - l_{k-1} \leq 2 \left(\lambda_k - l_0 \right)$. Observe
that \eqref{eq:bisecerror} immediately gives the desired result
\eqref{eq:outputbsearch}.

To show \eqref{eq:outputbsearch} when $|\eta_k ^* | > 2 \epsilon_{\lambda_k}$,
notice that \eqref{eq:errorbd} implies
$||\eta_k ^*| - |\eta_k|| \leq \epsilon_{\lambda_k}$. Hence, $|\eta_k ^* | > 2
\epsilon_{\lambda_k}$ entails that the
second termination condition ($u_k-l_k \leq \epsilon /B$) of Algorithm
\ref{alg:bsearch} was reached and $\eta_k \eta_k ^* \geq 0$. Then, by \eqref{eq:loopinvar},
\begin{equation}\label{eq:bisecerroratk2}
   \sup_{\lambda \in [l_0,u_0]} g(\lambda) - \psi ^*(\lambda) - \left(g(\lambda_k) - \psi
     ^*(\lambda_k)\right) \leq \sup_{\lambda \in [l_k,u_k]}\eta_k ^*(\lambda - \lambda_k) \leq \epsilon
\end{equation}
where the second inequality is a consequence of
$|\eta_k ^*| \leq \sup_{\lambda \in [l_0,u_0]} \left|\partial_-
  g(\lambda)\right| + \left| \partial_+ \psi ^*(\lambda)\right|\leq B$. The
desired result \eqref{eq:outputbsearch} then follows. \endproof
\begin{proposition}\label{prop:compcplexpb}
  Let $f$ be $L$-smooth \eqref{eq:smoothness}, $\scparam$-semiconvex, and assume
  that $\psi$ provides
  $C/\Ep_{\mu}\left[\norm{\nabla f(x)}^2\right]$-regularization
  \eqref{eq:dualgradcond2} at $\scparam + 1$ for $C \geq 8L^2$. If
  $\sup_{\lambda \in [l,u]} \left| \partial_+ \psi^*(\lambda) \right| \leq D$
  for $l$ and $u$ given by \eqref{eq:intvbnds2} then, for
  any $\delta > 0$, there exists an stochastic algorithm which returns a
  $\lambda ^*$ such that, recall \eqref{eq:gfunc},
  \begin{equation*}
      \sup_{\lambda \in \R} g(\lambda) - \psi ^*(\lambda) -
      \left(g(\lambda ^*) - \psi ^*(\lambda ^*)\right)  \leq \epsilon
  \end{equation*}
  with probability $1 - \delta$. Moreover, this algorithm draws at most
  \begin{equation}\label{eq:compcplexpbsamp}
  \widetilde{O}\left(\frac{\Ep_{\mu}\left[\norm{\nabla f(x)}^4\right]
    }{\delta \epsilon^2}\right)
  \end{equation}
  independent samples from $\mu$ and performs $
  \widetilde{O}\left(L^{1/2}\Ep_{\mu}\left[\norm{\nabla
        f(x)}^4\right]/\left(\delta \epsilon^2\right)\right)$ expected gradient evaluations of $f$--
  where $\widetilde{O}$ suppresses logarithmic factors in $L,C,D,
\Ep_{\mu}\left[\norm{\nabla f(x)}^2\right]$ and $\epsilon$.
\end{proposition}
\proof Similarly to the proof of Proposition \ref{prop:compcplexep}, it
is enough for an algorithm to return a value $\lambda^*$ which (with probability
$1 - \delta$) satisfies
\begin{equation}\label{eq:pbbound}
  \sup_{\lambda \in [l,u]} g(\lambda) - \psi ^*(\lambda) - \left(g(\lambda
    ^*) - \psi ^*(\lambda ^*)\right) \leq \frac{\epsilon}{2}
\end{equation}
where $l$ and $u$ are given by \eqref{eq:intvbnds2}. Again, without loss of generality,
it will be assumed that $u-l > 0$.

Apply Algorithm \ref{alg:bsearch} to the interval $[l,u]$ with the supergradient
oracle given by Proposition \ref{prop:sorcalgp}; where the error tolerance in
Proposition \ref{prop:sorcalgp} is set to $\epsilon / 4$ and the error probability
in is set to $\delta/(\log_2 \left(4B(u-l)/\epsilon\right) + 1)$ for
\begin{equation*}
  B := 2\Ep_{\mu}\left[\norm{\nabla f(x)}^2\right] + D
\end{equation*}
By Lemma \ref{lem:gdiff} and
the fact that $\psi ^*$ is non-decreasing, one has
\begin{equation*}
  \sup_{\lambda \in [l,u]} \left|\partial_- g(\lambda) \right| + \left|\partial_-
  \psi^*(\lambda) \right|\leq B
\end{equation*}
Consequently, Proposition \ref{prop:sbisec} guarantees that the value $\lambda ^*$ returned by
Algorithm \ref{alg:bsearch} satisfies \eqref{eq:pbbound} with
probability $1-\delta$.

To compute a bound on the number of samples from $\mu$ which are drawn under
this procedure note that, by definition of Algorithm \ref{alg:bsearch}, at most
$\left \lceil \log_2 \left(4B(u-l)/\epsilon\right)\right\rceil$ calls are made to
the supergradient oracle given by Proposition \ref{prop:sorcalgp}. Via
\eqref{eq:kconst}, it follows that at most
\begin{equation}\label{eq:totaloutercalls}
 \frac{\left(32(\log_2 \left(4B(u-l)/\epsilon\right) + 1)\right)^2
   \Ep_{\mu}\left[\norm{\nabla f(x)}^4\right]}{\delta \epsilon^2}
\end{equation}
invocations of Algorithm \ref{alg:sorc} are performed with an error parameter
which is at least $\epsilon/(8 \max(u - \scparam, 1))$. Since each invocation of
Algorithm \ref{alg:sorc} requires a single sample of $\mu$,
the above procedure therefore draws
\begin{equation*}
  \widetilde{O} \left( \frac{\Ep_{\mu}\left[\norm{\nabla f(x)}^4\right]}{\delta \epsilon^2} \right)
\end{equation*}
independent samples from $\mu$-- where
$\widetilde{O}$ suppresses logarithmic factors in $C,D,
\Ep_{\mu}\left[\norm{\nabla f(x)}^2\right]$ and $\epsilon$.

Finally, since the expected number of gradient evaluations of $f$ performed by
Algorithm \ref{alg:sorc} obeys the bound \eqref{eq:gradevalbnd} for an error
parameter of $\epsilon$, the expected number of gradient evaluations of $f$ used
by Algorithm \ref{alg:sorc} with an error parameter of at least $\epsilon/(8
\max(u - \scparam, 1))$ is $\widetilde{O}\left(L^{1/2}\right)$. In
combination with \eqref{eq:totaloutercalls}, it follows that
\begin{equation*}
  \widetilde{O}\left(\frac{L^{1/2}\Ep_{\mu}\left[\norm{\nabla f(x)}^4\right]}{ \delta \epsilon^2}\right)
\end{equation*}
gradient evaluations of $f$ are performed in expectation. \endproof
\section{Proof of Theorem \ref{thm:primaldual} and Corollary  \ref{cor:fwcompv2}}
\begin{lemma}\label{lem:oracleerror}
  Under the assumptions of Theorem \ref{thm:primaldual}, let $M$ be the
  smoothness parameter of $\psi ^*$ and let $\theta_{g}$ be a
  ($\epsilon, \delta$)-supergradient oracle with high probability for $g$ on the
  interval $[l,u]$. If $1 \leq l - \scparam$, $g'(u) - \left(\psi ^* \right)'(u) \leq 0 \leq g'(l) -
  \left(\psi ^*\right)'(l)$, and Algorithm \ref{alg:primaldual} is
  run with $B= \max \left(M, 4(g'(l))^2\right)$, the output
$\lambda ^*$ of satisfies (recalling \eqref{eq:pilamdef})
    \begin{equation}\label{eq:outputprimaldual}
      \int f \, d \pi_{\lambda ^*, \mu} + \psi\left(\int \frac{1}{2}\norm{y -
          x}^2 \, d\pi_{\lambda ^*, \mu}\right) -
      \left(g(\lambda ^*) - \psi ^*(\lambda ^*)\right)  \leq (4 + l)\epsilon
    \end{equation}
    with probability at least
    $1-\delta \left(\log_2 \left(B(u-l)/\epsilon\right) + 1\right)$.
\end{lemma}
\proof[Proof of Lemma \ref{lem:oracleerror}] Notice that
$g'$ and $\psi'$ can be written since $g'$ exists on the interval $[l,u]$ by Lemma
\ref{lem:gdiff} and $\psi'$ exists under the assumptions of Theorem
\ref{thm:primaldual}. The proof proceeds in similar
style to the proof of Proposition \ref{prop:sbisec} in Appendix
\ref{sec:proof-sbisec}. Indeed, as before, let $\lambda_i,u_i,l_i$ and $\eta_i$
denote the $i$th values of $\lambda, u, l$ and $\eta$ that are computed-- where
the index $0$ denotes the initial value of the variable. The natural number $k$
denotes the total number of iterations performed by the loop of Algorithm
\ref{alg:primaldual} and clearly
$k \leq \log_2 \left(B(u_0 - l_0)/\epsilon\right) + 1$. Thus, the same union
bound argument \eqref{eq:unionbd} implies it is sufficient to show
\eqref{eq:outputprimaldual} when
\begin{equation}\label{eq:graderrorbd}
  \left|\theta_g(\lambda_i) - g'(\lambda_i)
  \right| \leq \frac{\epsilon}{\max \left(\lambda_i - l_0, 1\right)} \hspace{ 0.2in } \forall \, i \leq k
\end{equation}
to guarantee that \eqref{eq:outputprimaldual} occurs with probability at least
$1 - \delta \left(\log_2 \left(B(u_0 - l_0)/\epsilon\right) + 1\right)$.

For brevity, denote $\epsilon_{\lambda_i} = \epsilon/ \max(\lambda_i - l_0, 1)$ and
let $\eta_i ^* := g'(\lambda_i) - \left(\psi ^*\right)'(\lambda_i)$ be
the true gradient of \eqref{eq:gfunc} which $\eta_i$ approximates.
As before, \eqref{eq:graderrorbd} provides
\begin{equation}\label{eq:biseckeyprop2}
   \eta_i \eta_i ^*\leq 0  \hspace{
    0.2in } \Rightarrow \hspace{ 0.2in } \max \left(\left| \eta_i \right|,\left|
      \eta_i ^* \right|\right) \leq \epsilon_{\lambda_i}
\end{equation}
which will produce the desired guarantee
\eqref{eq:outputprimaldual} on $u_k = \lambda ^*$.

Indeed, from Algorithm \ref{alg:primaldual}, it is clear that either
$u_{k} = \lambda_i$ for some $i>0$ such that $\eta_i < - \epsilon_{\lambda_i}$ or
$u_k = u_0$.
Similarly, $l_{k} = \lambda_j$ for some $j > 0$ such that $\eta_j > - \epsilon_{\lambda_j}$ or $l_k =
l_0$. One can assume, without loss of generality, that $l_{k} = \lambda_j$ for some
$j > 0$ and, in combination with
\eqref{eq:biseckeyprop2} and $g'(u_{0}) -
\psi'(u_0) \leq 0 \leq g'(l_{0}) - \psi'(l_0)$, this gives
\begin{equation}\label{eq:bisecinvar}
   -\epsilon_{\lambda_j} \leq g'(l_{k}) - \psi'(l_{k})  \hspace{ 0.1in } \text{and} \hspace{
     0.1in } g'(u_{k}) - \psi'(u_{k}) \leq 0
\end{equation}
Moreover, since $0 \in \dom(\psi)$, the inequality
$g'(u_{k}) - \left(\psi^*\right)'(u_{k}) \leq 0$ implies
$\int \frac{1}{2}\norm{y-x}^2 \, d\pi_{u_{k}, \mu} = g'(u_{k}) \in \dom
\left(\psi \right)$. Hence, denoting the primal-dual gap,
\begin{equation*}
  T := \int f \, d \pi_{u_{k}, \mu} + \psi \left(\int \frac{1}{2}\norm{y-x}^2 \,
    d\pi_{u_{k}, \mu}\right) -
  \left(g(u_{k}) - \psi ^*(u_{k})\right)
\end{equation*}
Lemma \ref{lem:pdgap} yields
\begin{equation*}
  T \leq
  \left(u_{k} - \lambda^*\right)\left(
    \left(\psi^*\right)'(u_{k}) - g'(u_{k}) \right)
\end{equation*}
for any $\lambda ^* \in \partial \psi \left(\int \frac{1}{2}\norm{y-x}^2 \,
  d\pi_{u_{k}, \mu}\right)$.
If $\psi$ is minimized at 0 then $\lambda^* \geq 0$ and
\begin{equation}\label{eq:basicpdbnd}
 T \leq u_{k}\left(\left(\psi^*\right)'(u_{k}) - g'(u_{k}) \right)
\end{equation}
Using \eqref{eq:gradholder} and that $\left(\psi ^*\right)'$ is $M$-Lipschitz on
$[l_0,u_0]$, one obtains
\begin{align}
  T &\leq u_{k}\left(\left(\psi^*\right)'(u_{k}) - g'(u_{k}) \right) \notag\\
    &\leq
  u_{k}\left(\left(\psi^*\right)'(l_k) + M (u_{k} - l_k) - \left(1 -
      \frac{2}{\sqrt{u_k - \scparam}}(u_k-l_k)^{1/2}\right)g'(l_k) \right) \notag\\
  &\leq u_{k}\epsilon_{\lambda_j}  + u_{k}\left(M (u_{k} - l_k) +
      \frac{2}{\sqrt{u_k - \scparam}}(u_k-l_k)^{1/2}g'(l_k) \right)\label{eq:finalpdgap}
\end{align}
where the last inequality also used \eqref{eq:bisecinvar}.
To bound the first term on the left side of \eqref{eq:finalpdgap}, observe that
$l_k \neq l_0$ implies there exists a minimal $t > 0$ such that $l_t \neq
l_0$. Clearly,
\begin{equation*}
  u_{k}\epsilon_{\lambda_j} = \frac{u_k \epsilon}{\max(\lambda_j -l_0, 1)} \leq
  \frac{u_k \epsilon}{\max(\lambda_t -l_0, 1)} \leq  \epsilon\frac{u_t -
    l_0}{\max(\lambda_t -l_0, 1)} + l_0 \epsilon \leq (2 + l_0) \epsilon
\end{equation*}
Combining this with the termination condition
\begin{equation*}
u_k - l_k \leq \frac{\epsilon}{B} \leq \frac{\epsilon}{\max \left(M ,
      4g(l_0)^2\right)} \leq \frac{\epsilon}{\max \left(M ,
    4g(l_k)^2/(u_k-\scparam)\right)}
\end{equation*}
to bound the second term of \eqref{eq:finalpdgap} and one obtains the desired result
\begin{equation*}
  T \leq (4 + l_0) \epsilon
\end{equation*}
 \endproof

\proof[Proof of Theorem \ref{thm:primaldual}] Let $l = \scparam + 1$ and
$u = \scparam + 1 + \sqrt{2C}$ and apply Algorithm \ref{alg:primaldual} to the
interval $[l,u]$ with the supergradient oracle given by Proposition
\ref{prop:sorcalgp}. Set the error tolerance used by Algorithm \ref{alg:primaldual}
to $\epsilon/ \left(4 + \scparam + 1\right)$ and the termination width
to
\begin{equation*}
  B := \max \left(M , 16 \left(\Ep_{\mu}\left[\norm{\nabla f(x)}^2\right] \right)^2 \right)
\end{equation*}
Likewise, the error tolerance used in Proposition \ref{prop:sorcalgp} should be
be $\epsilon/ \left(4 + \scparam + 1\right)$ and the error probability
should be $\delta/(\log_2 \left(B(u-l)\left(4 + \scparam + 1\right)/\epsilon\right) + 1)$.

Under this setting of parameters, Lemma \ref{lem:oracleerror} establishes that
the output of $\lambda ^*$ of Algorithm \ref{alg:primaldual} satisfies
\begin{equation}\label{eq:desres}
      \int f \, d \pi_{\lambda ^*, \mu} + \psi\left(\int \frac{1}{2}\norm{y -
          x}^2 \, d\pi_{\lambda ^*, \mu}\right) -
      \left(g(\lambda ^*) - \psi ^*(\lambda ^*)\right)  \leq \epsilon
\end{equation}
with probability $\delta$ so long as
\begin{equation}\label{eq:uandlbounds}
g'(u) - \left(\psi ^*\right)'(u) \leq 0 \leq g'(l) - \left(\psi ^*\right)'(l)
\end{equation}
To see that \eqref{eq:uandlbounds} is fulfilled for the chosen $l$ and $u$,
notice that, by Assumption \ref{asm:regularization},
\eqref{eq:dualgradcond} holds for $C \geq 8(\scparam+1)^2$. Hence, \eqref{eq:optatL} gives
\begin{equation*}
 g'(l) - \left(\psi ^* \right)'(l) \geq 0
\end{equation*}
Similarly, \eqref{eq:dualgradcond} combined with \eqref{eq:umin} provides
\begin{equation*}
  g'(u) - \left(\psi ^* \right)'(u) \leq \frac{2}{\left(u -
      \scparam\right)^2}\Ep_{\mu}\left[\norm{\nabla f(x)}^2\right] - \left(\psi ^*
  \right)'(u) \leq \frac{1}{C}\Ep_{\mu}\left[\norm{\nabla f(x)}^2\right] - \left(\psi ^*
  \right)'(l) \leq 0
\end{equation*}
Hence, \eqref{eq:uandlbounds} holds and the output $\lambda^*$ of Algorithm
\ref{alg:primaldual} obeys \eqref{eq:desres} with probability $\delta$.

It remains to compute a bound on the number of samples from $\mu$ which are
required by this procedure. Clearly, by the definition of Algorithm
\ref{alg:primaldual}, at most
$\left \lceil \log_2 \left(B(u-l)\left(4 + \scparam +
      1\right)/\epsilon\right)\right\rceil$ calls are made to the
supergradient oracle given by Proposition \ref{prop:sorcalgp}. Via
\eqref{eq:kconst}, this yields that at most
\begin{equation}\label{eq:outercalls}
  \frac{\left(8(4 + \scparam + 1)(\log_2\left(B(u-l)\left(4 + \scparam +
          1\right)/\epsilon\right) + 1)\right)^2\Ep_{\mu}\left[\norm{\nabla f(x)}^4\right]}{\delta \epsilon^2}
\end{equation}
invocations of Algorithm \ref{alg:sorc} are performed with an error parameter
which is at least
$\epsilon/(2(4 + \scparam + 1)(\max(u- \scparam,1)))$.
Since each invocation of Algorithm \ref{alg:sorc} requires a single sample of
$\mu$, it follows from \eqref{eq:outercalls} that
\begin{equation*}
  \widetilde{O} \left(\frac{\scparam^2 \Ep_{\mu}\left[\norm{\nabla f(x)}^4\right]}{\delta \epsilon^2}\right)
\end{equation*}
samples are used by Algorithm
\ref{alg:primaldual}-- where $\widetilde{O}$ suppresses logarithmic factors in
$\scparam, C, M, \Ep_{\mu}\left[\norm{\nabla f(x)}^2\right],$ and $\epsilon$.

Finally, as the expected number of gradient evaluations performed by a call to
Algorithm \ref{alg:sorc} obeys \eqref{eq:gradevalbnd} for an error parameter of
$\epsilon$, the expected number of gradient evaluations executed by each call to
Algorithm \ref{alg:sorc} is at most
$\widetilde{O}\left(L^{1/2}\right)$. In
combination with \eqref{eq:outercalls}, one obtains that at most
\begin{equation*}
  \widetilde{O}\left(\frac{\scparam^2L^{1/2}\Ep_{\mu}\left[\norm{\nabla f(x)}^4\right]}{\delta\epsilon^2}\right)
\end{equation*}
expected gradient evaluations are performed.
 \endproof
\proof[Proof of Corollary \ref{cor:fwcompv2}.]
Apply Theorem \ref{thm:primaldual} to \eqref{eq:2normprimal} with $\psi$ given by
\begin{equation}\label{eq:fwpenalty}
  \psi(x) =
  \begin{cases}
    0 & \text{if } 0 \leq x \leq \frac{\delta^2 }{2} \\
    \infty & \text{otherwise}
  \end{cases}
\end{equation}
The trust-region problem
\eqref{eq:trustregionprob} is obtained and, as
$\mu \in \Pc_2\left(\R^{d}\right)$, sufficient conditions for strong duality
\eqref{eq:domconds} hold. By Theorem \ref{thm:primaldual}, it suffices
to show that $\psi$ provides
$C \bigr / \left(\Ep_{\mu}\left[\norm{\nabla
      f(x)}^2\right]\right)$-regularization \eqref{eq:dualgradcond2} at $L+ 1$
for $C \geq 8L^2$. Indeed, the other suppositions of Theorem
\ref{thm:primaldual} are clearly true since: any $L$-smooth function $f$ is also
$L$-semiconvex, $\psi$ is minimized at 0, and
\begin{equation}\label{eq:fwpsidual}
  \psi ^*(\lambda) = \left(\delta^2/2\right) \left(\lambda\right)_+
\end{equation}
is trivially smooth on any interval not containing 0. From
\eqref{eq:fwpsidual} and the guarantee
$\delta \leq \norm{\nabla f}_{L^2(\mu)}/2L$, however, this
level of regularization is clear since
\begin{equation*}
  \partial_- \psi ^*(L + 1) \leq \frac{\Ep_{\mu}\left[\norm{\nabla f(x)}^2\right] }{8L^2}
\end{equation*}
 \endproof

\section{Proof of Proposition \ref{prop:wassprop}}
\label{sec:proofwassprop}
\proof[\nopunct] The first bullet is essentially a restatement of Theorem 7.2.2 in
\cite{Ambrosio05}. To verify the second bullet, we first establish the existence
of such a $v_t$. Let $\mu_t$ be the constant speed geodesic and define the
set of functions
\begin{equation*}
  A_{\mu} := \left\{ z \in L^2 \left([0,1]\right) : W\left(\mu_t, \mu_s\right) \leq
    \int_s^t z(r) \, dr \,\,\,\, \forall \,\, 0 \leq s \leq t \leq 1 \right\}
\end{equation*}
It is clear that the function $m(r) := \Wc(\mu_0, \mu_1)$ is in $A$ and
satisfies
\begin{equation}\label{eq:pnormbd}
  m = \argmin_{z \in A} \int_0^1 z^p(r) \, dr
\end{equation}
for any $p \geq 1$. Hence, the metric derivative $|\mu'|$ of $\mu_t$ fulfills
\begin{equation*}
  |\mu '|(t) = d(\mu_0, \mu_1) \hspace{ 0.2in } \text{Lebesgue almost everywhere
  for $t \in [0,1]$ }
\end{equation*}
By Theorem 8.3.1 in \cite{Ambrosio05}, there exists Borel vector field $v_t : [0,1]
\times \R^{d} \to \R^{d}$ satisfying the continuity equation \eqref{eq:continuity}
such that
\begin{equation}\label{eq:vequality}
  \norm{v_t}_{L^2(\mu_t)} = |\mu '|(t) = \Wc(\mu_0, \mu_1)
  \hspace{ 0.2in } \text{Lebesgue almost everywhere for $t \in [0,1]$}
\end{equation}
Combined with \eqref{eq:pnormbd}, this implies that $v_t$ is a solution of
\eqref{eq:minvectorfield}. Uniqueness of $v_t$ follows directly from the
third bullet.

For a constant-speed geodesic $\mu_t$ from $\mu$ to $\nu$.
Theorem 2.4 in \cite{Ambrosio08} gives that, for any
$\sigma \in \Pc_2(\R^{d})$,
\begin{equation}\label{eq:wassderiv2}
  \frac{d}{dt} \frac{1}{2}\Wc^2 \left(\mu_t, \sigma \right) = \int \left
    \langle v_t(x), x - y \right \rangle \, d \bar{\gamma}(x,y) \hspace{ 0.2in }  \forall\,\,
  \bar{\gamma} \in \Pi_o(\mu_t , \sigma)
\end{equation}
where $\Pi_o(\mu_t, \sigma) \subseteq \Pc(\R^{d} \times \R^{d})$ is the set of
optimal transport plans between $\mu_t$ and $\sigma$. Setting $\sigma = \nu$,
the fact that $\mu_t$ is a geodesic implies that there is a unique optimal
coupling $\gamma \in \Pc \left(\R^{d} \times \R^{d}\right)$ between $\mu$ and
$\nu$ such that
\begin{equation*}
  \left((1-t)x + t y, y\right)_{\#}\gamma \, \in \,\Pi_o(\mu_t, \sigma)
\end{equation*}
Hence, \eqref{eq:wassderiv2} gives
\begin{align}
  -(1-t)\Wc^2(\mu, \nu) &= \int \left \langle v_t(x), x - y \right \rangle \, d
  \bar{\gamma}(x,y) = -(1-t)\int \left \langle v_t((1-t)x + t y), y - x \right \rangle \, d
  \gamma(x,y)\nonumber \\
   \\
   \Rightarrow \,\,\,\,\Wc^2(\mu, \nu) &= \int \left \langle v_t((1-t)x + t y),
                                         y - x \right \rangle \, d \gamma(x,y) \label{eq:l2identity}
\end{align}
For $t$ satisfying \eqref{eq:vequality}, the fact that
$\norm{v_t}_{L^2(\mu_t)} = \Wc(\mu, \nu)$ and
$\norm{y-x}_{L^2(\gamma)} = \Wc(\mu, \nu)$ means that \eqref{eq:l2identity} gives
equality for Cauchy-Schwarz. Thus,
$v_t((1-t)x + t y) = y - x$, $\gamma$-almost surely and \eqref{eq:velexpr}
follows for Lebesgue almost every $t \in [0,1]$.

The final bullet is a direct restatement of the results of Section 8.4 in \cite{Ambrosio05}.
 \endproof

\section{Proof of Theorem \ref{thm:outerconvres}}
\begin{lemma}\label{lem:liptosmth}
  Let $\gamma \in \Pi (\mu, \nu)$ be an optimal transport plan between
  $\mu \in \Pc_2(\R^{d})$ and $\nu \in \Pc_2(\R^{d})$. If
  $\phi_{\mu} \in C^{1}(\R^{d})$ is $L$-smooth
  \begin{equation}\label{eq:phimusmooth}
    \norm{\nabla \phi_\mu (x) - \nabla \phi_\mu(y)} \leq L \norm{x-y}
  \end{equation}
  then
  \begin{equation}\label{eq:integradbd}
    \left | \int_{\R^{d}} \left \langle \nabla \phi_\mu(x), y-x \right \rangle \,
      d \gamma(x,y) - \left(\int_{\R^{d}} \phi_\mu  \, d\nu - \int_{\R^{d}} \phi_\mu  \, d\mu\right) \right | \leq
    \frac{L}{2}\Wc^2(\nu, \mu)
  \end{equation}
\end{lemma}
\proof
First, it will be shown that
  \begin{equation}\label{eq:velintbd}
    \left | \int_{\R^{d}} \left \langle \nabla \phi_\mu(x), y-x \right \rangle \,
      d \gamma(x,y) - \int_0^1 \left \langle \nabla \phi_\mu, v_t \right
     \rangle_{\mu_t} \, dt \right | \leq
    \frac{L}{2}\Wc^2(\nu, \mu)
  \end{equation}
  for $\mu_t$ and $v_t$ which correspond \eqref{eq:minvectorfield} to the
  unique-constant speed geodesic given by $\gamma \in \Pi(\mu, \nu)$
  \eqref{eq:geobiject}. Notice that, since $\nabla \phi_\mu $ has at most linear
  growth, therefore both terms in the left-hand side of \eqref{eq:velintbd} are
  finite. Moreover, by \eqref{eq:velexpr}, one has
\begin{equation}\label{eq:veldecomp}
  \int_0^1 \left \langle \nabla \phi_\mu, v_t \right \rangle_{\mu_t} \, dt =
  \int_0^1 \int_{\R^{d} \times \R^{d}} \iprod{\nabla \phi_\mu((1-t)x + t y)}{y-x}\, d \gamma(x,y) \, dt
\end{equation}
Thus, Cauchy-Schwarz and \eqref{eq:phimusmooth} give
\begin{align*}
    &\left | \int_0^1 \left \langle \phi_\mu, v_t \right \rangle_{\mu_t} \, dt -
    \int_{\R^{d} \times \R^{d}} \left \langle \nabla \phi_\mu(x), y-x \right
  \rangle \, d \gamma(x,y)\right | = \\
  &
    \left | \int_0^1 \int_{\R^{d} \times \R^{d}} \iprod{\nabla \phi_\mu((1-t)x + t y) - \nabla
                                                                                          \phi_\mu(x)}{y-x} \, d \gamma(x,y) \, dt \right |\leq \\
  &\int_0^1 \int_{\R^{d}} t L\norm{x - y}^2  \, d \gamma(x,y) \, dt =
    \frac{L}{2}\Wc^2(\nu, \mu)
  \end{align*}
To obtain \eqref{eq:integradbd}, it only remains to show that
\begin{equation}\label{eq:int2func}
 \int_0^1 \left \langle \nabla \phi_\mu, v_t \right
     \rangle_{\mu_t} \, dt = \int \phi_\mu \, d \nu - \int \phi_\mu \, d \mu
\end{equation}
Moreover, since $v_t$ satisfies \eqref{eq:continuity}, Lemma 8.1.2 in
\cite{Ambrosio05} gives
\begin{equation}\label{eq:int2funcres}
 \int_0^1 \left \langle \nabla \psi, v_t \right
     \rangle_{\mu_t} \, dt = \int \psi \, d \nu - \int \psi \, d \mu
\end{equation}
for every $\psi \in C_c^1(\R^{d})$-- where $C_c^1(\R^{d})$ denotes the space of
continuously differentiable functions on $\R^{d}$ with compact support. Hence,
\eqref{eq:int2func} will be obtained from \eqref{eq:int2funcres} by the
following approximation argument.

Define the functions:
\begin{equation*}
  \beta_{-}(x) := \left(\sqrt{\norm{x}^2 + 1} - \sqrt{2}\right)^{-1}
  \hspace{ 0.2in } \text{and} \hspace{ 0.2in }
  \beta_{+}(x) := \left(\sqrt{5} - \sqrt{\norm{x}^2 + 1}\right)^{-1}
\end{equation*}
and
\begin{equation*}
  \eta(x) := \begin{cases} 1 & \text{if $\norm{x} \leq 1$} \\
    \frac{e^{\beta_{-}(x)}}{e^{\beta_{-}(x)} + e^{\beta_{+}(x)}} & \text{if $1 < \norm{x} < 2$} \\
    0 & \text{if $\norm{x} \geq 2$} \\
            \end{cases}
\end{equation*}
It is easy to verify that $\eta \in C_c^\infty \left(\R^{d}\right)$ and
$\norm{\nabla \eta(x)} \leq B$ for all $x \in \R^{d}$ and some constant $B$.
Moreover, $\eta$ provides a sequence of mollified approximations of $\phi_\mu$
\begin{equation*}
  \psi_k(x) := \phi_\mu(x) \eta_k(x) \hspace{ 0.2in } \text{for} \hspace{ 0.2in } \eta_k (x) := \eta \left(\frac{x}{k}\right)
\end{equation*}
where $\psi_k \in C_c^1(\R^{d})$. Clearly, \eqref{eq:int2funcres} holds for
all such $\psi_k$. Thus, if
\begin{equation}\label{eq:rightcon}
  \lim_{k \to \infty} \int \psi_k \, d \nu - \int \psi_k \, d \mu =
  \int\phi_\mu \, d \nu - \int \phi_\mu \, d \mu
\end{equation}
and
\begin{equation}\label{eq:leftcon}
  \lim_{k \to \infty} \int_0^1 \iprod{\nabla \psi_k}{v_t}_{\mu_{t}} \, dt = \int_0^1 \iprod{\nabla \phi_\mu}{v_t}_{\mu_{t}} \, dt
\end{equation}
then \eqref{eq:int2func} will follow directly from \eqref{eq:int2funcres}.

The relations \eqref{eq:rightcon} and \eqref{eq:leftcon}
are straight-forward consequences of dominated convergence. Indeed, as $\eta_k \to 1$ and
$\nabla \eta_k \to 0$ (pointwise), clearly
\begin{equation}\label{eq:mollifiedprops}
  \psi_k \to \phi_\mu \hspace{ 0.2in } \text{and} \hspace{ 0.2in } \nabla \psi_k \to \nabla \phi_\mu
\end{equation}
Quadratic growth of $\phi_\mu$ yields $\phi_\mu \in L^2(\mu) \cap L^2(\nu)$
and combined with
\begin{equation*}
  |\psi_k(x)| \leq |\phi_\mu(x)| \hspace{ 0.2in }  \forall x \in \R^{d}
\end{equation*}
\eqref{eq:rightcon} clearly holds via dominated convergence. One also has
\begin{equation}\label{eq:gradbound1}
  \norm{\nabla \psi_k(x)} \leq \norm{\nabla \phi_\mu(x)} +
  \frac{B|\phi_\mu(x)|}{k} \one_{\left\{ \norm{x} < 2k \right\}}
\end{equation}
Using the quadratic growth of $\phi_\mu$, linear growth of
$\norm{\nabla \phi_\mu }$, and the bound
$\norm{x}\one_{\left\{ \norm{x} < 2k \right\}}/k \leq 2$, \eqref{eq:gradbound1}
yields
\begin{equation}\label{eq:gradbound2}
  \norm{\nabla \psi_k(x)} \leq \norm{\nabla \phi_\mu(x)} +
  C\norm{x} \one_{\left\{ \norm{x} < 2k \right\}} + D \leq E\norm{x} + F
\end{equation}
for some constants $C,D,E \in \R_+$. Recalling \eqref{eq:veldecomp},
\eqref{eq:gradbound2} provides
\begin{align}
  \int_{\R^{d} \times \R^{d}} \left| \iprod{\nabla \psi_k((1-t)x + t y)}{y-x}
  \right| \, d \gamma(x,y) &\leq \int_{\R^{d} \times \R^{d}} \norm{\nabla
                             \psi_k((1-t)x + t y)}\norm{y-x} \, d \gamma(x,y)
                             \notag \\
  &\leq \int_{\R^{d} \times \R^{d}} \left(E\norm{(1-t)x + ty} +
    F\right)\norm{y-x} \, d \gamma(x,y) \notag \\
                           &\leq H \label{eq:velboundfordc}
\end{align}
for some $H \in \R_+$; where the last inequality is a result of Cauchy-Schwarz.
The combination of pointwise convergence \eqref{eq:mollifiedprops} and
\eqref{eq:velboundfordc} then immediately yield \eqref{eq:leftcon} by dominated
convergence and \eqref{eq:veldecomp}. \endproof
\begin{lemma}\label{lem:seqconv}
  Let $r_i \in \R_+$ be a sequence of non-negative numbers satisfying
  \begin{equation}\label{eq:descseq}
    r_{i+1} \leq r_i - \kappa r_i^p
  \end{equation}
  for some constants $\kappa > 0$ and $p \geq 0$. Then,
  \begin{equation}\label{eq:descbounds}
    r_n \leq
    \begin{cases} e^{-\kappa n / r_0^{1-p}}r_0 & \text{if } p \leq 1 \\
                  \left(\kappa n + r_0^{1-p}\right)^{-1/(p-1)} & \text{if } p > 1 \\
    \end{cases}
  \end{equation}
\end{lemma}
\proof
If $p \leq 1$, then \eqref{eq:descseq} combined with the fact that $r_i$ is a non-increasing sequence implies
  \begin{equation*}
    r_i \leq \left(1 - \frac{\kappa}{r_0^{1-p}}\right) r_{i-1}
  \end{equation*}
  Iterating this inequality from 1 to $n$ yields the first part of
  \eqref{eq:descbounds}. Next, let $p > 1$ and notice that, by taking the
  reciprocals of both sides of \eqref{eq:descseq} and rearranging, one obtains
  \begin{equation*}
    \frac{\kappa r_{i-1}^{p-2}}{1-\kappa r_{i-1}^{p-1}} \leq r_i^{-1} - r_{i-1}^{-1}
  \end{equation*}
  Summing this inequality over $i$ (from 1 to $n$),
  \begin{equation*}
    \kappa n r_{k}^{p-2} \leq \sum_{i=1}^{n}\frac{\kappa r_{i-1}^{p-2}}{1-\kappa r_{i-1}^{p-1}} \leq r_k^{-1} - r_{0}^{-1}
  \end{equation*}
  where the first inequality is a result of $r_i$ being non-increasing.
  Algebraic manipulation then provides
  \begin{equation*}
    r_{n}\leq \left(\kappa n + r_0^{1-p}\right)^{-1/(p-1)}
  \end{equation*}
\endproof
\proof[Proof of Theorem \ref{thm:outerconvres}.]
Recall the parameters specified in Assumptions \ref{asm:fwsmooth} and \ref{asm:fwloja} and let $\epsilon$ be
the desired tolerance with which \eqref{eq:fwdesiredresid} should hold. Let Algorithm \ref{alg:frankwolfe} be run with the following parameters:
\begin{equation}\label{eq:algostepsz}
  \beta_1 = \min \left(\Delta_1, \Delta_2\right), \hspace{ 0.8em } \beta_2 = \alpha(4L)^{-1}, \hspace{ 0.8em }\beta_3 = (1-\alpha/2)^{1/\alpha}T^{-1/\alpha}
\end{equation}
and
\begin{equation}\label{eq:algoerrors}
  r= \tau \epsilon^\theta/2, \hspace{ 0.8em } \hat{\epsilon} = \left(2 \alpha ^*\right)^{-1}r,\hspace{ 0.8em }
  \bar{\epsilon} = \alpha r/2, \hspace{ 0.8em }
  \widetilde{\epsilon} = \left(4 \alpha ^*\right)^{-1}r, \hspace{ 0.8em } k= \lceil M \rceil
\end{equation}
where $\alpha ^* = (1+\alpha)/\alpha$ is the dual exponent of $1+\alpha$ and $M$
is defined in \eqref{eq:iterationlb}. It will be shown that the last iterate,
$\mu_l$, computed by Algorithm \ref{alg:frankwolfe} satisfies
\eqref{eq:fwdesiredresid}.

First, we bound the decrease in $J$ at each step of Algorithm
\ref{alg:frankwolfe}. Let $\delta_i$ be the $i$th value of $\delta$ that is
computed by Algorithm \ref{alg:frankwolfe} and let $s_i$ denote the $i$th value
of $s$. One has the relation
\begin{equation}\label{eq:deltadef}
  \delta_i = \min \left(\beta_1, \beta_2 s_i, \beta_3s^{\alpha ^* - 1}_i\right)
\end{equation}
and, since $\delta_i \leq \Delta_2$ for all $i$, $\mu_0 \in S$ implies $\mu_i
\in S$ for all $i$. Via the smoothness of $J$ on $S$ and $\delta_i
\leq \Delta_1$, it follows that
\begin{equation*}
  J(\mu_i) \leq J(\mu_{i-1}) + \int_{\R^{d} \times \R^{d}} \iprod{F(\mu_{i-1};x)}{y-x}\, d \gamma(x,y)  + \frac{T}{1+\alpha} \delta_{i}^{1 + \alpha}
\end{equation*}
for any optimal transport plan $\gamma \in \Pi(\mu_i, \mu_{i-1})$ between
$\mu_i$ and $\mu_{i-1}$. Recognizing \eqref{eq:gradoraclerror},
\begin{align*}
  \int_{\R^{d} \times \R^{d}} \iprod{F(\mu_{i-1};x) - \nabla
  \widehat{\phi}_{\mu_{i-1}}(x)}{y-x}\, d \gamma(x,y)
  \leq \norm{F(\mu_{i-1};x) - \nabla
  \widehat{\phi}_{\mu_{i-1}}}_{L^2(\mu_{i-1})}W(\mu_i, \mu_{i-1}) \leq
  \delta_i\hat{\epsilon}
\end{align*}
and therefore
\begin{equation*}
  J(\mu_i) \leq J(\mu_{i-1}) + \int_{\R^{d} \times \R^{d}} \nabla
  \widehat{\phi}_{\mu_{i-1}}(x)^T \left(y-x\right)\, d \gamma(x,y)  +
  \frac{T}{1+\alpha}\delta_{i}^{1 + \alpha} + \delta_{i}\hat{\epsilon}
\end{equation*}
Via Lemma \ref{lem:liptosmth},
\begin{equation}\label{eq:firsterrorub}
  J(\mu_i) \leq J(\mu_{i-1}) + \int \widehat{\phi}_{\mu_{i-1}} \, d\mu_i - \int \widehat{\phi}_{\mu_{i-1}} \, d\mu_{i-1} + \frac{T}{1+\alpha}\delta_{i}^{1 + \alpha} +
  \frac{L}{2}\delta_{i}^2 +  \delta_{i}\hat{\epsilon}
\end{equation}

Now, since $\widehat{\phi}_{\mu_{i-1}}$ is $L$-smooth, it is a
Kantorovich potential \cite[Section 6.1]{Ambrosio05} for $\mu_{i-1}$-- under the cost function
$L \norm{x-y}^2/2$. Thus, there exists a geodesic $\nu_t$ (Proposition
\ref{prop:wassprop}) such that: $\nu_0 = \mu_{i-1}$ and the transport plan
$\gamma_t \in \Pi(\mu_{i-1}, \nu_t)$ between $\mu_{i-1}$ and $\nu_t$ satisfies
\cite[Section 8.3]{Ambrosio05}
\begin{equation*}
  \int_{\R^{d} \times \R^{d}} \iprod{\nabla \phi_{\mu_{i-1}}(x)}{y-x} \, d
  \gamma_t(x,y) = - \frac{t}{L} \norm{\nabla
    \widehat{\phi}_{\mu_{i-1}}}^2_{L^2(\mu_{i-1})} \hspace{ 0.05in } \text{  and  } \hspace{ 0.1in }  \Wc \left(\nu_t,
    \mu_{i-1}\right) = \frac{t}{L}\norm{\nabla \widehat{\phi}_{\mu_{i-1}}}_{L^2(\mu_{i-1})}
\end{equation*}
for $0 \leq t \leq 1$. For the sake of notation, define
$g_{i-1} := \norm{\nabla \widehat{\phi}_{\mu_{i-1}}}_{L^2(\mu_{i-1})}$ and set
$t = L \delta_i / g_{i-1}$. Clearly, $t \leq 1$ since
$\delta_i \leq \beta_2 s_i \leq \beta_2 g_{i-1}$.

By construction, $\mu_i$ also satisfies
\begin{equation*}
  \int \widehat{\phi}_{\mu_{i-1}} \, d\mu_i - \int \widehat{\phi}_{\mu_{i-1}} \, d\mu_{i-1}
  \leq \int \widehat{\phi}_{\mu_{i-1}} \, d \nu_t - \int
  \widehat{\phi}_{\mu_{i-1}} \, d\mu_{i-1} + \zeta_i
\end{equation*}
for $\zeta_i = \delta_i\widetilde{\epsilon}$. Hence, with another application of Lemma \ref{lem:liptosmth},
one obtains
\begin{align}
  \int \widehat{\phi}_{\mu_{i-1}} \, d \mu_i - \int \widehat{\phi}_{\mu_{i-1}} \, d\mu_{i-1}
  &\leq \int_0^t \iprod{\nabla \widehat{\phi}_{\mu_{i-1}}}{v_s}_{\nu_s
    } \, ds + \zeta_i \notag \\[0.5em]
  &\leq \int_{\R^{d} \times \R^{d}} \iprod{\nabla \widehat{\phi}_{\mu_{i-1}}(x)}{y-x} \, d
    \gamma(x,y) + \frac{L}{2} \Wc(\nu_t, \mu_{i-1})^2 + \zeta_i\notag \\[0.5em]
  &= - \frac{t}{L}\left(1-\frac{t}{2}\right)g_{i-1}^2 + \zeta_i\label{eq:gnormdec}
\end{align}
Combining \eqref{eq:gnormdec}
with \eqref{eq:firsterrorub} and recalling $\delta_i = t g_{i-1}/L$ gives
\begin{equation}\label{eq:dummyerrorub}
  J(\mu_i) \leq J(\mu_{i-1}) -
  \frac{t}{L}\left(C -t - \frac{D}{1+\alpha}t^\alpha\right)g_{i-1} + \zeta_i
\end{equation}
for the values
\begin{equation*}
  C := 1 - \hat{\epsilon} \hspace{ 0.2in } \text{and} \hspace{ 0.2in }
  D := \frac{T}{L^\alpha g_{i-1}^{1-\alpha}}
\end{equation*}
Rewriting \eqref{eq:dummyerrorub} using the residual term
\begin{equation}\label{eq:fwresiddef}
  r(\nu) := J(\nu) - \inf_{\mu \in S} J(\mu)
\end{equation}
one obtains
\begin{equation}\label{eq:intermederrorub}
  r\left(\mu_{i}\right) \leq r\left(\mu_{i-1}\right) -
  \frac{t}{L}\left(C -t - \frac{D}{1+\alpha}t^\alpha\right)g_{i-1} + \zeta_i
\end{equation}
This relation will now be used to show that Algorithm
\ref{alg:frankwolfe} makes sufficient progress on $J$, prior to the
termination of it's loop.

Let $l$ be the index of the last iterate $\mu_i$ which is computed by Algorithm
\ref{alg:frankwolfe}. First, observe that if $s_{l+1} \leq r$, then early
termination of the loop in Algorithm \ref{alg:frankwolfe} has occurred. Using
\eqref{eq:lojaineq} and the definitions \eqref{eq:algoerrors}, it follows that
\begin{align}
  \tau \left(r\left(\mu_l\right)\right)^\theta &\leq \norm{F(\mu_{l})}_{L^2(\mu_{l})} \leq g_l + \hat{\epsilon} \notag\\[0.3em]
                                                   &\leq r +
                                                     \bar{\epsilon} +
                                                     \hat{\epsilon} \leq  \tau
                                                     \epsilon^\theta \label{eq:earlyterm}
\end{align}
and, hence, sufficient progress on $J$ has been made-- $\mu_l$ satisfies
\eqref{eq:fwdesiredresid}. Thus, we need only analyze the case where early
termination in Algorithm \ref{alg:frankwolfe} does not occur and $l = k$
\eqref{eq:algoerrors}.

If $l=k$, then $s_i > r$ for all $i \leq k$ and, by extension,
$g_{i-1} > r$ for all
$i \leq k$ since $s_i$ is a lower bound for
$g_{i-1}$. In this case, the
definitions of $\hat{\epsilon}$ and $r$ \eqref{eq:algoerrors} imply
$C \geq 1 - \alpha /\left(2(1+\alpha)\right)$ and the choices for $\beta_2$
and $\beta_3$ \eqref{eq:algostepsz} provide
\begin{equation*}
  t \leq \min \left(\frac{\alpha}{2(1+\alpha)}, \frac{\left(1- \alpha/2 \right)^{1/\alpha }}{D^{1/\alpha}}\right)
\end{equation*}
This gives
\begin{equation*}
   C - t - \frac{D}{(1+\alpha)}t^\alpha \geq (2 \alpha ^*)^{-1}
\end{equation*}
from which substitution into \eqref{eq:intermederrorub} yields
\begin{align}
  r\left(\mu_i\right) &\leq r\left(\mu_{i-1}\right) -
  \frac{t}{2L \alpha^*}g_{i-1} + \zeta_i \notag \\[0.5em]
  &\leq r\left(\mu_{i-1}\right) -
  \frac{\delta_i}{2 \alpha^*}g_{i-1} + \zeta_i \notag \\[0.5em]
  &\leq r\left(\mu_{i-1}\right) -
  \frac{\delta_i}{4 \alpha^*}g_{i-1}\label{eq:dlterrorub}
\end{align}
where the last inequality is a result of the definition of
$\widetilde{\epsilon}$ \eqref{eq:algoerrors}, $\zeta_i$, and
$g_{i-1} > r$. As $\delta_i$
is the minimum of three different terms \eqref{eq:deltadef},
\eqref{eq:dlterrorub} will be used to analyze the amount of progress, that is
made on the objective $J$, corresponding to each of these three terms. Note, the
following identities that will be used in the analysis of each term:
\begin{equation}
 \left(1- \frac{\alpha}{2}\right)g_{i-1} \leq g_{i-1} - \frac{\alpha r}{2} \leq  g_{i-1} - \bar{\epsilon} \leq s_i \label{eq:dltcases}
\end{equation}
and
\begin{align}
  -g_{i-1}^p &\leq -\left(\norm{F(\mu_{i-1})}_{L^2(\mu_{i-1})} -
                                      \hat{\epsilon}\right)^{p} \notag\\[0.2em]
  &\leq -
  \left(1 - \frac{\alpha }{2+\alpha}\right)^{p}\norm{F(\mu_{i-1})}^{p}_{L^2(\mu_{i-1})}
  \leq - \frac{1}{2e}\norm{F(\mu_{i-1})}^{p}_{L^2(\mu_{i-1})}\label{eq:gnormstabil}
\end{align}
for all $1 \leq p \leq \alpha ^*$. The relation \eqref{eq:dltcases} simply
observes that $s_i$ is a multiplicative approximation to
$g_{i-1}$ in Algorithm \ref{alg:frankwolfe}, while
\eqref{eq:gnormstabil} is a consequence of $r - \hat{\epsilon} \leq \norm{F
  \left(\mu_{i-1}\right)}_{L^2(\mu_{i-1})}$.

First, consider the case where $\delta_i = \beta_1$. Substitution into
\eqref{eq:dlterrorub}, coupled with \eqref{eq:gnormstabil}, provides
\begin{equation}\label{eq:preplineq}
  r\left(\mu_{i}\right) \leq r\left(\mu_{i-1}\right) -
  \frac{\beta_1}{8e \alpha ^*}\norm{F(\mu_{i-1})}_{L^2(\mu_{i-1})}
\end{equation}
Applying \eqref{eq:lojaineq} to \eqref{eq:preplineq} and defining $r_i := r\left(\mu_i\right)$ (for the sake of notation) yields
\begin{equation}\label{eq:dec1}
  r_i \leq r_{i-1} - \kappa^{(1)} r^\theta_{i-1} \hspace{ 0.2in } \text{for} \hspace{
    0.2in } \kappa^{(1)} := \omega\beta_1
\end{equation}
for the constant $\omega = (8e \alpha^*)^{-1}\tau $. In the cases
\eqref{eq:deltadef} corresponding to $\beta_2$ and $\beta_3$, similar
applications of the previous identities (along with \eqref{eq:dltcases}) give
\begin{align}
  r_i \leq r_{i-1} - \kappa^{(2)}r_{i-1}^{2 \theta} \hspace{ 0.2in } &\text{for} \hspace{
    0.2in } \kappa^{(2)} := \omega \tau(1-\alpha/2) \beta_2 \label{eq:dec2}\\[0.3em]
  r_i \leq r_{i-1} - \kappa^{(3)}r_{i-1}^{\alpha^*\theta}
\hspace{ 0.2in } &\text{for} \hspace{
                   0.2in } \kappa^{(3)} := \omega \left(\tau(1- \alpha/2)\right)^{1/\alpha}\beta_3 \label{eq:dec3}
\end{align}
Now, for the sake of notation, define the function
\begin{equation*}
  z(u,v) := u^{-1}\epsilon^{-(1-v)_-}\left(r_0\log^{1/(1-v)}(r_0 / \epsilon)\right)^{(1-v)_+}
\end{equation*}
where $(\cdot)_+ $ and $ (\cdot)_-$ denote the positive and negative parts. Using Lemma \ref{lem:seqconv}, it
follows that, if \eqref{eq:dec1} occurs for more than
$\omega^{-1}z(\beta_1, \theta)$ iterations of Algorithm \ref{alg:frankwolfe},
then $r_k \leq \epsilon$, where $k$ is the index of the last loop iteration in Algorithm
\ref{alg:frankwolfe}. Similar deductions for \eqref{eq:dec2} and \eqref{eq:dec3}
lead to the conclusion that, if
\begin{equation}\label{eq:iterationlb}
  k \geq \omega^{-1}\left( z(\beta_1, \theta) + z(\tau(1-\alpha/2) \beta_2, 2\theta) +
    z(\left(\tau(1-\alpha/2)\right)^{1/\alpha} \beta_3, \alpha^*\theta)\right) := M
\end{equation}
then either \eqref{eq:dec1}, \eqref{eq:dec2}, or \eqref{eq:dec3} has occurred
sufficiently many times during the execution of Algorithm \ref{alg:frankwolfe}
to guarantee $r_k \leq \epsilon$. As $k$ has been chosen exactly so that
$k = \left \lceil M \right \rceil$ \eqref{eq:algoerrors}, one obtains that
$\mu_k$ satisfies \eqref{eq:fwdesiredresid}. The desired complexity bound
\eqref{eq:fwcomplexity} on $M$ now follows by plugging in for
$\beta_1, \beta_2,$ and $\beta_3$ in \eqref{eq:iterationlb} and then, taking
asymptotic estimates as $\epsilon \to 0$; the term
$z(\left(\tau(1-\alpha/2)\right)^{1/\alpha} \beta_3, \alpha ^* \theta)$ clearly
dominates.

To obtain the stated sample complexities of Theorem \ref{thm:outerconvres}
notice that each iteration requires sampling from $\mu_{i-1}$ to estimate both
$s_i$ and $\mu_{i}$. Computing $s_i$ is a simple mean estimation task and can be
performed (with the necessary accuracy $\bar{\epsilon}$) using
$\widetilde{O}\left(\epsilon^{-2 \theta }\right)$ samples. From Corollary
\ref{cor:fwcompv2}, recall that a $\lambda_i$ such that
$\pi_{\lambda_i, \mu_{i-1}}$ \eqref{eq:pilamdef} yields $\mu_i$ can be computed
using $O(\zeta_i^{-2})$ samples from $\mu_{i-1}$-- where
$\zeta_i = \delta_i \widetilde{\epsilon}$. Utilizing the definition of
$\widetilde{\epsilon}$ and the previously computed lower bounds on $\delta$, it
follows that sample access to $\mu_i$ can be obtained with
$O\left(\zeta_i^{-2}\right) = O\left( \epsilon^{-2 \alpha ^* \theta}\right)$
samples from $\mu_{i-1}$. Clearly, this dominates the number of samples required
to estimate $s_i$ since $\alpha ^* \geq 2 $. Thus, an iteration of Algorithm
\ref{alg:frankwolfe} requires $O\left( \epsilon^{-2 \alpha ^* \theta}\right)$
samples from $\mu_{i-1}$.

To reduce this to a sample complexity in terms of $\mu_0$, notice that (provided a computed $\lambda_{i-1}$ and $\phi_{\mu_{i-2}}$) a
draw from $\mu_{i-1}$ can be obtained using accelerated gradient descent and a draw
from $\mu_{i-2}$; in only $O(\log\epsilon^{-1})$ gradient evaluations of
$\phi_{\mu_{i-1}}$ (see the proof of Proposition \ref{prop:sorcalge} for this
analysis). Chaining this procedure, it follows that, if $\lambda_j$ has been
computed for all $j \leq i-1$, a sample from $\mu_{i-1}$ can be produced using a
sample from $\mu_0$ and $O(i\log\epsilon^{-1})$ total gradient evaluations.
Hence, each iteration of Algorithm \ref{alg:frankwolfe} can be performed using
$O\left( \epsilon^{-2 \alpha ^* \theta}\right)$ samples from $\mu_0$.
\endproof
\section{Geodesic convexity and {\L}ojasiewicz inequalities}
\begin{lemma}\label{lem:linearlb}
  If $J : \Pc_2(\R^{d}) \to \bar{\R}$ is geodesically convex
  (Definition \ref{def:geoconv}) and Wasserstein differentiable (Definition
  \ref{def:wassdiff}) then the Wasserstein gradient field $F : \Pc_2(\R^{d}) \to
  \ctangt_{\Pc_2(\R^{d})}$ satisfies
  \begin{equation}\label{eq:geoconlb}
    J(\mu) + \int_{\R^{d}  \times \R^{d}} F(\mu;x)^T(y-x)  \, d \gamma(x,y) \leq J(\nu)
  \end{equation}
  where $\gamma \in \Pi(\mu, \nu)$ is any optimal transport plan between $\mu$
  and $\nu$. Consequently, if $J$ has a bounded (with respect to $\Wc$)
  level set with diameter $R$
  \begin{equation}\label{eq:bndlevel}
    \diam \left(Q_p\right) \leq R,\hspace{ 0.2in }
    Q_p := \left\{ \mu \in \Pc_2(\R^d) : J(\mu) \leq p \right\}
  \end{equation}
  then $J$ satisfies a {\L}ojasiewicz inequality \eqref{eq:lojaineq} on $Q_p$
  with parameters $\tau = R^{-1}$ and $\theta = 1$.
\end{lemma}
\proof
Let $\gamma \in \Pi(\mu, \nu)$ and let $\mu_t$ be the constant-speed geodesic \eqref{eq:geobiject} corresponding to $\gamma$. Rearranging the definition of geodesic convexity \eqref{eq:geoconv}, one obtains
\begin{equation*}
  \frac{J(\mu_t) - J(\mu)}{t} \leq J(\nu) - J(\mu)
\end{equation*}
Taking the limit as $t \to 0$ and applying \eqref{eq:wassderiv} provides
\eqref{eq:geoconlb}. To obtain a {\L}ojasiewicz inequality on a bounded level
set $Q$, simply recognize that \eqref{eq:geoconlb} and Cauchy-Schwarz imply
  \begin{equation*}
    J(\mu) -  J(\nu)\leq \norm{F(\mu)}_{L^2(\mu)}\Wc(\mu, \nu)
  \end{equation*}
  for any $\mu, \nu \in Q$.
\endproof

\printbibliography % if more than one, comma separated

\end{document}